\documentclass[11pt,a4paper]{article}
\usepackage{mathrsfs}
\usepackage{amsmath}
\usepackage{amssymb}
\usepackage{dsfont}
\usepackage{esint}

\usepackage{amsmath,amssymb,amsfonts,a4wide,graphicx,bm,times,psfrag,wrapfig,sidecap}
\usepackage{cite}
\usepackage[colorlinks=true,linkcolor=black, citecolor=black,
urlcolor=black]{hyperref}
\numberwithin{equation}{section}
\makeatletter \let\old@startsection=\@startsection
\renewcommand{\@startsection}[6]
{\old@startsection{#1}{#2}{#3}{#4}{#5}{#6\mathversion{bold}}}
\makeatother

\def\<{\langle}
\def\>{\rangle}
\def\sign{\rm sign}
\def\tr{{\rm   tr} }
\def\Im{{\rm Im}}
\def\Re{{\rm Re}}
\newcommand\encadremath[1]{\vbox{\hrule\hbox{\vrule\kern8pt
\vbox{\kern8pt \hbox{$\displaystyle #1$}\kern8pt}
\kern8pt\vrule}\hrule}} \def\enca#1{\vbox{\hrule\hbox{
\vrule\kern8pt\vbox{\kern8pt \hbox{$\displaystyle #1$} \kern8pt}
\kern8pt\vrule}\hrule}}
  \usepackage{bm}

\begin{document}

\thispagestyle{empty}

\begin{flushright}
  
\end{flushright}

\vspace{1cm}
\setcounter{footnote}{0}

\begin{center}

{\Large\bf Classical and Quantum Integrability in Laplacian Growth}

\vspace{20mm}

Eldad Bettelheim   \\[7mm]

Racah Inst.  of Physics, \\Edmund J. Safra Campus,
Hebrew University of Jerusalem,\\ Jerusalem, Israel 91904 \\[5mm]

\end{center}

\vskip9mm

\vskip18mm

\noindent{ We review here particular aspects of the connection between Laplacian growth problems and classical integrable systems. In addition, we put forth a possible relation between quantum integrable systems and Laplacian growth problems. Such a connection, if confirmed, has the potential to allow for a theoretical  prediction of the fractal properties of Laplacian growth clusters, through the representation theory of conformal field theory.     }

\newpage
\tableofcontents

\section{Introduction}
\subsection{Laplacian Growth}The phenomenon of Laplacian growth constitutes one of the basic paradigms for the appearance of fractal patterns in physical systems out of equilibrium. Laplacian growth models may be encountered in natural settings, in engineering applications and in the lab. To introduce Laplacian growth we describe briefly an instance of it found in the latter settings,  in laboratory experiment involving a Hele-Shaw cell. 

\begin{figure}[h!!]
\centering
\includegraphics[width=0.75\columnwidth]{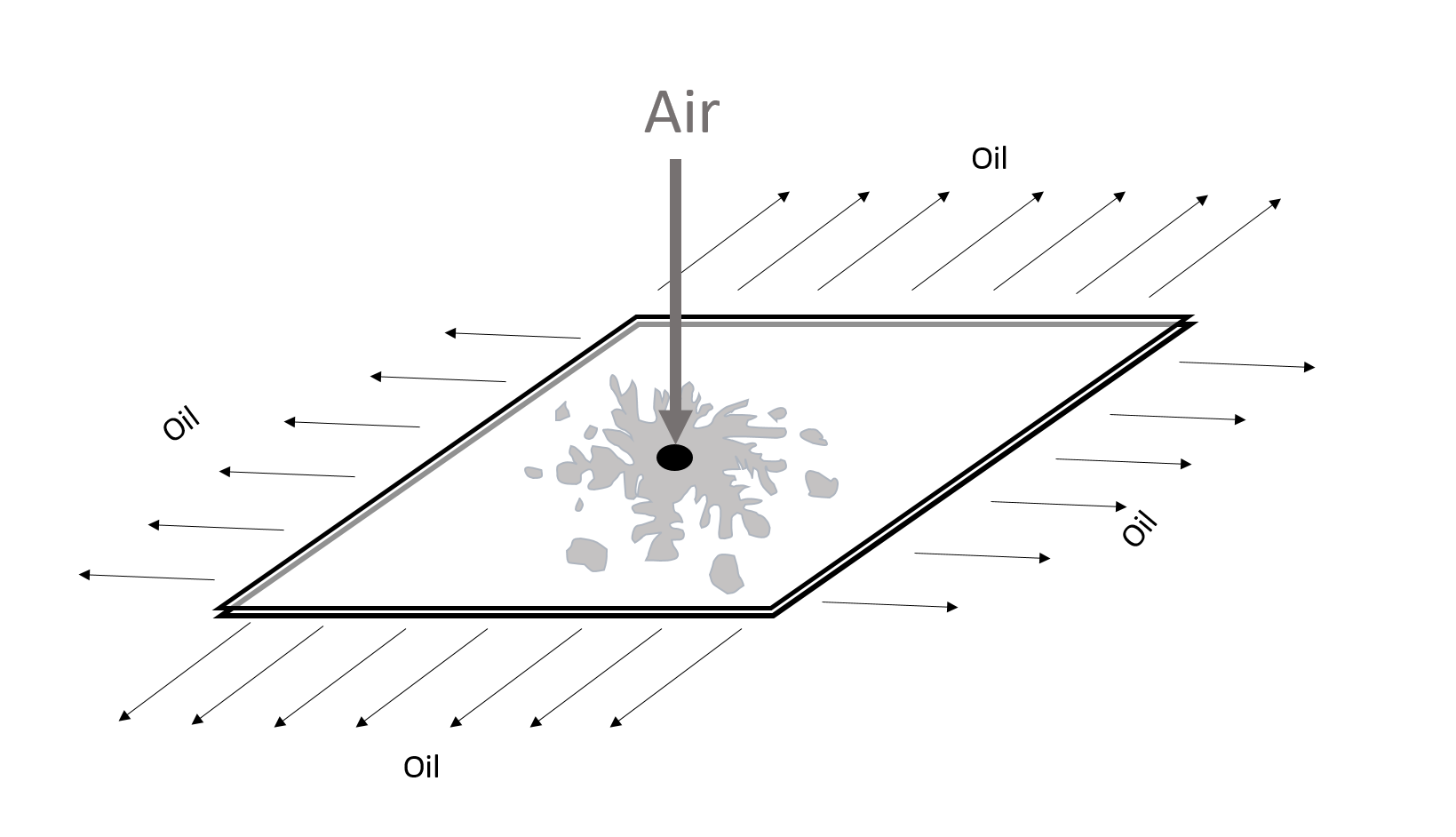}
\caption{A low viscosity fluid, say air, is pumped through a hole (black circle) in one of the glass plates constricting the gap of the Hele-Shaw cell, which is otherwise filled with a high viscosity fluid, say oil. As air is pumped, oil escapes through the sides of the cell, while a droplet with an intricate interface is formed. Depending on the protocol of injection, the interface may become multi-connected.   \label{Syringe}}
\end{figure}

A Hele-Shaw cell (see Fig.  \ref{Syringe}) is composed of two glass plates fixed in place such that a small gap, which may be filled with fluid, is allowed to form between them. One of the glass plates contains a hole. The glass plates are horizontal. To obtain a Laplacian growth pattern between the two plates a
high viscosity fluid is allowed to entirely  fill the gap between the plates. Then, fluid of low viscosity is  injected through the hole. The two fluids are immiscible. As the low viscosity fluid is injected, a bubble of the low viscosity fluid expands within the ambient high viscosity fluid, a portion of which, in turn, escapes the Hele-Shaw cell through the cell's perimeter.

The interface between the low viscosity bubble and the high viscosity fluid surrounding it  is seen to possess a intricate fractal pattern, which was intensely studied numerically and analytically. The interface  consists of fingers on different scales, somewhat reminiscent of a snow flake, though much less regular. The fractal dimension of the interface  has been established numerically\cite{Procaccia:Levermann:Correct:HEle-Shaw:D} and experimentally\cite{Swinney:Praud:Hele:Shaw:1.71}, and is found to be, say, $1.71\pm0.03$. In addition, the multi-fractal nature of the so-called harmonic measure of these clusters have been studied \cite{Duplantier:Halsey:Honda:Multifractal:DLA,Duplantier:Potential:Theory}. 

The rate of injection is immaterial to the shapes that appear, and as such we may assume this rate to be constant. Since then the area inside the interface increases linearly, we may measure time in units of the area of the low viscosity droplet. We may thus henceforth talk of the phsycal time and the area of the low viscosity droplet interchangeably. 

A theoretical prediction of the fractal dimension and other fractal properties of the cluster has long been at the focus of a fairly large body of scientific research (although interest may have waxed and waned). The purpose of the current article is twofold. First, we wish to review an approach to tackle the problem based on the relation of Hele-Shaw growth to classical integrability. Second, we wish to propose a tentative new direction of research based on quantum integrability. 

As a definition of the fractal dimension of the interface we shall take use relation between the linear size of the droplet $R$ and the area of the low viscosity droplet $t$. We define the fractal dimension through the observed power law relation between the two:
\begin{align}
R^D=t.\label{Ddef}
\end{align}
We leave open the question of how to define of the linear size of the droplet, $R$. In addition we shall assume that the relation (\ref{Ddef}) holds sharply, namely that $D$ defined above does not fluctuate. 

\subsection{The Hele-Shaw Probability Density}
One of the main goals within the theory of the Hele-Shaw problem is to find the  function giving the probability density to find Hele-Shaw interfaces of different shapes. Let us define $t(\mathcal{C})$ as the area inside the interface $\mathcal{C}$, and the probability distribution function , $  P(\mathcal{C})$, as the probability density to find an interface $\mathcal{C}$ at time $t(\mathcal{C})$, given some probability distribution on the initial condition   $P_0(\mathcal{C})$. Let $\mathcal{C}^{\Delta t}$ be the the interface $\mathcal{C}$  evolved  a time $\Delta t$. Namely, if $\mathcal{C}$ has area $t$, then  $\mathcal{C}^{\Delta t}$ has area $t+\Delta t$. The function $ P$ is defined as follows:
\begin{align}
 P(\mathcal{C})=  P_0(\mathcal{C}^{-t(\mathcal{C})})\label{PdefFromP0}
\end{align}
We shall specify the measure $d\mathcal{C},$ with respect to which the probability density $P(\mathcal{C})$ is defined, but shall only demand that it is time translation invariant, $d\mathcal{C}=d\mathcal{C}^{\Delta t}$ . Indeed, Eq. (\ref{PdefFromP0}) makes little sense otherwise.

 The probability density $P$  is time translation invariant. Indeed,
from (\ref{PdefFromP0}) one may easily show\begin{align}
P(\mathcal{C}^{\Delta t})=P(\mathcal{C}). \label{PTimeTranslation}
\end{align}   
The function  $P$ possesses  scale invariance, as well. Let us define, with some abuse of notations, $P(R)$ as follows:
\begin{align}
P(R)=\int P(\mathcal{C})\delta(R-R(\mathcal{C})) d\mathcal{C},
\end{align}
where $R(\mathcal{C})$ is the linear size of the interface. For Laplacian growth we assume Eq. (\ref{Ddef}) holds which, we shall see, leads to:
\begin{align}
P(R)dR= dR ^D \label{PScaleInvariance}.
\end{align}
Indeed,\begin{align}
&P(R)=\int P(\mathcal{C})\delta\left(R-t^{1/D}(\mathcal{C})\right) {d\mathcal{C}} =\\&=DR^{D-1}\int   P(\mathcal{C})\delta\left(R^D-t(\mathcal{C})\right)   d\mathcal{C}= \\
 &=DR^{D-1},    
\end{align}
where the last equality follows from the fact that the measure is time translation invariant.

We note that the time translation invariance of the probability density means that the probability measure can be thought as a probabilistic uniform superposition of the interface at different times. That the probability density is also scale invariant is a demonstration of the fractal property of the interface, namely Eq. (\ref{Ddef}), such the two properties together define the problem of finding the probability density which encodes the fractal dimensions.

A further important comment is that in searching for the probability density obeying scale and time translation invariance,  the problem may be somewhat generalized
in the following sense. The droplet in the Laplacian problem remains singly connected if the  injection of the low viscosity fluid is always fast enough to rule out surface tension effects. However, if the protocol of injection is altered the droplet may become multi-connected. Imagine, for example taking a large droplet, with a fully-developed fractal structure over many scales. The interface is self similar in that if the low viscosity fluid is extracted, allowing the droplet to shrink, but we re-scale the resulting interface by a proper factor, one obtains the original droplet. Assume now that before extracting the low viscosity fluid we wait some time and allow surface tension to change the shape. This has the effect of   coarsening the shape on the lowest scales as well as allowing for the possibility  for the droplet to  become multi-connected. Then some  low viscosity fluid is quickly extracted. We may still expect that the self-similarity, though perhaps in  a weaker sense, is preserved. It seems then that the single-connectedness of the droplet should not be imposed as a strict condition. As a result we  consider probability densities which give non-zero measure to multi-connected droplets. 

\subsection{Preview of the Paper}

The fact that the time translation  invariance property of the probability distribution function, Eq. (\ref{PTimeTranslation}), coexists with the scale invariance property, Eq. (\ref{PScaleInvariance}), is equivalent to the scaling law $R^D(\mathcal{C}) = t(\mathcal{C})$. We may formulate the problem of determining $D$, the fractal dimension, as finding the scaling law (\ref{PScaleInvariance}) for a probability distribution function obeying the time translation invariance property, Eq. (\ref{PTimeTranslation}).  The current paper puts forward the proposition that candidates for the probability distribution function may be found within the realm of quantum integrable systems and conformal field theory  . 

Let us briefly recount the connection between fractal structures and conformal field theory. The context in which the two concepts may be most successfully  linked is  within the framework of equilibrium two dimensional critical phenomena. There, scale invariance, which is a consequence of the divergence of correlation lengths leads to conformal invariance, and it is assumed that, at the critical point, the statistical mechanics problem is described by conformal field theory. The vacuum is a Gibbs state in which fractal objects, such as cluster boundaries, exist at all length scales. A fractal cluster may then be pinned to a particular point by applying a scaling field at that point\cite{Duplantier:Brother}. The dimension of the scaling field is determined by the power-law dependence of the scaling field on the small scale cut-off. Since this cut-off also determines the probability that the fractal can be found and pinned to  a small box of linear size given by the cut-off scale,  and this in turn is related to the notion of box-counting fractal dimension, there is a relation between the dimension of the scaling field and the fractal dimension of the object being pinned by it. 

Non-equilibrium problems, such as Laplacian growth, are unlikely to be linked in the same manner to conformal field theory. Indeed, it does not seem a reasonable assumption that the Laplacian growth interface may be pinned to a point by applying, to some vacuum containing already many fractal structures on all scales, a scaling field of conformal field theory. Such a scenario would mean that the fully developed large area interface can be picked  at once from a Gibbs state without a need to apply the nonlinear evolution from an initial interface. In other words, such a scenario would mean that there is a representation of the interface as an equilibrium object. 

Nevertheless, a connection to conformal field theory is an  appealing prospect, due to the fact that conformal field theory is inherently scale invariant, just as the fractal dimension of the interface is, and due to the fact that such a connection would alow for tools to predict the fractal dimensions, since conformal field theory naturally supplies a discrete spectrum of possible scaling dimensions, these in turn possibly determining the fractal dimension. 

Here we propose that instead of basing the connection  to conformal field theory on the notion of  scale invariance being promoted to conformal invariance, a relation to conformal field theory may be established making use of the fact that conformal field theory has a hidden integrable structure, a structure which can be viewed as a  quantization of the integrable structure hidden in Laplacian growth. At present, we are only able to make this connection within some limit, where only a small part of the interface, which happens to be long and narrow, is focused upon. Within this limit, the integrable structure of Laplacian growth becomes that of the Korteweg-de Vries equation, the quantum analogue of which is  the integrable structure hidden within conformal field theory. 

To flesh out this connection, we review first the integrable structure within Laplacian growth\cite{43:Zabr:Wiegm,Krichever:Mineev:Weinstein:Wiegmann:zabrodin:Laplacian:Growth:Whitham,MapsandToda,Richardson1972} in section \ref{LGClassicalIntegrability}. This integrable structure turns out to be the dispersionless limit of  a reduction of the two dimensional Toda lattice. Since we are eventually interested in the quantum integrable system, we must discuss how this dispersionless limit is obtained from a fully quantum system. To this end section \ref{AlgebraicGeometricLPG} and \ref{DispersiveQuantization} describe how the dispersionless limit is removed to obtain the full but reduced and classical two dimensional Toda lattice. The next logical step would be to pass to the quantum problem. However, before discussing how to  quantize the problem, we pass, on a classical level, from the two dimensional Toda lattice  to the Korteweg-de Vries equation\cite{KdVPaper}. This is done  in section \ref{KdVLimit}, and includes focusing on a long and narrow finger of the Laplacian growth. In section \ref{qKdVSection} we discuss the quantization of the problem\cite{Zamolodchikov:KdV:I,Zamolodchikov:KdV:II,Zamolodchikov:KdV:III,Fateev:Lukyanov:Poisson-Lie:PregenatorToqKdV,Fateev:Lukyanov:Vertex:PregenatorToqKdV}, which leads us to  the quantum Korteweg de-Vries equations. This quantum problem  lies at the heart of conformal field theory. This fact allows us, in section \ref{QCDostrbution}, to show how a certain  object, familiar within the framework of conformal and integrable field theory, may describe the classical Laplacian growth problem, through the probability distribution function $P(\mathcal{C})$. Section \ref{Discussion} contains  concluding remarks.

\section{Classical Integrability in Laplacian Growth\label{LGClassicalIntegrability}}
Here we review some of the mathematical structures underlying the problem of Laplacian growth. This includes the description through conformal maps and classical integrability\cite{43:Zabr:Wiegm,Krichever:Mineev:Weinstein:Wiegmann:zabrodin:Laplacian:Growth:Whitham}.
\subsection{Description Through Conformal Maps}Let us consider the Hele-Shaw setup, which was already described in the introduction. The  ambient, high viscosity fluid is described by a velocity field $\vec{v}$. This velocity is the average velocity across the gap between the two glass plates. Due to the high viscosity of the fluid, one must balance the internal viscous friction force experienced by the fluid, with the force supplied by the pressure, neglecting inertia, which is small because of the small separation of the plates. As the force of friction scale with the average velocity across the gap  $\vec{v}$ , with the magnitude of the viscosity, $\mu$, and inversely with the size of the gap squared, $\frac{1}{b^2}$ one obtain Darcy's law:
\begin{align}
\frac{12\mu}{b^2}\vec{v} =- \vec{\nabla}P.
\end{align}
The factor $12$ may be obtained by a more detailed, but basic calculation, which takes into account the no-slip conditions at the interface between the fluid and the glass plates, which results in a parabolic velocity profile.

The dynamics are dictated by the following conditions.  The motion of the interface between the inviscid and viscous fluid are determined by the normal velocity at the interface, which is proportional to the normal derivative of the pressure at the interface:
\begin{align}
v_n=-\frac{b^2}{12\mu}\partial_n P.\label{DarcyLaw}
\end{align}

Incompressibility, $\vec \nabla\cdot \vec v=0$,  requires, that there exists a real stream function, $\varphi$:
\begin{align}
\vec{v} =\frac{Q}{2\pi}  \hat z\times \vec{\nabla}\theta,\label{StreamFunction}
\end{align}
where $\hat z$ is the out-of-plane unit vector and $Q$ is the volume flux, namely the volume of fluid injected into the cell in unit time, inserted here for convenience, as to make $\varphi$ dimensionless.  

Defining the complex function $v$ by $v=v_x+\imath v_y$, and taking $\partial$, and $ \bar \partial$ to be a derivative with respect to $z$\ and $\bar z$, respectively ($\partial = \frac{\partial_x - \imath \partial y}{2},$ $\bar\partial = \frac{\partial_x + \imath \partial y}{2}$), we may write the relations between the pressure, the stream function and the velocity as:
\begin{align}
v=-\frac{b^2}{6\mu} \bar\partial P=\frac{Q}{\pi} \imath   \bar\partial\theta\label{vintermsofPandPhi}
\end{align}
We may define a complex analytic potential $\varphi_t$ (where in $\varphi_t$ the index denotes explicitly the time dependence):
\begin{align}
\varphi_t(z) = \theta -\imath\frac{\pi b^2}{6Q\mu}P.\label{psiDef}
\end{align}
Indeed, relation (\ref{vintermsofPandPhi}) suggests that $\varphi_t(z)$  is analytic $\bar \partial \varphi_t=0,$ Taking the complex conjugate of (\ref{vintermsofPandPhi}) also leads, after simple algebra, to:
\begin{align}
\bar v= -\frac{\imath Q}{2\pi} \partial \varphi_t.\label{barvispartialphi}
\end{align}

The pressure is constant inside the droplets of lower viscosity, since the fluid there has neither inertia (due to the narrowness of the gap) nor viscosity (by assumption).   We may assume this pressure to be zero. This means that we assume a type of dynamics in which, even after the low viscosity droplet breaks up into several droplets, the same pressure is maintained across the different droplets. This scenario cannot be attained in a Hele-Shaw cell, since only one of the droplets will be connected to the source of low viscosity fluid through the hole. One may imagine instead of a glass plate using a plate made of a material permeable to the low viscosity fluid but impregnable to the high viscosity fluid. The low viscosity fluid may then be supplied through the entire plate, and the different droplets will maintain the same pressure, which without loss of generality may be set to zero, since only gradients of the pressure play any role in the dynamics.  

We now consider the analytic properties of  $\varphi _t$. First note that we assume that the fluid is syphoned off in a radially symmetric fashion around infinity, which suggests $v(r) \sim \frac{ Q \hat r}{2\pi r}$ as $r\to\infty$. This leads to the following asymptotic behavior of $\varphi_t:$
\begin{align}
\varphi _t(z) \sim-\imath  \log(z), \quad \mbox{as }z\to\infty,\label{psiAsymptote}
\end{align} 
which leads to the fact that the imaginary part of $\varphi+t$, namely the stream function $\varphi$ winds by $2\pi$ as one encircles infinity.
In fact $\varphi_t$ is only well-defined if a branch cuts are introduced in the complex plane, such that from each droplet a branch cut extends to infinity.  This can be understood by considering that the winding by  $2\pi$  of $\varphi_t$  around infinity has its origin in the winding of $\theta$ around the droplets. Indeed, consider the integral over the velocity around the droplet. From simple  hydrodynamical kinematics, this integral is equal to volume flux of fluid entering the cell, on the one hand, and from the definition of the stream function, Eq. (\ref{StreamFunction}),  it is equal to the integral over the gradient of $\theta$ around the droplets:
\begin{align}
Q=\oint \vec{v} \times d\vec r=\frac{Q}{2\pi}    \oint \vec\nabla\theta\cdot d\vec r,
\end{align}
leading to the conclusion that the winding of $\varphi$ around the droplets is $2\pi$:
\begin{align}
\oint \vec \nabla\theta = 2\pi.
\end{align}
The the variable $\theta$ can thus be used as a coordinate along the droplets, whose range is $[0,2\pi]$. 

A central role will be given to the inverse function of $\imath\varphi_t$, which we shall denote by $f_t$:
\begin{align}
f_t(\imath\varphi_t(z))=z
\end{align}
$f_t(\imath\varphi)$ is defined such that for $\varphi\in[0,2\pi]$, the function $f_t(\imath\varphi)$ gives the complex coordinate of a point on the interface of the droplets, with given  $\theta$ and at time $t$.
The function  $f_t$ is defined with period  $\imath 2\pi$:
\begin{align}
f_t(\imath\varphi+2\pi \imath) = f_t(\imath\varphi), 
\end{align}
 For complex $\varphi$, the function $f_t(\imath\varphi)$ is defined by analytic continuation.   As a result of the periodicity, the function $f_t$ is defined on a cylinder, $\mathcal{S}$, where points displaced by $2\pi\imath$ are identified, $\mathcal{S}=\mathds{C}/2\pi\imath \mathds{Z}$. The function $f_t(\imath\varphi)$  has the following expansion at $\Im(\varphi) \to -\infty:$
\begin{align}
f_t(\imath\varphi) =a_1^{(1)} e^{\imath \varphi}+a_0^{(1)}+a_{-1}^{(1)}e^{-\imath\varphi}+\dots=\sum_{j=-1}^{\infty} a^{(1)}_{-j} e^{-\imath j\varphi}, \label{mapExpansion}
\end{align} 
which results from (\ref{psiAsymptote}). 

Since the  interface is described by the map $f_t(\imath\varphi):[0,2\pi]\to z$ and since $f_t$ is periodic in $2\pi\imath$, there is an arbitrariness in choosing $f_t(\imath\varphi)$ given the shape of the interface. Indeed the functions $f_t(\imath\varphi)$ and $f_t(\imath(\varphi+c))$ describe the same interface,  the only difference being that the reference point on the interface where the angle $\theta$ is considered as $0$ is moved from $f_t(0)$ to $f_t(\imath c)$. Thus, when describing the dynamics of the interface as the dynamics of $f_t$, one has to bear in mind that the dynamics of $f_t$ which are equivalent to shifting the reference point of $\varphi$ are meaningless and arbitrary. It is common in the literature to remove this arbitrariness by requiring $a_1^{(1)}$ to be real. We shall not follow that convention, and rather leave this arbitrariness in place.    

 An useful mathematical object to study interfaces in two dimensions is is the Schwarz function, $S(z)$ -- the an analytic continuation of a function giving   $\bar z$ on the boundary. Namely for $z\in\mathcal{C}$, we have $S(z)=\bar z$, with $S(z)$ being analytic in a neighborhood of $\mathcal{C}$. One may easily ascertain that the Schwarz function can be written as 
\begin{align}
S(z) = \bar f_t(-\imath\varphi _t(z)),\label{SchwarzDef} 
\end{align}  
where, as usual, for any function $g(z)$, the function $\bar g(z)$ is defined as $\overline{g(\bar z)}$. To see that (\ref{SchwarzDef}) indeed gives the Schwarz function, consider that for $z$ on the boundary, $\varphi_t(z)$ , is purely real and so,  $\bar f_t(-\imath\varphi_t(z))=\overline {f_t(\imath\varphi_t(z))}=\bar z.$  The Schwarz function, $S(z),$ obeys the `unitarity' condition:
\begin{align}
\bar S(S(z))=z.\label{Unitarity}
\end{align} 
This relation is proven by considering it on the interface, where it simply follows from the fact that $S(z)=\bar z$\ on the interface, and then trivially analytically continuing the relation away from the interface.

\subsection{Richardson Moments}As was found by Richardson \cite{Richardson1972},  during the course of the evolution an infinite set of parameters, namely the Richardson moments remain constant. The $k$'th Richardson moment is defined as:
\begin{align}
t_k = \int_{\mathcal{D}_+} \frac{dz\wedge d\bar z}{2\pi\imath  k z^k},\label{tkAsAreaIntegral}
\end{align}
where $\mathcal{D}_+$ denotes the domain which is exterior to the droplet. To see that these are conserved consider $\dot t_k$:
\begin{align}
\dot t_k =\hat{z}\cdot \oint_{\mathcal{C}} \frac{d\vec{r} \times \vec{v}}{2\pi\imath  kz^k}.
\end{align}
We may write  $\hat z \cdot d\vec{r}\times \vec{v} =\frac{\imath}{2}(dz \bar v - d\bar z v)$. We may re-write the right hand side of this equation using (\ref{barvispartialphi}) as  $\frac{Q}{4\pi} (dz \partial \varphi +d\bar z\bar \partial \varphi)   =\frac{Q}{4\pi} d\varphi_t$, one obtains:
\begin{align}
\dot t_k =\frac{Q}{8\imath \pi^2 k}  \oint\frac{d\varphi_t}{z^k},
\end{align}
As the integrand is analytic in $\varphi_t$, the contour of integration may be deformed to a large contour surrounding infinity in the $z$-plane, where the asymptotic relation $\varphi_t \sim-\imath  \log z$ holds. This yields: 
\begin{align}
\dot t_k =\frac{Q}{8\imath \pi^2 k}\oint_\infty\frac{dz}{z^{k+1}}  =0.
\end{align}

Note also that the Schwarz function encodes the harmonic moments. Indeed, Stokes theorem states that (\ref{tkAsAreaIntegral}) can be written as a contour integral over the boundary of the droplets $t_l=\oint_{\mathcal{C}} \bar z z^{-l}\frac{dz}{2\pi\imath  l} $ and since one the contour $S(z)=\bar z$, this can be written as:
\begin{align}
t_l = \oint S(z)z^{-l} \frac{dz}{2\pi\imath  l}, \quad t=\oint S(z)\frac{dz}{2\pi\imath }.\label{tlAsContouIntegral}
\end{align}

One may construct from knowledge of the Richardson moments, $t_k$,  and the area of the droplets, $t$  its shape, given some additional data, such as the number of connected components. The Richardson moments are thus reminiscent of the conserved quantities of completely integrable systems, since the knowledge of which is enough to determine completely the dynamics.  The results of Refrs. \cite{43:Zabr:Wiegm,Krichever:Mineev:Weinstein:Wiegmann:zabrodin:Laplacian:Growth:Whitham,MapsandToda} show, however, that the actual integrable structure behind the zero surface-tension Hele-Shaw problem treats the Richardson moments rather as the `times' conjugate to the conserved quantities, or `Hamiltonians', which may be denoted by $H_k$. In addition it is appropriate to treat the complex conjugate of the Richardson moments (which are themselves complex quantities), namely the quantities $\bar t_k$, as additional independent  times, with which one associated a set of Hamiltonians, $\bar H_k$.

To be able to define the Hamiltonians, $H_k$, $\bar H_k$, we must first define a Poisson bracket. This is given by:
\begin{align}
\{\imath\varphi,t\}=1.
\end{align}
With this definition of the Poisson bracket, the evolution of $f_t$ is encoded in the following relations:
\begin{align}
\{f_t(\imath\varphi),\bar f_t(-\imath \varphi)\}=1\label{StringEq}, 
\end{align}
which, due to $\bar f_t(-\imath\varphi) =S(f_t(\imath\varphi)),$ can also be written as:
\begin{align}
\{f_t(\imath \varphi), S(f_t(\imath\varphi))\}=1\label{StringEqS}.
\end{align}

 That the evolution of $f_t$ is determined by (\ref{StringEq}) can be shown as follows. First note that (\ref{StringEq}) can be written as:
\begin{align}
\Im\partial_t f_t\partial_\varphi \bar f_t=1.\label{Imfft}
\end{align}
Since $\partial_t f_t$ is the velocity of a point on the interface with given $\varphi$, we may write it in usual vector notation as $\vec{\tilde{v}}$, the tilde denoting that it may not be equal to the velocity of the fluid. Nevertheless, its normal component must be equal to the velocity of the fluid so we have $\tilde v_n= v_n$.   Furthermore, $\partial_\varphi f_t$ may be written in vector notations as $\frac{d\vec r}{d\varphi}$. Taking into account that if $a=a_x+\imath a_y$ and $b=b_x + \imath b_y$ are the complex notations for $\vec a$ and $\vec b$, then the vector product is given by $\vec a \times \vec b = \hat z \Im(a\bar b)$, we may rewrite   (\ref{Imfft}) as:
\begin{align}
\vec{\tilde{v}}\times \frac{d \vec r}{d \varphi}  =1.
\end{align}
This relation implies $v_n \frac{dl}{d\varphi}=1$, or $v_n = \frac{d\theta }{dl}.$ Now, since $\theta$ and $\frac{\pi b^2}{6Q\mu}P$ are conjugate harmonic functions (Eq. (\ref{psiDef})), one may use the Cauchy-Riemann relations to obtain $\frac{d\varphi}{dl} =-\frac{b^2}{6Q\mu} \frac{dP}{dn} $, yielding the equation governing the evolution of the droplet, Eq. (\ref{DarcyLaw}).

The evolution with respect to the physical time, is  given by relation (\ref{StringEq}). Alternatively, we can write:
\begin{align}
\partial_t f=\{\imath \varphi,f\},\label{varphiisH}
\end{align}
showing that  the  designation of the Hamiltonian that generate the time translations as $\imath \varphi  $ is appropriate. We may thus write  $H=\imath \varphi$. Note however, that Eq. (\ref{varphiisH}) is almost an empty statement, in the sense that the evolution of $f$ cannot be extracted from it, as the right hand side is identically equal to the left hand side irrespective of the time evolution of the droplet. 

Another way to describe the evolution is through the Schwarz function. Writing the Poisson bracket in  (\ref{StringEqS}) explicitly and dividing by $\frac{\partial f_t}{\partial \varphi}$, one obtains:
\begin{align}
\frac{\partial S(f_t(\imath\varphi))}{\partial t} = \frac{1}{\frac{\partial f_t(\imath\varphi))}{\partial \varphi}}\left(\imath+ \frac{\partial f_t(\imath \varphi)}{\partial t} \frac{\partial  S(f_t(\imath\varphi))}{\partial\varphi}\right).
\end{align}
Evaluating this equation at $\varphi=\varphi_t(z)$ and making use of \begin{align}\left.\frac{\partial f_t(\imath\varphi))}{\partial \varphi}\right|_{\varphi=\varphi_t(z)}=\left( \frac{\partial \varphi_t(z)}{\partial z}\right)^{-1} \mbox{    and}\quad  \frac{\partial \varphi_t(z)}{\partial t} \frac{\partial f_t(\imath\varphi)}{\partial \varphi} +\left.\frac{\partial f_t(\imath\varphi)}{\partial t} \right|_{\varphi=\varphi_t(z)}=0,\label{BasicParital}\end{align} yields:
\begin{align}
\frac{\imath\partial \varphi_t(z)}{\partial z}= \frac{\partial \varphi_t(z)}{\partial t} \frac{\partial  S(f_t(\imath\varphi))}{\partial\varphi}  + \left.\frac{\partial S(f_t(\imath\varphi))}{\partial t}\right|_{\varphi=\varphi_t(z)}.
\end{align}
The right hand side can be identified with $\frac{\partial S(z)}{\partial t}$ such that we have:
\begin{align}
\frac{\partial S(z)}{\partial t}=\frac{\imath\partial \varphi_t(z)}{\partial z}\label{dSdtISdphidz},
\end{align}
which constitutes a simple evolution equation for the Schwarz function. 

Eq. (\ref{dSdtISdphidz}) is useful to show that the dynamics prescribed by (\ref{StringEq}) is such that the pressure is equal on all interfaces. This needs to be shown because above we have only assumed that the pressure is constant on any connected component of the interface.  Suppose then that points $a$ and $b$ are on the interface, perhaps on different connected components of which. Eq. (\ref{dSdtISdphidz}) together with (\ref{psiDef}) allows us to write:
\begin{align}
\frac{P(a)-P(b)}{\frac{\pi b^2}{6Q\mu}} = \Re  \frac{\partial }{\partial t }\int_a^b S(z)dz.\label{paritaltDeltaOmegaIsDeltaP}
\end{align}
On the other hand we have:
\begin{align}
\int_a^b S(z)dz= -\overline{\int_{\overline{S(a)}}^{\overline{S(b)}} S(y)dy }+\left .zS(z)\right|^b_a,\label{OmegaAlmostImaginary}
\end{align}
which is obtained by first changing the integration variable from $z$ to $y=\overline {S(z)}$, on the left hand side,  this change of variables being facilitated by using unitarity,  Eq.  (\ref{Unitarity}), to obtain $z=\overline{S(y)}$, then integrating  by parts the result. Taking the real part of Eq. (\ref{OmegaAlmostImaginary}) and a time derivative, one obtains:
\begin{align}
2\Re \int_a^b \partial_tS(z)dz =\Re\left( z \partial_tS(z)-\bar z \partial_t \overline{S(z)}\right)=0,
\end{align}
which when compared to (\ref{paritaltDeltaOmegaIsDeltaP}) yields:
\begin{align}
P(a)=P(b),
\end{align}
for any two points $a$, $b$ on the interface, including points on different connected components. 

\subsection{Integrability of Laplacian growth}   

Integrability of Laplacian growth is the statement that the evolution of the Hele-Shaw interface can be described by a Hamiltonian, $H$, which is in involution with an infinite set of other Hamiltonians (conserved quantities)
, $H_k,$ with respect to some Poisson bracket. As usual, the other Hamiltonians, in Laplacian growth can be used to generate flows with respect to new times $t_k,$ in one to one correspondence to the Hamiltonians, $H_k$. These Hamiltonians in Laplacian growth can be chosen such that $t_k$ are  the Richardson moments of the interface. Namely, the Hamiltonians, $H_k,$ generate deformations of the interface corresponding to changing the Richardson moments $t_k$, respectively. Due to the fact that these moments are complex, the Hamiltonians are also  complex and to each $H_k$ we may associate a  $\bar H_k$, which generates, formally, deformations which correspond to changing $\bar t_k$. Of course one cannot change $t_k$ without changing $\bar t_k$, yet, as usual, they may be treated as independent variables, just as $H_k$ and $\bar H_k$ may be treated as independent conserved quantities or Hamiltonians. 

The Hamiltonians read explicitly:
\begin{align}
H_k (\varphi,t)=  \left({f_t^k(\imath\varphi)} \right)_+  + c^{(k)},\label{HkWithStillToBeDetermined}
\end{align}
where $c^{(k)}$ is a an arbitrary function of the times only, which (as shall be seen below) does not affect the dynamics, while the index $+$ denotes taking the positive (polynomial) part of the Laurent expansion around infinity in the variable $e^{\imath\varphi}$, using the following convention. Due to Eq. (\ref{mapExpansion}) we may expand $f^k_t$ as
\begin{align}
f_t^k = \sum_{j=-\infty}^k a_j^{(k)} e^{\imath j \varphi}.\label{fkExpansion}
\end{align}
The positive part of this series is defined as follows: 
\begin{align}
 \left({f_t^k(\imath\varphi)} \right)_+ =\frac{a^{(k)}_0}{2}+\sum_{j=1}^k a_j^{(k)} e^{\imath j \varphi},\label{pluspartdef}  
\end{align}
which is natural, except the zeroth order term, defined with a factor $1/2$ for later convenience.

The Hamiltonian $\bar H_k$ is defined as  follows:\begin{align}
\bar H_k(\varphi,t)= \overline{H_k(\bar \varphi,t)} =  \left({\bar f_t^k(-\imath\varphi)} \right)_+  + \bar c^{(k)} .\label{barHasrefection}
\end{align}

We now show that the evolution generated by the Hamiltonians,\begin{align}
&\frac{\partial f_t(\imath\varphi)}{\partial t_k} = \{f_t(\imath\varphi),H_k\},\label{mapEvolution}\\
&\frac{\partial f_t(\imath\varphi)}{\partial \bar t_k} =- \{f_t(\imath\varphi),\bar H_k\}, \label{mapBarEvolution}
\end{align}
effect a deformation of the contour corresponding to changing the Richardson moments. Namely, we show that if (\ref{tkAsAreaIntegral})\ holds at any moment in times, then it holds for all subsequent times, for the  evolution prescribed by (\ref{mapEvolution}), (\ref{mapBarEvolution}).

First note that the complex conjugate of Eq. (\ref{mapBarEvolution}) can be also cast as the evolution of the Schwarz function. Indeed, the complex equation of that equation,  has the following form:
\begin{align}
\frac{\partial \bar f_t(-\imath\varphi)}{\partial t_k} = \{\bar f_t(-\imath\varphi),H_k\},\label{SofphiEvolution}
\end{align}
which is the evolution of the Schwarz function, $S(z)=\bar f_t(-\imath\varphi_t(z))$, evaluated at $z=f_t(\varphi)$. We proceed  by first showing:
\begin{align}
\frac{\partial S(z)}{\partial t_k} = \frac{\partial H_k}{\partial z},\quad \frac{\partial S(z)}{\partial \bar t_k} = \frac{\partial \bar H_k}{\partial z} \label{SofZevolution}.
\end{align}
We do this by writing:
\begin{align}
&\frac{\partial S(z)}{\partial t_k } = \frac{\partial\bar f_t(-\imath \varphi)}{\partial t_k}+\left. \frac{\partial \varphi_t(z)}{\partial t_k} \frac{\partial \bar f_t(-\imath \varphi)}{\partial \varphi}\right|_{\varphi=\varphi_t
(z)}\label{dSdtkFirstStep}
\end{align}
Using (\ref{mapEvolution}) and the first equation in (\ref{BasicParital}) along with $$\frac{\partial \varphi_t(z)}{\partial t_k} = -\left.\frac{\partial f_t(\imath\varphi)}{\partial t_k}\left(\frac{\partial f_t(\imath\varphi)}{\partial\varphi} \right)^{-1}\right|_{\varphi=\varphi_t(z)}, $$
one may re-write (\ref{dSdtkFirstStep}) as: 
\begin{align}
\frac{\partial S(z)}{\partial t_k } =\frac{\partial H_k}{\partial \varphi} \left.\left(\frac{\partial S(f_t(\imath \varphi))}{\partial t} +\frac{\partial S(z)}{\partial z} \frac{\partial f_t(\imath\varphi)}{\partial t} \right)\right|_{\varphi = \varphi_t(z)} =\frac{\partial S(z)}{\partial t}\left.\frac{\partial H_k}{\partial \varphi}\right|_{\varphi=\varphi_t(z)} . 
\end{align}
Making use now of (\ref{dSdtISdphidz}) , the first equation in (\ref{SofZevolution})  follows. The second equation in (\ref{SofZevolution}) may be derived by the analogous manipulations.
 
Given (\ref{SofZevolution}), and the definition of $H_k$, we may write \begin{align}
\frac{\partial S(z)}{\partial t_k} =k z^{k-1} +O(1/z^2),\quad \frac{\partial S(z)}{\partial \bar t_k} = O(1/z^2). \label{SderivExpansion}
\end{align}
Indeed, substituting into the first equation in (\ref{SofZevolution}) the explicit expression for $H_k$ and evaluating at $\varphi=\varphi_t(z)$, leads to the first equation. For the second equation note that  $\bar H_k$ contains only negative powers of $e^{\imath \varphi}$. These, in turn, when evaluated at $\varphi=\varphi_t(z)$ can be written as negative powers of $z$,  due to (\ref{mapExpansion}), leading to the second equation in (\ref{SderivExpansion}). Equation (\ref{SderivExpansion}) immediately implies: \begin{align}
 \oint_{\mathcal{C}} z^{-l}\frac{\partial S(z)}{\partial t_k} \frac{dz}{2\pi \imath k}= \delta_{kl},\quad \oint_{\mathcal{C}} z^{-l}\frac{\partial S(z)}{\partial \bar t_k} \frac{dz}{2\pi \imath k}=0
.\label{tkAsContourDifferentialForm}\end{align}
Taking a derivative with respect to  $t_k$  and $\bar t_k$ of Eq. (\ref{tlAsContouIntegral}), one realizes that Eq. (\ref{tkAsContourDifferentialForm}) is just the differential form of these equations.  Thus we have proven the required statement. 

Note that $c^{(k)}$ in (\ref{HkWithStillToBeDetermined}) are indeed arbitrary, in the sense that we did not need to assume anything about them to prove that (\ref{tlAsContouIntegral})\ follow from (\ref{mapEvolution}), (\ref{mapBarEvolution}). In fact, $c^{(k)}$ just generates  transformations that shift the reference point for the angle $\varphi$. To see this consider that one always applies the Hamiltonians in a real combination. Namely any deformation of the droplet is effected by:
\begin{align}
\delta f_t=\sum_k \{f_t,\delta t_k H_k-\delta \bar t_k\bar H_k\}.
\end{align}
The free ($\varphi$-independent) term in $\delta t_k H_k-\delta \bar t_kH_k$ reads $2\imath\Im(\delta t_k c^{(k)})$, while the latter generates the following deformation of $f_t$:
\begin{align}
\{f_t,2\imath\Im(\delta t_k c^{(k)})\} =2\Im(\delta t_k \dot c^{(k)}) \frac{\partial f_t}{\partial \varphi},
\end{align}   
where the dot signifies a $t$ derivative. Thus,  $2\imath\Im(\delta t_k c^{(k)})$ indeed is the generator of real, times-dependent $\varphi$ translations, and thus do not deform the interface. 
As a result, we 
may
 choose $c^{(k)}$ freely. 

We make the choice of $c^{(k)}$ such that it is consistent with the involution of the Hamiltonians. Consider first the indefinite integral  $\int S(z)dz= \sum_k t_k z^k + t \log(z) +c +O(1/z),$ where $c$ is an arbitrary function of the times, the arbitrariness of $c$ being due to the indefiniteness of the integral. We may choose $c$ such that $\partial_t c=-\log(a_0^{(1)}),$ and define the result, which is now definite up to a time and $z$ independent additive factor as the integral of $S(z)$ :
\begin{align}
\int S(z)dz = \sum_k t_k z^k + t \log(z) -\int_0^t \log(a_0^{(1)})+O(1/z).\label{OmegaExpansion}
\end{align}    
The integral over $S$ is a generating function for the Hamiltonians. Indeed due to (\ref{SofZevolution}), we have:
\begin{align}
\left.\frac{\partial }{\partial t_k}\int S(z)dz\right|_z = \alpha^{(k)}+H_k,\label{dOmegaWithck}
\end{align}
for some $z$-independent constant, $\alpha^{(k)}$. We may choose $c^{(k)}$ such that this constant is zero, and write:\begin{align}
\left.\frac{\partial }{\partial t_k}\int S(z)dz\right|_z=H_k, \quad \left.\frac{\partial }{\partial t}\int S(z)dz\right|_z =H.
\end{align}

This last equation (which is derived from  the choice of $c^{(k)}$) states that $\int S(z)dz$ is a generating function for the Hamiltonians and, as shall be seen below, insures that the Hamiltonians are in involution. Indeed,
The identities  $\frac{\partial^2 }{\partial t_k \partial t_l}=\frac{\partial^2}{ \partial t_l\partial t_k}$,  $\frac{\partial^2 }{\partial t_k \partial\bar t_l}=\frac{\partial^2}{ \partial\bar  t_l\partial t_k}$, leads to:
\begin{align}
\left.\frac{\partial H_k}{\partial t_l}\right|_z  =\left.\frac{\partial H_l}{\partial t_k} \right|_z, \left.\frac{\partial H_k}{\partial \bar t_l}\right|_z  =\left.\frac{\partial \bar H_l}{\partial t_k} \right|_z\quad \label{WhithamEqsforHs}
\end{align}   
and this yields the involution of the Hamiltonians, when written in terms of derivatives at fixed $\varphi$ rather than at fixed $z$; We first write: \begin{align}
&\left.\frac{\partial H_k}{\partial t_l} \right|_{z} =\frac{\partial H_k}{\partial t_l } +\frac{\partial\varphi_t(z)}{\partial t_l} \left. \frac{\partial H_k}{\partial\varphi}
 \right|_{\varphi=\varphi_t(z)}= \\ &=\frac{\partial H_k}{\partial t_l } -\left. 
 \frac{\partial f_t(\imath\varphi)}{\partial t_l} 
 \left(\frac{\partial f_t(\imath\varphi)}{\partial \varphi} \right)^{-1}  \frac{\partial H_k}{\partial\varphi} \right|_{\varphi=\varphi_t(z)},  
\end{align}  
then subtract the equation with $k$ and $l      $ interchanged and use (\ref{mapEvolution}) to obtain:\begin{align}
\frac{\partial H_k}{\partial t_l} - \frac{\partial H_l}{\partial t_k} - \{H_k,H_l\}=0 .\label{InInvolution}
\end{align}  
The latter being the required statement of the Hamiltonians being in involution. Of course one may easily obtain the following as well:
\begin{align}
\frac{\partial H_k}{\partial \bar t_l} - \frac{\partial \bar H_l}{\partial t_k} - \{H_k,\bar H_l\}=0 .\label{InInvolutionBar}
\end{align}

\section{Algebraic Solutions to Laplacian Growth\label{AlgebraicGeometricLPG}}
The interface may be described as a curve, namely as a relation between $x$ and $y,  $ the Cartesian coordinates on the interface. This may be done by finding some function $\tilde F(x,y)$ such that $\tilde F(x,y)=0$ gives the interface. The ellipse is the simplest example except the circle. For the ellipse we have $\tilde F(x,y) = \frac{x^2}{r_1^2}
+\frac{y^2}{r_2^2}-1$. Using complex coordinates one may alternatively describe the curve by a relation between $z$ and $\bar z$, for example, the ellipse may be described by $F(z,\bar z)=\tilde F\left(\frac{z+\bar z}{2} ,\frac{z-\bar z}{2\imath}\right)=0,$ where  $F(z,\bar z) =\frac{1}{2} \left( \frac{1}{r_1^2} + \frac{1}{r_2^2}\right)|z|^2+\frac{1}{2} \left( \frac{1}{r_1^2} - \frac{1}{r_2^2}\right)(  z^2 +\bar z^2). $ One may solve this equation for $\bar z$. This gives the Schwarz function, $S(z)$, which satisfies $F(z,S(z))=0. $  If, as is true for the ellipse, the function $F(z,\bar z)$ is a polynomial in both arguments, then $S(z),$ as a function of $z,$ is an algebraic function. As an example, for the ellipse we obtain:
\begin{align}
S(z)=\frac{(r_1^2+r_2^2)z+2r_1 r_2 \sqrt{z^2+r_2^2-r_1^2}}{r_1^2-r_2^2}.\label{Sellipse}
\end{align}
Namely $S(z)$ contains roots and polynomials. As an algebraic function, $S(z)$ may be defined on a algebraic Riemann surface. 

 Defining the Schwarz function on the Riemann surface arises due to the fact that $S(z)$ is generically a multi-valued function, such as the function in  (\ref{Sellipse}), where the square roots makes $S(z)$ multi-valued. One is lead to resolve this multi-valuedness, by considering $m$ copies of the complex plane, one copy for each possible value of $S(z)$ for given $z$. Let us denote the $m$ possible values of $S(z)$ by $S^{(i)}(z),$ with $i=1,\dots,m$. Each copy of the complex plane must also be  provided with branch cuts, in order to fully resolve the ambiguity. One should imagine  cutting the  sheets along the branch cuts. After cutting the sheet, each branch cut has the topology of a closed curve.  On copy $i$ of the complex plane, the Schwarz function is designated the value $S^{(i)}(z)$ for given $z$. Then the different copies of the complex plane are glued together along the branch cuts, corresponding to the fact that as $z$  crosses the branch cut on sheet $i$  the function $S^{(i)}(z)$ will coincide smoothly with the value of function $S(z)$ on some other sheet, $j$. 

For the example of the ellipse there are two sheets and 
\begin{align}
S^{(1,2)}(z)=\frac{(r_1^2+r_2^2)z\pm2r_1 r_2 \sqrt{z^2+r_2^2-r_1^2}}{r_1^2-r_2^2}.
\end{align}    
The square root is taken in a way dictated by convention $\sqrt{r e^{\imath \theta}}= \sqrt{r} e^{\imath \theta/2}, $ for $ -\pi\leq \theta<\pi.$ A branch cuts are drawn on both sheets between $\sqrt{r_1^2 - r_2^2}$  (assuming $r_1>r_2$). And the two sheets are glued along the branch cut, the upper bank of the branch cut on first sheet to the lower bank of the branch cut on the lower sheet.

The algebraic Riemann surface, $\mathcal{A,}$ that is produced by considering the complex curve $F(z,S(z))=0$, is also called an algebraic curve\cite{ST:why:becomes:o4} . It is composed of $m$ copies of the complex plane $\mathds{C}^{(i)}$ called sheets equipped with branch cuts. On each sheet, away from the branch cuts, one defines a holomorphic local parameter given by $z$. On the branch cuts, but away from the branch points, the local parameter spans more than one sheet. If $z_0$ is a simple branch point on sheet $i$, then the local parameter around the point is $\sqrt{z-z_0}$, spanning the two sheets that are connected at the branch point.

We may define a function on the algebraic curve $\mathcal{A}$ by specifying its values for any $z$ and any sheet $i$. It is thus convenient to introduce the notation $\bm z$ for a point on the $\mathcal{A}$, where $\bm z=\{z,i\}$, specifying the complex coordinate, $z$, of the point, and the sheet, $i$. For example, $S(\bm z)$ is naturally defined as:
\begin{align}
S(\bm z) = S(\{z,i\}) = S^{(i)}(z).
\end{align}
Finally, note that we shall called first sheet, $i=1$, also the 'upper sheet', and all other sheets, collectively as 'lower sheets'.

The anti-analytic function $\tau(z)$, defined as:
\begin{align}
\tau(z) = \overline{S(z)},
\end{align}
is, in fact, a one-to-one and onto map from the Riemann surface to itself. This is due to the fact that it is its own inverse $\tau(\tau(z))=z$, which is a consequence of unitarity (Eq. (\ref{Unitarity})). The set of points invariant with respect to $\tau$ is  the interface of the droplet, $\tau(z)=z \Leftrightarrow S(z)=\bar z$.   The anti-holomorphic involution, $\tau$, allows us to equivalently describe the Riemann surface as a Schottky double\cite{Gustafsson:Schottky:Double}. 
\begin{figure}[h!!]
\centering
\includegraphics[width=0.75\columnwidth]{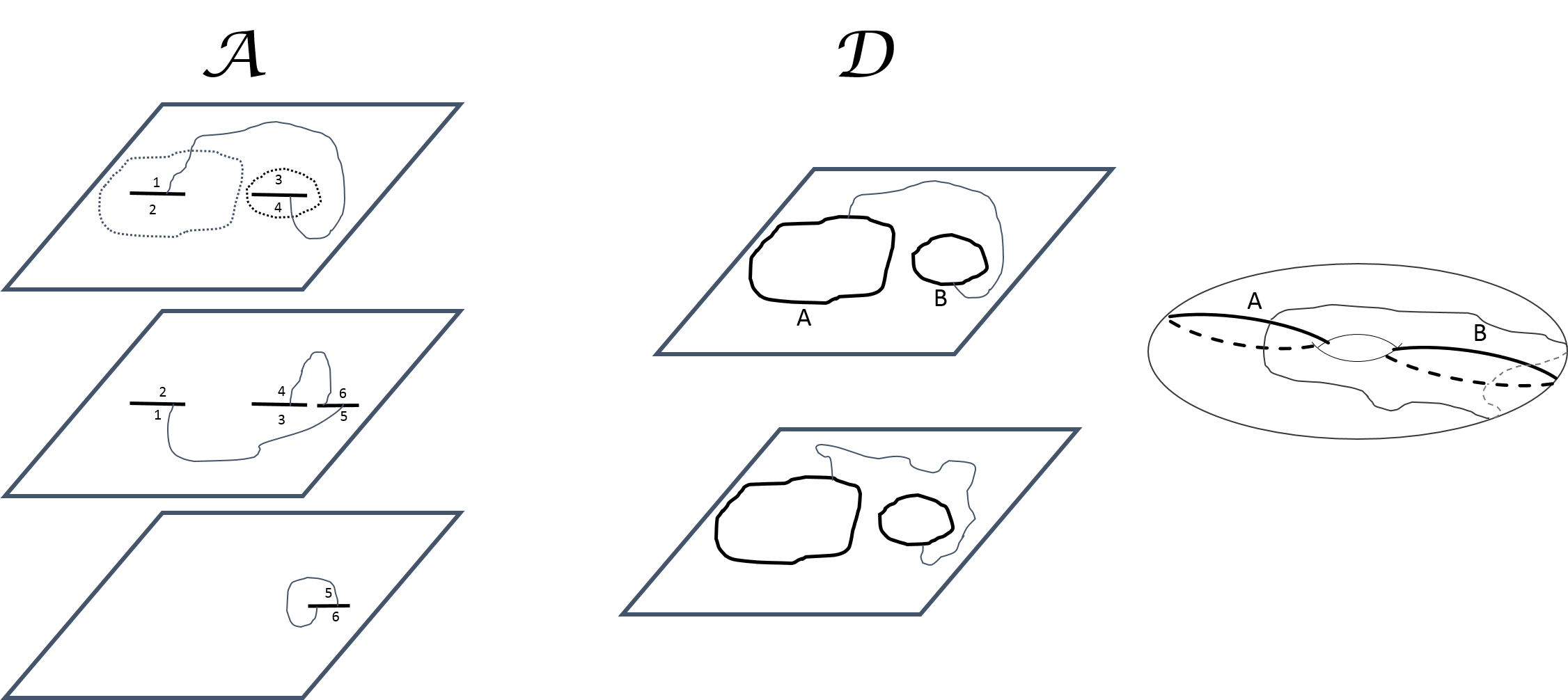}
\caption{The construction of the Riemann surface as an algebraic curve, $\mathcal{A,}$ is drawn on  the left. The surface is three sheeted. The two banks of each of the branch cuts are numbered, such that banks that are glued together on different sheets are given the same number. A meandering closed path, through the Riemann surface is also drawn. The dotted line is the interface. The algebraic curve $\mathcal{A}$ is mapped onto the Schottky double $\mathcal{D}$, in the middle drawing. The map of the meandering line is also drawn. The two connected components of the interface are now labelled $A$ and $B$. On the right is a torus of genus $1$ drawn as a doughnut. The interface with its two connected components, $A$ and $B,$\ are drawn there, along with the meandering path. The torus is topologically equivalent both to the algebraic Riemann surface, $\mathcal{A}$ and the Schottky double $\mathcal{D}$. \label{RiemannSurfaceConstruction}}
\end{figure}
        
To describe the Schottky double, consider two copies of that part of the complex plane that lies outside the droplet. Let us denote these two copies by  $D^{(\uparrow) }$  and  $D^{(\downarrow)}$, respectively. The Schottky double, $\mathcal{D}$ is defined to consist of the two copies of the exterior of the droplet, $D^{(i)}$, with local holomorphic  coordinates inherited from the algebraic curve  $\mathcal{A},$ through a one to one mapping $v,$ defined below, from the algebraic curve $\mathcal{A}$ to  the Schottky double $\mathcal{D}$. The mapping $v$ is defined as follows:
\begin{align}
v(\{z,i\}) = \left\{\begin{array}{lr} 
\{z,\uparrow\} &  \mbox{$i=1$ and $z$ outside $\mathcal{C}$}  \\ 
\{\tau(z),\downarrow\} & \mbox{otherwise}
\end{array} \right.
\end{align}

The upper sheet of the Schottky double, $D^{(\uparrow)}$ is usually called `the front side', while the lower sheet, $D^{(\downarrow)},$ is  commonly known, apparently,  as `the back side'.

Note that since $\tau(z)$ is anti-holomorphic, the holomorphic coordinate on $D^{(\downarrow)}$ is $\bar z$, rather than $z$. Thus, by definition, an analytic on the Schottky double, is an analytic  function of the variable $z$ on $D^{(\uparrow)}$ and an analytic function of the variable $\bar z$ on $D^{(\downarrow)}$. The function must also be analytic across the interface. Another way to describe an analytic function, $f$, on the Schottky doubleis to say that $f$ is analytic on the Schottky double, $\mathcal{D}$,   if $g=f\circ\ v$ is analytic on the algebraic curve, $\mathcal{A}$. 

As examples of analytic functions on the Schottky double, consider the function $id(\bm z)$  defined as $id(\{z,\uparrow\})=z$, $id(\{z,\downarrow\})=\bar z$ . Another example (with some abuse of notations) is the function $S(\bm z)$, defined as $S(\{z,\uparrow\})=S(z)$ and $S(\{z,\downarrow\})=\bar z$, with apologies for using the same symbol $S$, for the Schottky double function, just defined, and the Schwarz function. The two functions are distinguished by the the fact that the former takes the combined notation $\{z,i\}$ as an argument while the latter takes a complex number as the argument. In that respect, we shall often use what we shall term as `front-side notations'. Namely, we shall denote a function on the Schottky double by the same notation defined on the algebraic curve, if both functions agree on $D^{(\uparrow)}$. Thus, for example, the function $id$ above, would also be denoted by $z$. If the function is analytic there is no ambiguity introduced in these notations, since there is a unique continuation of the function from $D^{(\uparrow)}$ to $D^{(\downarrow)}$ on the Schottky double just as there is a unique analytic continuation from the upper sheet of the algebraic curve to all lower sheets.

\begin{figure}[h!!]
\centering
\includegraphics[width=0.25\columnwidth]{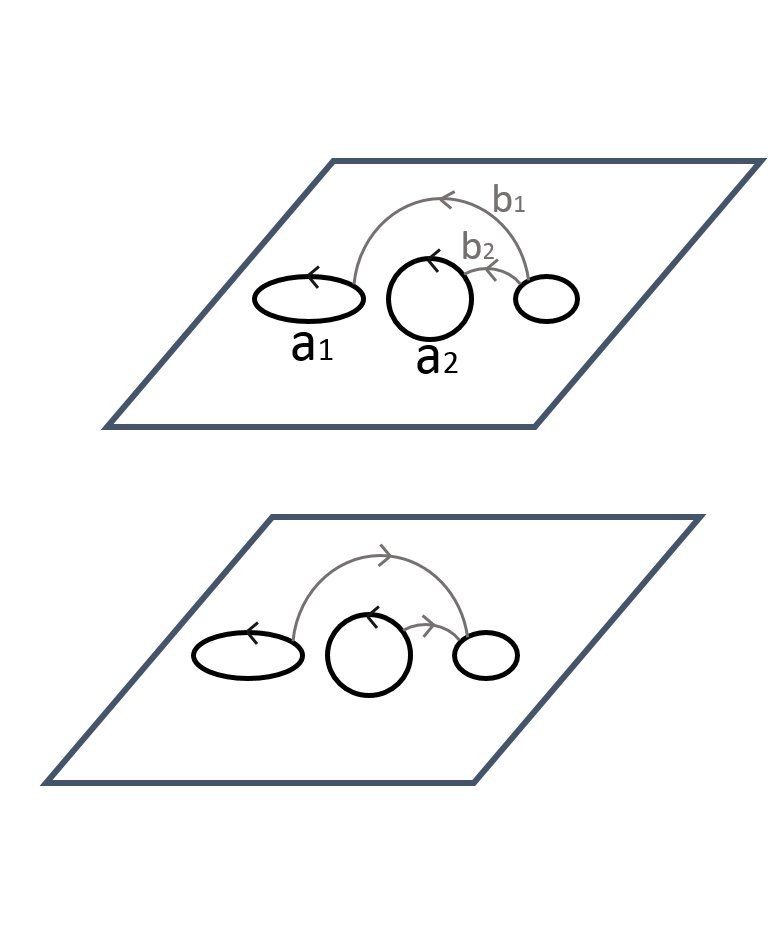}
\caption{The choice of cycles for a genus $2$ Riemann surface. The $b$ cycles draw a path, one part of which is on the front side and the other on the back side. The $a$ cycles are drawn exactly on the contours connecting the front and back sides. \label{SchottkyCycles}}
\end{figure}

Let us make note of the non-trivial cycles on the different models of  Riemann surfaces we have defined. If the the interface has more than one connected component, the topology of the Riemann surface becomes non-trivial, in that there are cycles which cannot smoothly be contracted to a point. Two cycles are inequivalent if they cannot be deformed one onto the other smoothly. If the interface has $g+1$ connected components, then we may define $g$ such non-trivial inequivalent cycles as $g$ of the $g+1$  connected components. The last connected component is actually equivalent to the union of the first $g$, and thus may not be considered independent. These $g$ cycles are denoted by $a_j$, with $j=1,\dots,g$.    In addition, $g$  independent cycles, $b_j$,  may be defined, which intersect the cycles $a_j$, respectively, with intersection number $1$. This fact is written formally as:
\begin{align}
a_i\cdot b_j = \delta_{i,j}.
\end{align} 
On the Schottky double, the path of the  cycle $b_j$ may be considered to start on the front side on the $g+1$ connected component of the interface, then continues to  the back side intersecting cycle $a_j$ finally connecting to the starting point through the back side. The situation is shown in Fig. \ref{SchottkyCycles}. 

Finally let use describe how the Hamiltonians, $H_k$, $\bar H_k$,  are determined by the Riemann surface $\mathcal{D}$ (or equivalently $\mathcal{A}$). The Hamiltonians,  $H_k,$ are  multi-valued functions on the Riemann surface. Their derivatives with respect to $z,$ are single-valued, due to   Eqs. (\ref{SofZevolution}), (\ref{dSdtISdphidz})  and the  single-valuedness of the Schwarz function.
Let us define 
\begin{align}
H^{(R)}_k= \frac{H_k + \bar H_k}{2}, \quad H^{(I)}_k= \frac{H_k - \bar H_k}{2\imath},
\end{align}
which are the Hamiltonians conjugate to the real and imaginary parts of $t_k$, respectively. Denote the real and imaginary parts of $t_k$ by $t_k^{(R)}$ and $t_k^{(I)},$ respectively. We shall use the convention that $H_0=H/2$, where by $H_0^{(R)}=0,$  $H_0^{(I)}=H/2.$ 

The following are well-defined (single-valued) meromorphic differentials on the Riemann surface: 
\begin{align}
dH_k^{(R/I)} = \frac{\partial H_k^{(R/I)}}{\partial z} dz=  \frac{\partial S}{\partial t_k^{(R/I)}}dz.\label{dHkisdSdtk}
\end{align}
Due to (\ref{SderivExpansion}) and (\ref{SofZevolution}), the differnetial $dH^{(R/I)}_k$ has a pole of order $k+1$ at infinity. Indeed:
\begin{align}
dH_k^{(R)} =-\frac{(k-1)}{2} \frac{ds}{s^{k+1}}+O(1/s^2ds),\quad dH_k^{(I)} =\mp\frac{(k-1)}{2\imath} \frac{ds}{s^{k+1}}+O(1/s^2ds)\label{HkSingularities}  
\end{align}
where $s=z^{-1}$ is the local parameter around infinity and the $\pm$ sign in the expansion of $dH^{(I)}_k$ is to be taken respectively with whether the expansion is around $\infty_{\uparrow}$ or $\infty_\downarrow$. The differential of the Hamiltonians have no further singularities. This last fact can be surmised from (\ref{dHkisdSdtk}) and (\ref{tlAsContouIntegral}). 

A meromorphic differential with specified singularities, such as prescribed by Eq. (\ref{HkSingularities}), is unique up to an addition of a  holomorphic differential, of which there are $g$ on a genus $g$ Riemann surface. Since we can add to   $dH_k^{(R/I)}$ any linear combination of them with complex coefficients, the space of differentials obeying (\ref{HkSingularities}) is $2g$ dimensional over $\mathds{R}$. To fix $dH_k^{(R/I)}$ completely, one may explore the integral over any cycle of such a differential. We first note that Eq. (\ref{OmegaAlmostImaginary}) implies that $\oint S(z) dz$ over any cycle is purely imaginary (set $a=b$ in that equation). This fact, when combined with (\ref{dHkisdSdtk}) gives:
\begin{align}
\Re \oint_{a_j} dH^{(R/I)}_k = \Re \oint_{b_j} dH^{(R/I)}_k =0.
\end{align} 
This constitutes $2g$ real constraints on $dH^{(R/I)}_k,$ which completely fix the ambiguity in determining them for any $k$. This means that $dH_k$ can also be fixed uniquely given $\mathcal{D}$ or equivalently, $\mathcal{A}$. As a consequence we shall write in the following $H_k(\mathcal{A})$ in occasions where we shall want to stress that the Hamiltonians are determined from the Riemann surface.

\section{Dispersive Quantization and the  two dimensional Toda lattice\label{DispersiveQuantization}}

We shall now want to proceed and quantize the Laplacian growth dynamics, eventually obtaining quantum field theory, with the aim of using methods of integrable and conformal field theory to gain better understanding of the fractal patterns in Laplacian growth. The program of quantization requires in this case two steps. The first step is removing the dispersionless limit from the two dimensional reduced dispersionless Today hierarchy that was presented above as the hierarchy determining Laplacian growth. This gives a classical nonlinear integrable system, the (dispersionful) two dimensional Toda lattice. Formally, the procedure is very similar to quantization, thus it is dubbed here as `dispersive quantization', although it connects to classical problems. It is only then that we can apply the second step, that of physical quantization, to obtain a quantum field theory.   

\subsection{The two dimesional Toda Hierarchy}
The two dimensional Toda hierarchy \cite{58:Takasaki}, is an integrable system of nonlinear equations generalizing the more familiar Toda lattice. The latter being a one dimensional system of springs, having exponential force law $F = F_0 \cdot (e^{-\Delta x/x_0}-1)$, where $F$ is the force that the string exerts when it is extended a distance $\Delta x.$ The nonlinear equation is the Newton equation for the springs:
\begin{align}
m\frac{d^2x_n }{dt^2} =F_0 \cdot  \left(e^{\frac{x_{n-1}-x_{n}}{x_0} } - e^{\frac{x_{n}-x_{n+1}}{x_0} } \right).
\end{align} 
The two dimensional Toda lattice, despite the name, does not generalize the problem of exponential springs to two dimensions, but is rather a formal generalization, in which time itself is treated as a complex variable:
\begin{align}
\frac{\partial x_n}{\partial t_1\partial \bar t_1}  =  e^{{x_{n-1}-x_{n}} } - e^{{x_{n}-x_{n+1}} } .\label{2DTL}
\end{align} 
Here we have set $m$, $F_0$ and $x_0$ to unity, as the physical interpretation in terms of springs was already lost when we complexified time, and so there is no real use of keeping track of physical units. 

The equation of motion for Laplacian growth, when treated as equations determining the evolution of a algebraic Riemann surface, are identical to the equations that are obtained by considering the slow time modulation of the two dimensional Toda lattice hierarchy.  Here one assumes that a solution to the two dimensional Toda lattice has the property that it is rapidly oscillating in time (which itself a complex parameter, denoted here by $t_1$), in such a way that the envelope of the oscillations varies on a much larger scale than the period of oscillations. If one parametrizes the envelope of oscillations using the correct variables, the equations for these parameters acquire an elegant form, which resemble a semiclassical limit of the full hierarchy. In the present case these semiclassical equations are Eqs. (\ref{InInvolution}), defining the Laplacian growth evolution, while the equations for the two dimensional Toda hierarchy are obtained by replacing the Hamiltonians $H_k$ by differential operators and the Poisson brackets by commutators. This section, together with  appendices \ref{BAexplicitSection},\ref{BADefPropAppen}, \ref{WhithamAppendix}, is devoted to showing how this connection, described above in heuristic terms, may be derived rigorously.

We must first define the whole two dimensional Toda lattice hierarchy, with its infinite number of complex times, and the respective Hamiltonians. The hierarchy may be defined\cite{58:Takasaki} making use of a Lax operator matrix, $L. $ For the two dimensional Toda lattice, $L$ is an infinite matrix depending on the times,  having the property:
\begin{align}
L_{nm}=0 \quad{\rm for}\quad m>n+1.\label{semiinfiniteness}
\end{align}
  Each time $t_l$ has associated with it a Hamiltonian $L^{(l)}$ with matrix elements:\begin{align}
L^{(l)}_{mn} = \left\{\begin{array}{lr}
(L^l)_{mn} & m>n \\ 
\frac{1}{2}(L^l)_{mn} & m=n \\
0 & m< n 
\end{array} \right. , \label{Ltothepowerl+}
\end{align}
here, e.g., $(L^l)_{mn}$ is the $mn$ element of the matrix $L$\ raised to the power $l$.
 The time $\bar t_l$, has the $-L^{(l)\dagger}$ as the associated Hamiltonian. Note the similarity of (\ref{Ltothepowerl+}) with (\ref{HkWithStillToBeDetermined}), (\ref{pluspartdef}). 

The nonlinear equations of the hierarchy may be obtained by demanding:
\begin{align}
\frac{\partial L}{\partial t_k } = \left[L,L^{(k)}\right],\quad \frac{\partial L}{\partial \bar t_k } =- \left[L,L^{(k)\dagger}\right] .\label{LEvolution}
\end{align}
 As a consequence\cite{35:Gerasimov:Matrix:Toda} of (\ref{LEvolution}) and the definition of $L^{(l)}$, the zero-curvature conditions hold:
\begin{align}
\frac{\partial L^{(l)}}{\partial t_k } - \frac{\partial L^{(k)}}{\partial t_l }-[L^{(l)},L^{(k)}]&=0\label{ZeroCurvatureDispersionfull} \\
\frac{\partial L^{(l)\dagger}}{\partial t_k } + \frac{\partial L^{(k)}}{\partial \bar t_l }-[L^{(l)\dagger},L^{(k)}]&=0 \label{ZeroCurvatureDispersionfullDagger}.
\end{align}
We shall be interested in a special reduction of the two dimensional Toda hierarchy. Namely, we shall assume  the the Lax operator obeys the following condition:
\begin{align}
[L,L^\dagger]=1\label{QStringEq}.
\end{align} 
In the dispersionless limit this equation becomes (\ref{StringEq}).

To demonstrate how (\ref{ZeroCurvatureDispersionfull}), (\ref{ZeroCurvatureDispersionfullDagger}), give rise to a whole hierarchy of nonlinear integrable partial differential equations, let us derive the first member of the hierarchy, Eq. (\ref{2DTL}), from them. First note that any matrix satisfying (\ref{semiinfiniteness}) may be written as:
\begin{align}
L=\sum_{m=-\infty}^1 a^{(1)}_m(\hat t)e^{m\hat p},\label{LasShifts}
\end{align}
where $\hat t$ acting on a vector $\vec \psi$, whose elements are $\psi_n$ gives a vector whose elements are $n\psi_n$, while the operator $e^{m\hat p}$ when acting on the same vector gives a vector whose elements are $\psi_{n+m}$:
\begin{align}
(\vec\psi)_n = \psi_n,\quad (\hat t \vec \psi)_n = n\psi_n,\quad (e^{m\hat p}\vec \psi)_n = \psi_{m+n}.
\end{align}

Note the similarity between Eq. (\ref{LasShifts}) with (\ref{mapExpansion}). In fact the latter equation is the semiclassical limit of the former if we assume that in the limit $\hat p\to\imath \varphi$. The similarity between the dispersionless limit (the limit in which one describes the modulations of fast oscillating solutions) and a semiclassical limit, will present itself repeatedly below. The rigorous basis for the similarity will be treated in the next subsections, supplemented by the appendices. 

From Eqs. (\ref{LasShifts}), (\ref{Ltothepowerl+}) one may obtain $L^{(1)}$, as follows:
\begin{align}
L^{(1)} =a^{(1)}_1(\hat t) e^{\hat p } + \frac{a^{(1)}_0(\hat t)}{2}.
\end{align}
Setting $k=l=1$ in (\ref{ZeroCurvatureDispersionfullDagger}) one obtains the equation:
\begin{align}
&\frac{1}{2}\left( \frac{\partial a^{(1)}_0(\hat t)}{\partial \bar t_1 }+ \frac{\partial \bar a^{(1)}_0(\hat t)}{\partial t_1 }\right)+e^{-\hat p }\frac{\partial \bar a^{(1)}_1(\hat t) }{\partial t_1 } + \frac{\partial a^{(1)}_1(\hat t)}{\partial \bar t_1 }e^{\hat p }+=\left|a^{(1)}_1(\hat t-1)\right|^2-\left|a^{(1)}_1(\hat t)\right|^2\label{2DTLfromOperator}+\\&+e^{-\hat p }\frac{\bar a^{(1)}_1(\hat t)a^{(1)}_0(\hat t)-\bar a^{(1)}_1(\hat t)a^{(1)}_0(\hat t+1)}{2}+\frac{a^{(1)}_1(\hat t)\bar a^{(1)}_0(\hat t)-a^{(1)}_1(\hat t)\bar a^{(1)}_0(\hat t+1)}{2}e^{ \hat p }.\nonumber
\end{align}
For the operator on the left hand side to be equal to the operator on the right hand side, the free terms (the terms in the brackets on the left hand side and the first two terms on the right hand side) on the right and left hand sides must be equal, as well as the coefficient of $e^{\hat p}$ and $e^{-\hat p}$, on both sides of the equation, respectively. 

To obtain (\ref{2DTL}), one writes:
\begin{align}
a^{(1)}_1(n)=e^{\imath \theta_n}e^{\frac{x_{n}-x_{n+1}}{2}}.
\end{align}
The the equation relating the  coefficient of   $e^{\hat p}$ and $e^{-\hat p}$ on both sides of (\ref{2DTLfromOperator}) lead to the same equation:
\begin{align}
-\imath2 \frac{\partial \theta}{\partial t_1}+\frac{\partial }{\partial t_1}(x_{n}-x_{n+1})&=a^{(1)}_0(n)-a^{(1)}_0(n+1)\label{2DTLFromOP2}\\
\frac{\partial a^{(1)}_0(n)}{\partial \bar t_1 }+ \frac{\partial \bar a^{(1)}_0(n)}{\partial t_1 }&=e^{x_{n-1}-x_{n}}-e^{x_{n}-x_{n+1}}
\end{align}
Adding the $\bar t_1$ derivative of (\ref{2DTLFromOP2}) to its complex conjugate, one obtains:
\begin{align}
\frac{\partial^2}{\partial t_1 \partial \bar t_1}(x_n-x_{n+1})=e^{x_{n-1}-x_{n}}+e^{x_{n+1}-x_{n+2}}-2e^{x_{n}-x_{n+1}},\label{2DTLFromOP3}
\end{align}
which is just what is obtained by subtracting Eq. (\ref{2DTL}) from the same equation when $n$ is shifted up by $1$,  (that is $n\to n+1$). Namely, if we assume that $x_n\to 0$ fast enough as $n\to\pm\infty$, then Eq. (\ref{2DTLFromOP3}) is equivalent to (\ref{2DTL}), since in that case we can use, e.g., $x_n = \sum_{m=n}^\infty x_m-x_{m+1}.$

\subsection{Multi-Periodic Solutions}

As mentioned above, the reduced dispersionless two dimensional Toda lattice hierarchy, which determines the evolution Laplacian growth can be obtained by taking the dispersionful reduced two dimensional Toda lattice hierarchy and considering slowly modulated waves. In order to be able to discuss these modulated waves, we must introduce the unmodulated periodic solutions. In particular we must discuss how these are constructed. To do this it is useful to make use of the Baker Akhiezer function\cite{Baker}.

The Baker Akhiezer function, is an eigenfunction of the Lax operator, with complex eigenvalue $z$:
\begin{align}
L\psi = z \psi.
\end{align}        
The time dependence of $\psi$ is defined naturally using the Hamiltonians:
\begin{align}
\frac{\partial \psi}{\partial t_k} = L^{(l)} \psi, \quad\frac{\partial \psi}{\partial\bar t_k} = -L^{(l)\dagger} \psi .\label{BAsimplistic}
\end{align}

For each $z$ there may be several eigenfunctions of $L$. Suppose that, generically,  there are $m$ of these eigenfunctions for each $z$. We may take $m$ copies of the complex plane $\mathds{C}$ and assign to each copy one of the eigenfunctions. It turns out that there is a large class of $L$'s such that by drawing branch cuts on the different sheets   and gluing the sheets together along the branch cuts, the function $\psi$ becomes a smooth function of $z$, except at infinity, and at a set of discrete poles. In fact, Krichever \cite{132:Krich:Intgr:AlgGeo} has shown that an effective way to construct multi-periodic solutions is to start with an algebraic Riemann surface, and define the Baker-Akhiezer function according to some analytic conditions (see also Ref. \cite{84:Dubrovin:Algebr:Geome}
for an overview of this subject). The Lax matrix $L$, can then be constructed from the knowledge of the Baker-Akhiezer function. Since the elements of the matrix $L$\ are the nonlinear fields of of the two dimensional Toda hierarchy,   the lax matrix also encodes solutions to the nonlinear integrable equation. 

If the algebraic Riemann surface in section admits an anti-holomorphic, $\tau$, then the Riemann surface can be given the structure of a Schottky double. Anticipating the connection to the Riemann surfaces we encountered in Laplacian growth, we assume the existence of such a anti-holomorphic involution and the problem of constructing multi-periodic solutions thus becomes the problem of constructing Baker-Akhiezer functions on a\  given Schottky double. 

The details of how to define a Baker-Akhiezer function    from a set of  analytic properties is given in appendix \ref{BAexplicitSection}. The fact that such a function then solves (\ref{BAsimplistic}) is then shown in appendix \ref{BADefPropAppen}. The information pertinent to the sequel is that for each Schottky double a procedure yields a multi-periodic solution of the form:
\begin{align}
u_m(t_0)= f_m\left( \sum_{k=0}^\infty  t_k\vec \omega_k +  \bar t_k\vec{\bar\omega}_k \right),\label{MultiPeriodicus}
\end{align}
where  $\vec \omega_k$ are $g$ dimensional complex vectors  and $f_m$ are functions from $\mathds{C}^g$ to $\mathds{C}$, which are periodic with periods $\hat e_i$ and $\bm B \hat e_i$:
\begin{align}
f_m(\vec v+\hat e_i)=f_m(\vec v), \quad f_m(\vec v+\bm B\hat e_i) = f_m(\vec v). 
\end{align}
 Here $\hat e_i$ is the unit vector in the $i$-th direction, $(\hat e_i)_j=\delta_{ij}$, and $\bm B$ is a $g$ by $g$ matrix. Many more details are given in  appendices \ref{BAexplicitSection} and \ref{BADefPropAppen}. However, the main message of Eq.(\ref{MultiPeriodicus}) is that a complicated multi-periodic solution of this form may be obtained from a given Schottky double. In fact, $f_m$ is given as a combination of Riemann theta functions associated with the Riemann surface.

\subsection{Modulations Equations}

Given a multi-periodic solution, corresponding to some Schottky double, or equivalently, to a algebraic Riemann surface, one may consider the effect of applying  any perturbation to the nonlinear equations. The effect of such a perturbation would be to slowly change the nature of the multi-periodic solution. Namely, the amplitude, phase, average value and frequency would change. After a while, the solution would be described by a different Schottky double, than the one describing the solution at initial times. We may describe the modulation of the multi-periodic wave as dynamics of the Schottky double or the algebraic curve. 

The form of the dynamics of the Schottky double naturally depends on perturbation applies. It turns out, however, that if the strength of the perturbation is sent to zero, the original, unperturbed, solution is {\it not} recovered. In other words, the limit of vanishing perturbation is singular. Nevertheless, the limit of zero perturbation is universal, in that for sufficiently well behaved perturbations, taking the amplitude of the perturbation to zero yields the same universal Whitham dynamics on Schottky doubles. This leads to the interesting situation in which there is a natural dynamics defined on  Schottky doubles or, equivalently,  algebraic curves. For the case of the reduced two dimensional Toda lattice, the universal dynamics obtained is that of Laplacian growth, which was already described as the dynamics of the algebraic Riemann surface $\mathcal{A}$. 

Whitham \cite{Carroll:Remarks:On:Whitham,Whitham:Book,Whitham:1966:NDW,Whitham:Lagrangian,Fucito:Gamba:NonlinearWKB} developed some of the first tools to describe such universal modulated dynamics for nonlinear wave. Flaschka, Forest and McLaughlin connected this evolution to concepts in algebraic geometry, while Krichever  \cite{83:Krichever:Averaging}
was able to derive the dynamics rigorously making use of an averaging method that he had devised. The resulting dynamics has an integrable structure  (as we have seen in Laplacian growth above). Such dynamics are sometimes called dispersionless hierarchies or Whitham hierarchies. Appendix \ref{WhithamAppendix} is devoted to a review of  Krichever's work\cite{83:Krichever:Averaging}, following the more accessible presentation  of Ref. \cite{Fucito:Gamba:NonlinearWKB,Carroll:Remarks:On:Whitham}. 

The modulation equations obtained by this procedure  are Eqs. (\ref{WhithamEqsforHs}). They are obtained by essentially averaging the zero curvature conditions, Eqs. (\ref{ZeroCurvatureDispersionfull}), (\ref{ZeroCurvatureDispersionfullDagger}), with the Baker-Akhiezer function, and its adjoint over many periods of the multi-periodic solution, but on a scale much smaller than the typical scale of modulation. The details are given in appendix \ref{WhithamAppendix}. 

Let us note that Eqs. (\ref{WhithamEqsforHs}) can be solved by taking a Riemann surface which is times independent. Indeed, as was describe in  section \ref{AlgebraicGeometricLPG},  the Hamiltonians $H_k$ can be found by considering that $dH_k$ is a unique differential on the Riemann surface, such that if the Riemann surface does not change then both left and right hand side of (\ref{WhithamEqsforHs}) are zero. Nevertheless, such a solution is obviously not a solution of the Laplacian growth problem. Mathematically, we see this by noting that we are seeking solutions in which there is a Schwarz function that generates this Hamiltonian (Eq. (\ref{SofZevolution})) and obeys the unitarity condition (Eq.  (\ref{Unitarity})).
The fact that given initial conditions two solutions exist, is related to the fact that the Whitham equations are a singular limit obtained by adding a perturbation to the equations and sending this perturbation to zero. The solution which corresponds to Laplacian growth is the non-trivial solution in which the Riemann surface is time dependent and in which the unitarity condition is preserved by the evolution.

\section{The Classical Korteweg-de Vries  Limit\label{KdVLimit} }
After discussing how to remove the dispersionless limit from the integrable structure of Laplacian growth to obtain the (dispersionful) reduced two dimensional 
Toda lattice hierarchy, we would like to quantize the latter. A lot more is known, though, on the quantization of  Korteweg-de Vries equation than the quantization of the two dimensional Toda lattice. In fact, quantum Korteweg-de Vries is the integrable structure underlying conformal field theory, which makes this quantum  integrable system very appealing, due to the rich structure and the applicability of results of the representation theory of the Virasoro algebra. 

Due to these consideration, we opt to first treat a limit of Laplacian growth, which has the integrable structure of Korteweg-de Vries associated with it
 \cite{KdVPaper}. Only then will we consider quantization. The advantage of this approach is that we will be able to use results from conformal field theory after quantization, the drawback is that it is not clear to what extent the fractal nature of the growth extends to the Korteweg-de Vries limit. \

\subsection{Narrow Finger Limit of Hele-Shaw Flows}
\begin{figure}[h!!]
\centering
\includegraphics[width=0.6\columnwidth]{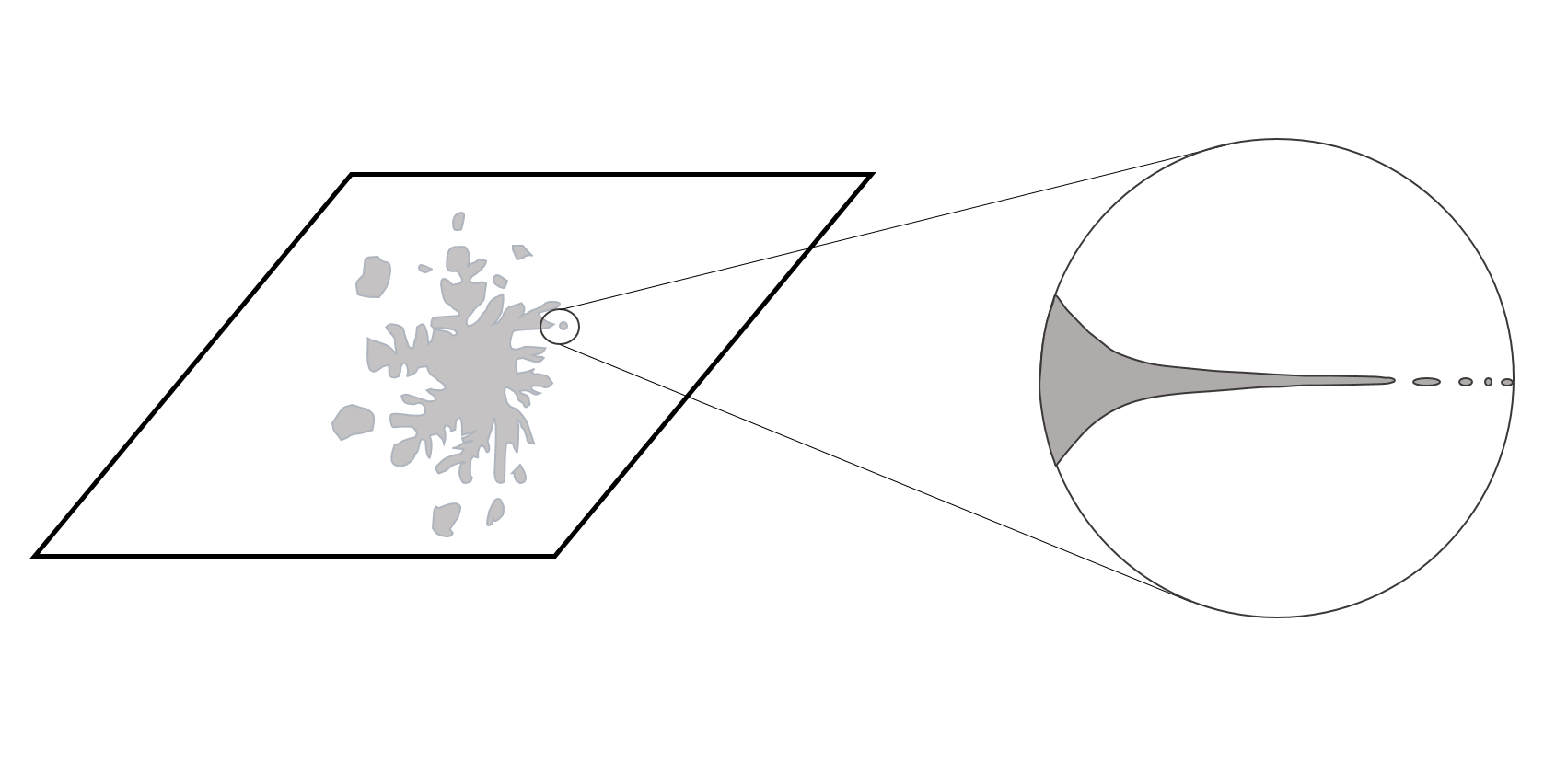}
\caption{A small part of the droplet is magnified, to reveal a long and narrow, possibly multi-connected, finger, symmetric with respect to reflections about the real axis. \label{KdVLimitPic}}
\end{figure}

If one focuses on a region around a Hele-Shaw finger, the integrable structure of the two dimensional Toda lattice  simplifies in the limit of  a very thin and long finger. Let us orient the $x$ axis to point in the long direction of the finger (see Fig. \ref{KdVLimitPic}). The Korteweg-de Vries limit occurs when the finger is symmetric to reflection across the $x$-axis and when the curve is described at intermediately  large and negative $x$ as \begin{align}y= y_0  \left(\frac{x}{x_0}\right)^{J-3/2}+\dots,\label{LongNarrowAsymptiote} \end{align}
where $x_0\gg y_0$, so that the finger is long and narrow. Eq.
(\ref{LongNarrowAsymptiote}) describes the shape of the finger at $x\sim x_0$, and may have  more complicated features at $x\ll x_0$. Nevertheless, it is assumed that throughout the region, for all points on the interface, $y\ll x$. Due to this $f_t(\imath\varphi)$ is approximately real in the entire region.

The Korteweg-de Vries description of the finger holds for any rapid evolution of the interface at $|x|\ll x_0$, while the asymptotic behavior at $x\sim x_0$, Eq. (\ref{LongNarrowAsymptiote}), does not change. Namely, in Eq. (\ref{LongNarrowAsymptiote}), we may assume that $x_0$ and $y_0$ are time independent. Furthermore, since the Korteweg-de Vries description only holds for the confined region $x\ll x_0,$ we may assume that Eq. (\ref{LongNarrowAsymptiote}) describes the asymptotic behavior of the not only at $x\sim x_0$ but all the way to $x\to-\infty.$ 

To describe  such a finger, we first define a re-scaled and shifted $\varphi$ and $t$ given by  $\tilde \varphi  $ and $\tilde t$, respectively. Namely, we define $\tilde \varphi(\varphi) = a\varphi+b$, $\tilde t= \alpha t +\beta$, where  the way $a, b, \alpha, \beta$ are chosen will be described below. Let:
\begin{align}
l_{\tilde{t}}(\tilde\varphi)=\frac{f_{t(\tilde t)}(\imath\varphi(\tilde\varphi))+\bar f_{t(\tilde t)}(-\imath\varphi(\tilde\varphi))}{2}, \quad  h_{\tilde t}(\tilde\varphi)=\frac{f_{t(\tilde t)}(\imath\varphi(\tilde\varphi))-\bar f_{t(\tilde t)}(-\imath\varphi(\tilde\varphi))}{2\imath} .\label{l-hdef}
\end{align}
We may now  expand $l_{\tilde t}$, $h_{\tilde t}$ as a Laurent series in $\tilde \varphi$ around infinity, such that the asymptotics described in Eq. (\ref{LongNarrowAsymptiote}) is recovered.The simplest choice is to take:
\begin{align}
l_{\tilde t}= \tilde \varphi^2+\sum_{j=-\infty}^0 \tilde a^{(1)}_j \tilde \varphi^{2j},\label{mapExpansionKdV}
\end{align}
and\begin{align}
h_{\tilde t}=\frac{y_0}{x_0^{J-3/2}} \tilde \varphi^{2J-3} + O\left(\tilde\varphi^{2J-1}\right).\label{hExpansion}
\end{align} 
The choice of $a$ and $b$ in  $\tilde \varphi(\varphi) = a\varphi+b$, is made such that the free term in (\ref{mapExpansionKdV})\  is missing and the highest order term has unit pre-factor.   The fact that $l_{\tilde t}$ has only even powers in an expansion in $\tilde\varphi$ and $h_{\tilde t}$ only odd powers, corresponds to the symmetry of the finger with respect to reflections across the real axis. Indeed, together with (\ref{hExpansion}), the reflection symmetry is effected by $\tilde \varphi \to-\tilde \varphi$, whereby $l_{\tilde t}\to l_{\tilde t}$, while $h_{\tilde t}\to -h_{\tilde t}$. The fact that this way to respect the reflection symmetry is actually implemented by the conformal map $f_{t}$ in the asymptotic region, is not guaranteed. Nevertheless, it may be expected to hold fairly generally, when the shape of the interface away from the finger does not induce a strongly asymmetric pressure field on the finger. We shall not dwell on this point, as we are mainly interested in the existence of the Korteweg-de Vries limit under some conditions, rather than demanding that it hold  under generic conditions.

Instead of the Schwarz function it is more appropriate to define the following combination:
\begin{align}
 \quad y(x)=h_{\tilde t}(\tilde \varphi_{\tilde t}(x)) ,\quad 
\end{align}
where $\tilde \varphi_{\tilde t}(x)$ is the inverse function of $l_{\tilde t}$:
\begin{align}
l_{\tilde t}(\tilde\varphi_{\tilde t}(x))=x.
\end{align}

The function $y(x)$ is the analytic continuation in $x-$plane of a function which gives the $y$ coordinate of the finger as a function of the $x$ coordinate, for $x$ that belongs to the droplet. Such $x$'s that below to the finger form segments on the real axis. These segments are branch cuts for the function $y(x)$. If one approaches the branch cut from above or below  one obtains the $\pm y(x)$, respectively. Namely as $x$ approaches the real axis we have $y(\bar x)=-y(x)$. Since $y$ is real we can take the complex conjugate of the left hand side of this equation and obtain:
\begin{align}
\bar y(x)=-y(x). \label{yreflection}
\end{align}
This equation holding for $x$ approaching the real axis. 

Around the contour we have ostensibly $y(x)=\sum_n a_n (\sqrt{x-\lambda_i})^n$. To obey (\ref{yreflection}), $a_{2n}$ have to be purely imaginary, but $y$ is real on the real axis and so $a_{2n}=0$. Thus, $y(x)$ has only odd powers of $\sqrt{x-\lambda_i}$ which implies that $y^2(x)$ is regular, as a function of $x$,  everywhere in the complex plane. Namely, it is an entire function. In addition, $y^2(x)\sim x^{2J-3}$ as $x\to \infty,$ which, together with the fact that it is entire, suggests that it is a  polynomial. Considering that the zeros of this polynomial  are the branch points, one obtains: 
\begin{align}
y^2(x) = \prod_{i=1}^{2g+1}(x-\lambda_i).\label{yofxhyperelliptic}
\end{align} 
This equation describes a Riemann surface, through a complex curve. Curves which are defined through $y^2=P_n(x)$, where $P_n$ is a polynomial, are called hyper-elliptic. Thus, in the Korteweg-de Vries limit, the Riemann surface is a hyper-elliptic curve.

Given the definition, Eq. (\ref{l-hdef}), of $l_{\tilde t}$ and $h_{\tilde t}$, Eq. (\ref{StringEq}) implies the following Poisson bracket for these objects:\begin{align}
\{l_{\tilde t},h_{\tilde t}\}=1,\label{StringEql-h}
\end{align}
where here the choice of $\alpha$ in $\tilde t=\alpha t+\beta$ is made as to have $1$ on the right hand side. The choice of $\beta$ remains arbitrary.
Eqs. (\ref{mapExpansionKdV}),  (\ref{hExpansion}) leads to the following expansion for $y$:
\begin{align}
y(x) = \sum_{k=1}^J \tilde{t}_{2k+1} x^{k-1/2} +O(z^{-3/2}).\label{yofxexpansion}
\end{align}
with $\tilde t_1=\tilde t$.   All the $\tilde t_{2k+1}$ are real parameters, as the symmetry of the finger dictates.

In a manner following closely to the way in which Eq. (\ref{dSdtISdphidz})
was derived from (\ref{StringEq}),  one may  derive from (\ref{StringEql-h}) the following equation:

\begin{align}\frac{\partial y(x)}{\partial\tilde  t}=\frac{\partial \tilde \varphi_{\tilde t}(x)}{\partial x}.
\end{align}

Define now the Hamiltonians 
\begin{align}
\tilde H_{2k+1} =\left( x_{\tilde t}^{k+1/2} \right)_{\tilde{+}},\label{HkDispersionless}
\end{align}
where the definition of $g(\tilde \varphi)_{\tilde +}$ , for any function $g(\tilde\varphi) = \sum_{k=-\infty}^\infty \alpha_k \tilde\varphi^{k}$ is as follows:
\begin{align}
g_{\tilde +}=\frac{\alpha_0}{2}+\sum_{k=1}^\infty \alpha_k \tilde \varphi^k.\\ 
\end{align}
Assume 
\begin{align}
\frac{\partial l_{\tilde t}}{\partial \tilde t_k } = \{x_{\tilde t},\tilde H_k\},\quad \frac{\partial h_{\tilde t}}{\partial \tilde t_k } = \{h_{\tilde t},\tilde H_k\},\label{DispersionlessKdV}  
\end{align}
for $k$ odd. Then, 
\begin{align}
&\frac{\partial y(x)}{\partial \tilde t_k } =\left.  \frac{\partial l_{\tilde{t}}(\tilde \varphi)}{\partial \tilde t_k} +\frac{\partial l_{\tilde{t}}(\tilde \varphi)}{\partial \tilde \varphi}  \frac{\partial \tilde\varphi_t(x)}{\partial \tilde t_k} \right|_{\tilde\varphi=\tilde \varphi_t(x)} = \\ 
&=\left.\{ l_{\tilde{t}},\tilde H_k\}-\left( \frac{\partial \tilde l_t(\tilde \varphi) }{\partial\tilde\varphi}\right)^{-1}\frac{\partial l_{\tilde{t}}(\tilde \varphi)}{\partial \tilde \varphi}  \frac{\partial \tilde l_t(\tilde \varphi)}{\partial \tilde t_k}  \right|_{\tilde\varphi=\tilde \varphi_t(x)}=\nonumber \\
&=\left. \frac{\partial \tilde H_k}{\partial \tilde\varphi}/\frac{\partial \tilde l_t(\tilde \varphi) }{\partial\tilde\varphi}\right|_{\tilde\varphi=\tilde \varphi_t(x)} = \frac{\partial \tilde H_k}{\partial x}
\end{align}
This equation is an evolution equation in time $\tilde t_k$ for $y$, and has the property that it retains the form (\ref{yofxexpansion}). This shows that the times $\tilde t_k$ may be viewed as the analogue of the harmonic moments in the narrow finger limit. First, each $\tilde t_k$ is conserved when any on of the other times is changed and secondly, the $\tilde t_k$ are quantities characterizing the shape of the interface. Indeed, the function $y(x)$, for which the $\tilde t_k$ are Laurent coefficients, can be found by directly from the shape of the interface, as it is the analytic continuation in the complex $x-$plane  of the function which gives the $y$ coordinate of the interface given the $x$ coordinate.  
\subsection{Dispersive Quantization of Korteweg-de Vries}
Just as we have applied `dispersive quantization' the the dispersionless limit of the two dimensional Toda lattice equations, we shall now be interested in applying the same procedure the the dispersionless Korteweg-de Vries equations (\ref{DispersionlessKdV}).  We shall at this point do away with the tildes over the time, over the $\varphi$ variable, and the Hamiltonians, since we shall not treat the two dimensionless Toda anymore, and as such there should be no confusion about whether the objects discussed belong to the Korteweg-de Vries or the two dimensional Toda hierarchies.

The Korteweg-de Vries hierarchy may be defined (See Ref. \cite{Das:Solitons} and references therein) by considering the Lax operator, $L$,  a differential operator in $t_1$,  given by:
\begin{align}
\hat L= \partial_{t_1}^2+\frac{u(t_1)}{6}  .
\end{align}
One may write the square root of $\hat L$ as a formal power series in $\partial_{t_1}$:
\begin{align}
\hat L^{1/2} = \partial_{t_1} +\sum_{m=0}^\infty b_m (t_1)\partial_{t_1}^{-m},
\end{align}
where $\partial_{t_1}^{-m}$  is the formal inverse of $\partial_{t_1}$ raised to the power $m$. The function $b_m(t_1)$ can be found, order by order, by demanding $\hat L^{1/2}\hat L^{1/2} = \hat L$. 

A set of Hamiltonians are then given by 
\begin{align}
\hat H_{2k+1}=\left(\hat L^{1/2}\hat L^{k}\right)_+,
\end{align}
where the $(\hat L^{k+1/2})_+$ denotes only positive powers of $\partial_{t_1}$ in the expansion of $L^{1/2}\hat L$, which  itself contains both positive and negative powers. The equations for the hierarchy follow from the Lax equations:
\begin{align}
\partial_{t_{k}} \hat L = [\hat L,\hat H_k]1\label{LaxKdV}
\end{align}

The dispersionless limit of this problem becomes the problem that has been discussed in the previous subsection. To show this one may follow much the same as described in for the two dimensional Toda lattice. In fact the result may be obtained by taking an appropriate limit of that procedure. Namely, the long and narrow finger limit, in which the Riemann surface becomes hyperelliptic, and only real odd times describe the evolution. As such, we shall not repeat those steps. We note, that the original literature cited here on the matter \cite{83:Krichever:Averaging,80:Krichever:Spect:Theory,88:Flaschka:KdV:Avging}
either deals directly with the Korteweg-de Vries problem, or with close relatives, which are easily reduced to the Korteweg-de Vries problem -- mainly the Kadomtsev-Petviashvili hierarchy. 

The first member of the hierarchy gives the tautology $\partial_{t_1}u=\partial_{t_1}u. $The first non-trivial member of the hierarchy is given by:\begin{align}\label{KdV}
4\partial_{t_3}u = \partial_{t_1}^3u + u \partial_{t_1}u.
\end{align}
Alternative to the Lax construction, this equation may be obtained by  defining the following Poisson brackets:
\begin{align}
\{u(x),u(y)\} = 2(u(x) + u(y))\delta'(x-y) + \delta'''(x-y),\label{ClassicalPoissonT}
\end{align}
and allowing for evolution with the Hamiltonian:
\begin{align}
H_3= \frac{1}{8} \oint u^2 (t_1) dt_1,
\end{align}
upon which, Eq. (\ref{KdV}) becomes equivalent to:
\begin{align}
\partial_{t_3} u = \{u,H_3\}.
\end{align}
The tautological equation $\partial_{t_1}u=\partial_{t_1}u   $   is generated by $H_1$ given by
\begin{align}
H_1=\int u(t_1) dt_1.
\end{align}
All higher order flows defined by (\ref{LaxKdV}) may be obtained using the Poisson bracket (\ref{ClassicalPoissonT}) by defining appropriate  Hamiltonians (See Ref. \cite{Das:Solitons} and references therein) .

\subsection{Inverse Scattering\label{KdVInveseScatteringSection}}

The method of inverse scattering was first invented for the Korteweg-de Vries equation\cite{Gardner:Greene:Kruskal:Miurac}
and later expanded to become a wide-ranging subject (see, e.g., 
Ref. \cite{Das:Solitons} and references therein).  It is useful to review briefly the inverse scattering method, since it provides a link between the classical multi-periodic solutions to the quantum states one obtains in solving the integrable quantum Korteweg-de Vries equation.  

To present the method of inverse scattering, first let us consider the periodic problem. Namely, we consider the problem in which the field $u(t_1)$ is periodic with period $2\pi$. Next, we consider the Miura transformation:
\begin{align}u = -\left(\frac{\partial\phi}{\partial t_1}\right)^2 - \frac{\partial^2\phi}{\partial t^2_1}.\label{Miura}\end{align}
The Poisson bracket for the field $\phi$ is then given by:
\begin{align}
\{\phi(t_1),\phi(t_1')\} =\frac{1}{2} \sign (t_1-t_1'), \label{BosonicPoissonTheta}
\end{align}
and can be shown to be equivalent to (\ref{ClassicalPoissonT}).  Namely,  substituting (\ref{Miura}) into (\ref{BosonicPoissonTheta}) one obtains (\ref{ClassicalPoissonT}) up to full derivatives. The latter are determined by requiring that the Poisson bracket of $\phi$ must be obey the Jacobi identity. 

The Korteweg-de Vries equation, Eq. (\ref{KdV}), translates into the following equation for $\phi$:
\begin{align}
\frac{\partial \phi}{\partial t_3} = \frac{\partial^3\phi}{\partial t^3_1}-2\left(\frac{\partial\phi}{\partial t_1}\right)^3   , \label{mmKdV}
\end{align}
which is a version of the modified Korteweg-de Vries equation (the standard modified Korteweg-de Vries equation is written for the field $v=\phi'$). 

The equation for $\phi$, Eq. (\ref{mmKdV}) can be obtained for a zero curvature condition for a  Lax pair. Consider the operators 
\begin{align}
\bm A_\lambda &=  \phi'  \sigma_z +\sqrt{\lambda}  \sigma_x, \label{AMatrixmKdV}\\
\bm B_\lambda&=\sqrt{\lambda}(4\lambda-2 \phi'^2)\sigma_x+(\phi'''+\phi'(4\lambda-2\phi'^2))\sigma_z - 2\imath\phi'' \sqrt{\lambda} \sigma_y, \label{BMatrixMKdV}  
\end{align}
where the prime denotes derivative with respect to $t_1$. 
Then the equation 
\begin{align}
\left[\frac{\partial}{\partial t_1} -\bm  A_\lambda   , \frac{\partial}{\partial t_3}-\bm B_\lambda\right]=0,\label{ZeroCurvatureMKdV}
\end{align}
becomes equivalent to (\ref{mmKdV}) as can be ascertained by computing the commutator explicitly.

 Eq. (\ref{ZeroCurvatureMKdV}), which is equivalent to the dynamic equation of $\phi$, Eq. (\ref{mmKdV}), has exactly the required form to serve as the consistency condition required for there to be a solution, $\vec{\psi}$, to the equations:
\begin{align}
\left(\frac{\partial}{\partial t_1} -\bm  A_\lambda \right)\vec{\psi}=0 \label{xEquationMKdV} \\
\left( \frac{\partial}{\partial t_3}-\bm B_\lambda \right)\vec{\psi} =0 \label{tEquationMKdV} 
\end{align} 
where $\vec{\psi}$  is a two dimensional vector. The significance of this fact is that one may approach the problem of finding solutions of the modified Korteweg-de Vries equation, Eq. (\ref{mmKdV}), or equivalently the Korteweg-de Vries equation, Eq. (\ref{KdV}), as the problem of constructing solutions to (\ref{xEquationMKdV}) and (\ref{tEquationMKdV}) as Baker Akhiezer functions as done above. In this section we shall show how the Baker-Akhiezer function is related to the method of separation of variables.
 
The solution of (\ref{xEquationMKdV}) is in fact simple. We may write:
\begin{align}
\vec\psi(t_1) = \mathscr{P}\left[e^{\int_0^{t_1} \bm A_\lambda } \right] \vec{\psi}(0),
\end{align}
where $\mathscr{P}$ denotes a path-ordered product. One is lead to define the shift operator:
\begin{align}
\bm M_\lambda(t'_1,t_1)=e^{\int_{t_1}^{t_1'} A_\lambda }.\label{MDef}
\end{align}

The problem of finding Bloch  wave-functions $\vec{\psi}(t_1+2\pi) = e^{\imath \theta(\lambda)} \vec\psi(t_1),$ satisfying (\ref{xEquationMKdV}) and (\ref{tEquationMKdV}) is then the problem of finding eigenvectors of the matrix $\bm M_\lambda(2\pi,0)$.  The object $ e^{\imath \theta(\lambda)} $ is called the Bloch multiplier of the Bloch wave-function $\vec \psi(t_1).$ Note that the first element of $\vec \psi$, denoted by $\psi_1$, is also  Bloch wave-functions for the following equation:  
\begin{align}
(\partial_{t_1}^2 + u (t_1))\psi_1 =\hat L\psi_1= \lambda \psi_1,\label{KdVSpectrakSimple} 
\end{align}
an equation which is easily derived making use of Eqs. (\ref{xEquationMKdV}) and (\ref{Miura}). A Bloch eigenfunction  of the Lax operator
is the Baker-Akhiezer function. Indeed, the Baker-Akhiezer function is a solution to the spectral problem, where the coefficients appearing in the Lax operator are multi-periodic functions. 

A special case in which the eigenvectors  of $\bm M_\lambda(2\pi,0)$ are apparent presents itself for values of $\lambda$ at which $\bm M_\lambda(2\pi,0)$ happens to be  diagonal.
The the Bloch wave-functions are given by the standard basis
\begin{align}
\left( \begin{array}{c} 1 \\0 \end{array} \right), \left( \begin{array}{c} 0 \\1 \end{array} \right),
\end{align}
up to a times dependent factor. We see that the first element of one of the Bloch function vanishes. Namely the Baker-Akhiezer function vanishes at this $\lambda$ on one of the sheets. The matrix $\bm M_\lambda$ depends on the times and as such the points $\lambda$ at which this matrix is diagonal also depend on time. Suppose that there are $g$ such points. Indeed, on  a Riemann surface of genus $g$ the Baker-Akhiezer function has $g$ zeros (see appendix \ref{BAexplicitSection}). These points form a set of $g$ dynamical variables denoted by $\gamma_i$.  Another set of dynamical parameter may be taken as the Bloch multiplier associated with the non-vanishing Baker-Akhiezer function at $\gamma_i$. Namely we take the upper left element, $\left(\bm M_{\gamma_i}\right)_{11},$ as a set of $g$ additional dynamical variables.    As shown in Ref.\cite{Flaschka:McLaughlin:Canonical:Conjugate:Variables} and reviewed below, these parameters have a Poisson structure that allows one to treat them as separated variables. We review the computation of  these Poisson brackets in the following, by  computing the Poisson brackets of the elements of  $\bm M_\lambda(2\pi,0)$ for any $\lambda$ and then specializing to the points where  $\bm M_\lambda(2\pi,0)$ is diagonal.  

We wish to find the Poisson brackets of the different elements of $\bm M_\lambda(2\pi,0).$ Since we are interested in finding the Poisson bracket of any two matrix elements of $\bm M_\lambda(2\pi,0)$, it is useful to consider the simultaneous operation of taking the tensor product and the Poisson bracket. Namely for any two matrices, $\bm a$ and $\bm b$, one may consider the object $\{\bm a \overset{\otimes}{,}\bm b\}$, which is an object which lives in the same space as $\bm a \otimes \bm b$, and has matrix elements given by:
\begin{align}
\{ \bm a \overset{\otimes}{,} \bm b\}_{\alpha \beta \gamma \delta} =\{\bm a_{\alpha \gamma},\bm b_{\beta \delta} \},  
\end{align}  
just as $(\bm a \otimes \bm b)_{\alpha \beta \gamma \delta} = \bm a_{\alpha \gamma} \bm b _{\beta \delta}$. 

From now on we shall denote 
\begin{align}
\bm M_\tau \equiv \bm M_\tau(2\pi,0)\label{TotalMDef}
\end{align} 
With this definition we compute:
\begin{align}
\{\bm M_\lambda\overset{\otimes}{,}\bm M_\mu\}\label{MMotimesPoisson1st}=\iint dx dx' \bm M_\lambda(2\pi,x)\otimes \bm M_\mu(2\pi,x') \{\bm A_\lambda(x)\overset{\otimes}{,} \bm A_\mu(x')\}\bm M_\lambda(x,0)\otimes \bm M_\mu( x',0)
\end{align}
This equation is a  consequence of the definition of $\bm M$ in terms of $\bm A$ (Eq. (\ref{MDef})) only, and may be derived by writing $\bm M$  in terms of a path ordered product:
\begin{align}
\bm M_\lambda = \lim_{N \to \infty} \mathscr{P}\prod_{i}^{N} \left(\mathds{1}+\int_{\frac{2\pi}{N}(i-1)}^{\frac{2\pi}{N}i}\bm A_\lambda   \right),
\end{align}
and computing the Poisson brackets that appear in (\ref{MMotimesPoisson1st}).

An explicit calculation
gives
\begin{align}
 \{\bm A_\lambda(x)\overset{\otimes}{,} \bm A_\mu(x')\} =\delta'(x-x') \sigma_z \otimes\sigma_z
\end{align}
Inserting this into (\ref{MMotimesPoisson1st}), writing $\delta'(x-y) = \frac{\partial_x - \partial_y}{2} \delta(x-y)$ and integrating by parts, one obtains:
\begin{align}
&\{\bm M_\lambda\overset{\otimes}{,}\bm M_\mu\}\label{MMotimesPoisson2nd}=\\
&\nonumber=\int dx \bm M_\lambda(2\pi,x)\otimes \bm M_\mu(2\pi,x)\left[ \sigma_z \otimes\sigma_z,\frac{\bm A_\lambda(x) \otimes\mathds{1}-\mathds{1\otimes}\bm A_\mu(x)}{2} \right]\bm M_\lambda(x,0)\otimes \bm M_\mu( x,0),
\end{align}
where here we used the fact that $\bm M_\tau,$   obeys similar equations to those that  $\vec \psi$ obeys:
\begin{align}
\partial_x \bm M_\tau (x,x')= \bm A_\tau (x)\bm M_\tau(x,x'), \quad \partial_{x'} \bm M_\tau (x,x')= -\bm M_\tau(x,x')\bm A_\tau (x'), 
\label{Mequations}
\end{align} 
whereupon one may substitute $\mu$ or $\lambda$ for $\tau$. 

Finally, we need the following identity, which may be obtained by an explicit calculation:
\begin{align}
\left[ \sigma_z \otimes\sigma_z,\bm A_\lambda(x) \otimes\mathds{1}- \mathds{1\otimes}\bm A_\mu(x)\right] = \left[\bm R_{\sqrt{\lambda/\mu}},\bm A_\lambda(x) \otimes\mathds{1}+ \mathds{1\otimes}\bm A_\mu(x)\right],\label{ZwischenPoisson}
\end{align}
where
\begin{align}
\bm R _\tau =\frac{\tau+\tau^{-1}}{\tau-\tau^{-1}} \frac{\sigma_z\otimes\sigma_z}{2} + \frac{2}{\tau-\tau^{-1}}(\sigma_+ \otimes \sigma_- + \sigma_-\otimes\sigma_+).
\end{align}
Substituting (\ref{ZwischenPoisson}) into (\ref{MMotimesPoisson2nd}) and making use  Eq. (\ref{Mequations}), one may write:
\begin{align}
&\{\bm M_\lambda\overset{\otimes}{,}\bm M_\mu\}=\nonumber\\
&=\int dx\partial_x \left( \bm M_\lambda(2\pi,x)\otimes \bm M_\mu(2\pi,x)\bm R_{\sqrt{\lambda/\mu}}\bm M_\lambda(x,0)\otimes \bm M_\mu(  x,0)\right) =\nonumber \\ &= \left[\bm R_{\sqrt{\lambda/\mu}},\bm M_\lambda\otimes\bm M_\mu\right] ,\label{Sklyanin}
\end{align}
This is the desired result. In the present, classical, terms, the equation encodes the Poisson brackets between the elements of $\bm M$ at different spectral parameters in terms of bilinears of $\bm M$ itself. The quantization of these relations will allow to compute the commutations relations for such elements in the same manner. 

An almost immediate consequence of (\ref{Sklyanin}) is that the trace of $\bm M_\lambda$, denoted as $\bm T_\lambda$:
\begin{align}
\bm T_\lambda  = \tr \bm M_\lambda,
\end{align}
Poisson commutes
\begin{align}
\{\bm T_\lambda , \bm T_\mu \} = 0,\label{TsCommute}
\end{align}
an equation which is immediately obtained from (\ref{Sklyanin}) by noting that $\tr \left[\{\bm a\overset{\otimes}{,}\bm b\} \right]= \{\tr [\bm a ], \tr [\bm b]\}$, just as $\tr \left[\bm a \otimes \bm b\right] = \tr[\bm a]\tr[ \bm b]$. 

The relation (\ref{TsCommute}) encodes the infinite number of conserved quantities in involution. Indeed, it turns out that the expansion of $\bm T_\lambda$ around $ \lambda \to \infty$, yields the conserved quantities as the coefficients in the Laurent expansion. This proves that they are in involution. 

It is customary to define the following matrix elements of the matrix $\bm M_\tau$:
\begin{align}
\bm M_\tau  = \left(\begin{array}{cc} 
a_\tau & b_\tau  \\
c_\tau & d_\tau
\end{array}\right).
\end{align} 
 To obtain separated variables one first notes that from (\ref{Sklyanin}) the following relation may be obtained by taking the appropriate matrix element:
\begin{align} 
\{b_\lambda,b_\mu\}=0.\label{bPoissonCommutes}
\end{align}
Consider the points at which $\bm M_\tau$ is diagonal. At these points, denoted by $\gamma_i$ we have $b_{\gamma_i}=0$. The following is then a consequence of (\ref{bPoissonCommutes}):
\begin{align}
\{\gamma_i,\gamma_j\}=0.
\end{align}
We have found a set of variables which are in involution. As further dynamical variables one may consider $\Lambda_i=a_{\gamma_i}$. Referring back to (\ref{Sklyanin}) and taking the appropriate matrix elements one concludes:
\begin{align}
\{\Lambda_i,\Lambda_j\}=0,
\end{align}
and
\begin{align}
\{\Lambda_i,b_{\gamma_j}\} = 4\delta_{i,j} \Lambda_i \gamma_i b'_{\gamma_i},\label{LambdabPoisoon}
\end{align}
where $b'_{\gamma_i}=\left.\frac{db_\tau}{d\tau} \right|_{\tau=\gamma_i}$. Eq. (\ref{LambdabPoisoon}) may be obtained if we postulate the following Poisson brackets:
\begin{align}
\{\Lambda_i,\gamma_j\} =4 \delta_{i,j}\Lambda_i \gamma_i.
\end{align} 
We thus have:
\begin{align}
\{\Lambda_i,\Lambda_j\} = \{\gamma_i,\gamma_j\}=0,\quad \{\log \Lambda_i,\log \gamma_j\} = \delta_{i,j}\label{ClassicalGammagammaPoissons}.
\end{align}

The variables $\log(\gamma_i)$ and $\log(\Lambda_i)$ are thus canonical. To show that these are also separated variables, one  needs to show that each pair $(\gamma_i, \Lambda_i)$ traces out a one dimensional loop, whose shape is independent of all the rest of the canonical variables. Namely, one must show that $\Lambda_i$ is a function of $\gamma_i$, the function depending only on the conserved quantities. The existence of such a representation is a consequence  of the fact that $\Lambda_i$ is the Bloch multiplier associated with the Baker-Akhiezer function at $\gamma_i$. An explicit expression for the Baker-Akhiezer function is given in appendix \ref{BAexplicitSection}, from which $\Lambda(z)$ may be found. We thus have $\Lambda_i= \Lambda(\gamma_i)$, as required. 

\section{ Quantum Korteweg-de Vries\label{qKdVSection}}
Having reviewed the classical inverse scattering method, we are able to present the quantum version of which,  discovered in Refrs.
\cite{Zamolodchikov:KdV:I,Zamolodchikov:KdV:II,Zamolodchikov:KdV:III} based on  results of Refrs. \cite{Fateev:Lukyanov:Poisson-Lie:PregenatorToqKdV,Fateev:Lukyanov:Vertex:PregenatorToqKdV}. More context on the quantum inverse scattering can be found in Ref. \ \cite{Faddeev:Book:Hamiltonian:Methods}
and is also treated briefly in Ref.  \cite{Das:Solitons}. We then go on to review the quantization of the Korteweg-de Vries problem based on quantizing the field $u(t_1)$, which leads to the connection to conformal field theory. An object which bridges the two approaches is given by the form factors, which has a central role in the proposed connection to Laplacian growth. 

\subsection{Quantum Separation of Variables\label{QSOV}}

The classical separation of variables method for the Korteweg-de Vries problem was achieved by considering the monodromy matrix. Quantum Korteweg-de Vries
can be defined by quantizing in the separate variables, in a method suggested by Sklyanin \cite{Sklyanin:SoV}.
On defines the quantum analogues of $\gamma_i$ and $\Lambda_j$ as  a set of operator $\hat \gamma_i$, $\hat \Lambda_i$,  with the following commutation relations: \begin{align}
[ \hat \Lambda_i, \hat \Lambda_j] = [ \hat \gamma_i, \hat \gamma_j]=0,\quad \hat\Lambda_i \hat\gamma_j = e^{\imath \pi \hbar\delta_{i,j}}\hat\gamma_j \hat\Lambda_i.\label{QuantumGammagammaCommutations}
\end{align}To see that indeed (\ref{QuantumGammagammaCommutations})  is the classical limit of the last equation of (\ref{ClassicalGammagammaPoissons}), we write the last equation in (\ref{QuantumGammagammaCommutations})\ as:
\begin{align}
e^{\log(\hat \Lambda_i) + \log(\hat \gamma_k)+\frac{1}{2}[\log(\Lambda_i), \log(\hat \gamma_k)] + \dots}=e^{\imath \pi \hbar\delta_{i,j}+\log(\hat \Lambda_i) + \log(\hat \gamma_k)-\frac{1}{2}[\log(\Lambda_i), \log(\hat \gamma_k)] + \dots},
\end{align}
obtained by using Baker-Campbell-Hausdorff Lemma. Taking the logarithm of both sides, one obtains:
\begin{align}
[\log \hat \Lambda_i,\log \gamma_j] = \delta_{i,j} \imath \pi \hbar + \dots . 
\end{align}
The semiclassical limit of this equation is indeed (\ref{ClassicalGammagammaPoissons}), as desired. 

 One may find a representation of this algebra by postulating some Hilbert space spanned by the joint eigenvectors of all the  $\hat \gamma_i$'s, which we shall denote by $|\vec \gamma\>$, with $\hat \gamma_i |\vec \gamma\> = \gamma_i | \vec \gamma\>$. The operator $\hat \Lambda_i$ can then be represented on such states as:
\begin{align}
\hat \Lambda_i |\vec \gamma\>= \epsilon_j e^{\imath \pi \hbar \gamma_j \frac{\partial}{\partial \gamma_j}} |\vec \gamma\>.
\end{align}

To make contact with a more usual field theory description of quantum Korteweg-de Vries, namely, a description in which the field $u(x)$ is quantized, the variables $\hat\Lambda_i$, $\hat\gamma_i,$ must be written through the quantum field $\hat u(x)$ . In fact it is more advantageous to start with the field Miura field, $\phi$ (see Eq. (\ref{Miura})), and quantize it, and then perform a quantum Miura transformation to the field $\hat u.$ To  this aim, we  define an operator $\hat \phi(x)$, analytic in the upper half $x$-plane, which satisfies the commutation relations:
\begin{align}
[\hat \phi(x),\hat \phi(y)] =\frac{\sign(x-y)}{\hbar}.\label{BosonCommutation}
\end{align}
The latter being an analogue of (\ref{BosonicPoissonTheta}). 

In analogy with  (\ref{MDef}), one may define an operator valued matrix wave-function, $\hat\Psi_\lambda(x)$ depending on the  spectral parameter $\lambda$, as follows:
\begin{align}
\hat \Psi_\lambda(x) =e^{\hat \phi(x)\tilde\sigma_z} \mathscr{P} \exp\left[ \lambda\int_0^x \left(e^{-2\hat\phi(x')}\tilde \sigma_+ + e^{2\hat\phi(x')}\tilde \sigma_-\right)dx'\right],\label{hatPhiDef}
\end{align}
where $\tilde \sigma_\pm$, $\tilde \sigma_z$ are matrices satisfying:
\begin{align}
[\tilde \sigma_z,\tilde \sigma_\pm]=\pm2\tilde\sigma_\pm,\quad [\tilde \sigma_+,\tilde \sigma_-]=\frac{e^{\imath \hbar \tilde \sigma_z}-e^{-\imath \hbar \tilde \sigma_z}}{e^{\imath \hbar}-e^{-\imath \hbar}},\label{su2q}
\end{align} 
which in the limit $\hbar\to0$ yields the usual Pauli matrices. $\hat \Psi_\lambda(x)$ is a operator-valued matrix  solution to a quantum analogue of Eqs. (\ref{xEquationMKdV}), (\ref{tEquationMKdV}). 

The rather unusual way in which the parameter $\hbar$ enters into (\ref{hatPhiDef}) through (\ref{su2q}) is necessary in order for  $\hat \Psi_\lambda(x)$ to
produce, through the monodromy matrix, a set of quantum variables satisfying (\ref{QuantumGammagammaCommutations}). Indeed, define the quantum monodromy matrix, $\bm M_\lambda$  as  the matrix $\hat \Psi_\lambda(x)$ at $x=2\pi$:
\begin{align}
\bm M_\lambda  =  \hat \Psi_\lambda(2\pi).\label{bmMdef} 
\end{align} 
It can be shown\cite{Zamolodchikov:KdV:III}, using the commutation relations (\ref{BosonCommutation}), the the following relation holds for $\bm M$: 
\begin{align}
R\left(\sqrt{\lambda/ \mu}\right) (\bm M_\lambda\otimes \mathds{1} )(\mathds{1}\otimes \bm M_\mu)=(\mathds{1}\otimes \bm M_\mu)(\bm M_\lambda\otimes \mathds{1} )R\left(\sqrt{\lambda/ \mu}\right) .\label{SklyaninQuantum}
\end{align}
 with
\begin{align}
R(\lambda)= \left( \begin{array}{cccc}
q^{-1}\lambda-q \lambda^{-1}  & 0 & 0  &0 \\
0 & \lambda-\lambda^{-1}& q^{-1} - q  & 0 \\
0 &q^{-1} - q  & \lambda-\lambda^{-1} &0 \\
0& 0 & 0  &q^{-1}\lambda-q \lambda^{-1}  
\end{array}\right)
\end{align}
This relation is the quantum analogue of (\ref{Sklyanin}). Indeed,  we set $q=e^{\imath \hbar}$ and write:
\begin{align}
\frac{R\left(\lambda\right)}{\lambda - \lambda^{-1}}=\mathds{1}-\imath \hbar\left(r(\lambda)+\frac{1}{2}\frac{\lambda+\lambda^{-1}}{\lambda-\lambda^{-1}}\mathds{1}  \right) + O(\hbar^2)\label{Rintermsofr}
\end{align}
If one substitutes (\ref{Rintermsofr}) into (\ref{SklyaninQuantum}) and uses the usual semiclassical expansion of the product of operators, $\hat A \hat B= A B+ \frac{\imath \hbar}{2}\{A,B\} + \dots $ one obtains (\ref{Sklyanin}) to first order in the expansion. Namely, the commutation relations between the matrix elements of $\bm M$, which are encoded in (\ref{SklyaninQuantum}) can be viewed as a quantization of the corresponding Poisson brackets, which are encoded in (\ref{Sklyanin}). Thus, one may proceed with a quantum version of the separation of variables method, based on (\ref{SklyaninQuantum}).

The proof of (\ref{SklyaninQuantum}) is rather cumbersome and Ref.  \cite{Zamolodchikov:KdV:III}
(the third from the series of papers \cite{Zamolodchikov:KdV:I,Zamolodchikov:KdV:II,Zamolodchikov:KdV:III}) focuses on this aspect of the quantum Korteweg-de Vries integrability, the calculations are based on insights gained from Ref. \cite{Fateev:Lukyanov:Poisson-Lie:PregenatorToqKdV,Fateev:Lukyanov:Vertex:PregenatorToqKdV}. 

Let us write:
\begin{align}
\bm M_\lambda = \left(\begin{array}{cc}
\mathcal{A}(\lambda)  & \mathcal{B}(\lambda)\\
\mathcal{C}(\lambda)  & \mathcal{D}(\lambda)
\end{array} \right).
\end{align}
We are now ready to write $\hat \gamma_i$ and $\hat \Lambda_i$ through $\hat \phi$, albeit rather indirectly, since we shall write these in terms of $\mathcal{A}$, $\mathcal{B}$, $\mathcal{C}$ and $\mathcal{D}$, which can, in turn, be written through $\hat\phi$, due (\ref{hatPhiDef}), (\ref{bmMdef}). The following expansion of the operators may be shown:
\begin{align}
\mathcal{B}(\lambda) = \mathcal{B}(0) \prod_j \left(1- \frac{\lambda^2}{\hat \gamma_j^2} \right) , \quad \mathcal{A}(\lambda) = \sum_{n=0}^\infty \lambda^{2n}\mathcal{A}_n,\mathcal{\quad D}(\lambda) = \sum_{n=0}^\infty \lambda^{2n}\mathcal{D}_n, 
\end{align}
which defines $\hat \gamma_i$, while $\hat \Lambda_i$, is defined by $\mathcal{A}(\hat \gamma_i)=\hat \Lambda_i$. In the substitutions   $\mathcal{A}(\hat \gamma_i)$, the operator $\hat \gamma_i^{2n}$ is to be placed to the left of $\mathcal{A}_n$ respectively. The commutation relations (\ref{QuantumGammagammaCommutations}) follow from the definition of $\hat \gamma_i$, $\hat \Lambda_i$, and from Eq. (\ref{SklyaninQuantum}).

\subsection{Conserved Quantities }

We now pass to a description of quantum Korteweg-de Vries that stresses the quantization of the field $u(x)$.  
As the quantum analogue of the Miura transformation (\ref{Miura}) one may take:
\begin{align}
\hat T(x) = -\frac{\pi}{\hbar} :\!\hat\phi'^2(x)\!:+\left(1-\frac{\pi}{\hbar}\right)\hat\phi''(x) - \frac{1}{24}.
\end{align}
The field $\frac{\hbar}{\pi}\hat T(x)$ is the quantum analogue of the Korteweg-de Vries field $u$. It is designated as $\hat T$ since, as defined, it has the commutation relations of the stress energy tensor of two dimensional conformal field theory. Indeed,
the commutation relations for $\hat\phi$, Eq. (\ref{BosonCommutation}), become the following commutation relations for the field $\hat T$:
\begin{align}
\imath [\hat T(x),\hat T(y)] = -\left(\hat T(x) + \hat T(y) \right) 2\pi\delta'(x-y) +\frac{\pi c}{6}\delta'''(x-y),\label{StreeTensorVirasoroAlgebra}
\end{align}
which are both familiar from conformal field theory and constitute a quantization of the commutation relations (\ref{ClassicalPoissonT}), where 
\begin{align}
c=13-6\left(\frac{\pi}{\hbar} + \frac{\hbar}{\pi} \right).
\end{align}
The commutation relations (\ref{StreeTensorVirasoroAlgebra})\ becomes the commutation relation (\ref{ClassicalPoissonT}) in the limit $\hbar \to 0$ if we assume that in this limit
\begin{align}
\hat T(x)\to\frac{\pi}{\hbar}u(x), \quad [\cdot , \cdot ] \to \imath \hbar \{\cdot,\cdot\}. 
\end{align}

Given Eq. (\ref{StreeTensorVirasoroAlgebra}), it is possible to construct an infinite set of mutually commuting quantities on the cylinder,\begin{align}
\hat I^{(cyl)}_1 &= \oint \frac{dx}{2\pi } \hat T \\
\hat I^{(cyl)}_3 &= \oint \frac{dx}{2\pi } :\hat T ^2: \\
\hat I^{(cyl)}_5 &= \oint \frac{dx}{2\pi } : \hat T^3: +\frac{c+2}{12}: \frac{\hat T'^2}{2}:\\
&\dots,\nonumber
\end{align}
where the normal ordering is defines as:
\begin{align}
:\hat A \hat B:(w) =\frac{1}{2\pi \imath} \oint \frac{dz}{z-w} \hat A(z) \hat B(w),
\end{align}
and the integral is over a small circle surrounding $w$. 

That these are mutually commuting quantities may be ascertained\cite{Sasaki:Yamanaka:QKdV:Conserved} by finding the commutation between them using (\ref{StreeTensorVirasoroAlgebra}). To construct these and higher conserved quantities, one may start by simply replacing the field $u$ of  the classical conserved quantities by $\hat T$, take the normal ordered products, and finally add correction terms such that the resulting quantum conserved quantities commute among themselves. 

The operator $\hat T(x)$\ may be Fourier transformed:
\begin{align}
\hat T(x) = \sum_n L_n e^{-\imath nx} - \frac{c}{24},\label{Texpansion}
\end{align}
where the constant $\frac{c}{24}$ is introduced for convenience. The commutation relations between the $L_n$'s may be inferred from (\ref{StreeTensorVirasoroAlgebra}) and is the well known Virasoro algebra:
\begin{align} 
[L_m,L_n] = (m-n)L_{m+n}+\frac{c}{12}(m^3-m)\delta_{m,-n}.\label{Virasoro}
\end{align}

The mutually commuting quantities then take the form:
\begin{align}
\hat I^{(cyl)}_1 &= \hat L_0 -\frac{c}{24} \\
\hat I^{(cyl)}_3 &= 2 \sum_{n=1}^\infty\hat L_{-n}L_n - \frac{c+2}{12} \hat L_0 +\frac{c(5c+22)}{2880}\frac{c}{24} \\
&\dots
\end{align}
 
The commuting variables, $\hat I_{2k+1}^{(cyl)}$ are a quantization of the Korteweg-de Vries Harmiltonians $\hat H_{2k+1}$ (up to unimportant numerical pre-factors) \cite{Sasaki:Yamanaka:QKdV:Conserved,Zamolodchikov:KdV:I,Zamolodchikov:KdV:II,Zamolodchikov:KdV:III}, in the sense that, in order to construct them, one may use the classical conserved quantities, which can be shown to Poisson commute given (\ref{ClassicalPoissonT}), and add to them quantum corrections, such as they would commute under the quantum commutation relations (\ref{StreeTensorVirasoroAlgebra}). To fill in the picture of how the classical algebro-geometrical solutions, with their separated variables, $\gamma_i$, $\Lambda_i$, are quantized, one must   show that the quantum variables $\hat \gamma_i$, $\hat \Lambda_i$ separate the problem of simultaneously diagonalizing the Hamiltonians $\hat I^{(cyl)}_{2k+1}.$ This is, in fact the case, and the mutual eigenfunctions of $\hat I^{(cyl)}_{2k+1}$ have a factorized form:
\begin{align}
\psi(\gamma_1,\gamma_2,\dots,\gamma_N) =\prod_{j=1}^N Q_j(\gamma_j), 
\end{align} 
while the eigenvalues of the operators  $\hat I^{(cyl)}_{2k+1}$ and the separate wave-functions $Q_j$ may be found using the algebraic Bethe ansatz. We shall, however, not delve into this subject or provide any of the technical details, which is a subject of a large body of work, which for the specialized problem of quantum Korteweg-de Vries was first treated in Refrs. \cite{Zamolodchikov:KdV:I,Zamolodchikov:KdV:II,Zamolodchikov:KdV:III}. The message of this section being primarily that the algebro-geometrical solutions may be viewed as the semiclassical analogues of the states produced by the algebraic Bethe ansatz. The limited set of technical details provided here serve merely to show on what basis such a connection can be made.

\subsection{Form factors}  

We now introduce the object that will be central to the proposition of how to connect the classical problem of Laplacian growth to conformal field theory. This object is known as the 'form factor'. To introduce it we shall consider the transformation properties of $\hat T$.  

The commutation relation (\ref{StreeTensorVirasoroAlgebra}) continue to hold if we replace $\hat T(x)$ by $\alpha^{-2} \hat T(\alpha x)$  and  $\hat T(y)$ by $\alpha^{-2} \hat T(\alpha y)$ . In fact for any analytic function $z(x),$ the commutation relations are invariant under the substitution of $\hat  T(x)$ with $\hat{\tilde{T}}(z)$ with \begin{align}
\hat{\tilde{T}}(z) = z'(x)^{-2}  \left[\hat T(x(z)) -\frac{c}{12} \{z|x\}\right],
\end{align} where $\{z|x\}$ is the Schwarzian derivative
\begin{align} 
\{z|x\} =\frac{z'''}{z'}- \frac{3}{2}\left(\frac{z''}{z'} \right)^2.
\end{align} 

This fact allows one to define an operator $\hat{ \tilde{ T}} (z)$ on the Riemann sphere (namely on $\mathbb{C}\cup\{\infty\}$) by applying the exponential map $z(x) = e^{ \imath x}$. We obtain:
\begin{align}
\hat{ \tilde{ T}} (z) =z^{-2}\left( \hat T (\log(z))+\frac{c}{24}\right).
\end{align} 
Thus, for example, we obtain:
\begin{align}
\hat I^{(cyl)}_1=   \oint \frac{dx}{2\pi } \hat T(x)=\frac{\imath}{2\pi}\oint dz z \hat{\tilde{T}}=L_{0}.
\end{align}
The common eigenstates of $\hat I_{2k+1}^{(cyl)} $ are the primary fields of conformal field theory and their descendants.\begin{figure}[h!!!]
\centering
\includegraphics[width=0.6\columnwidth]{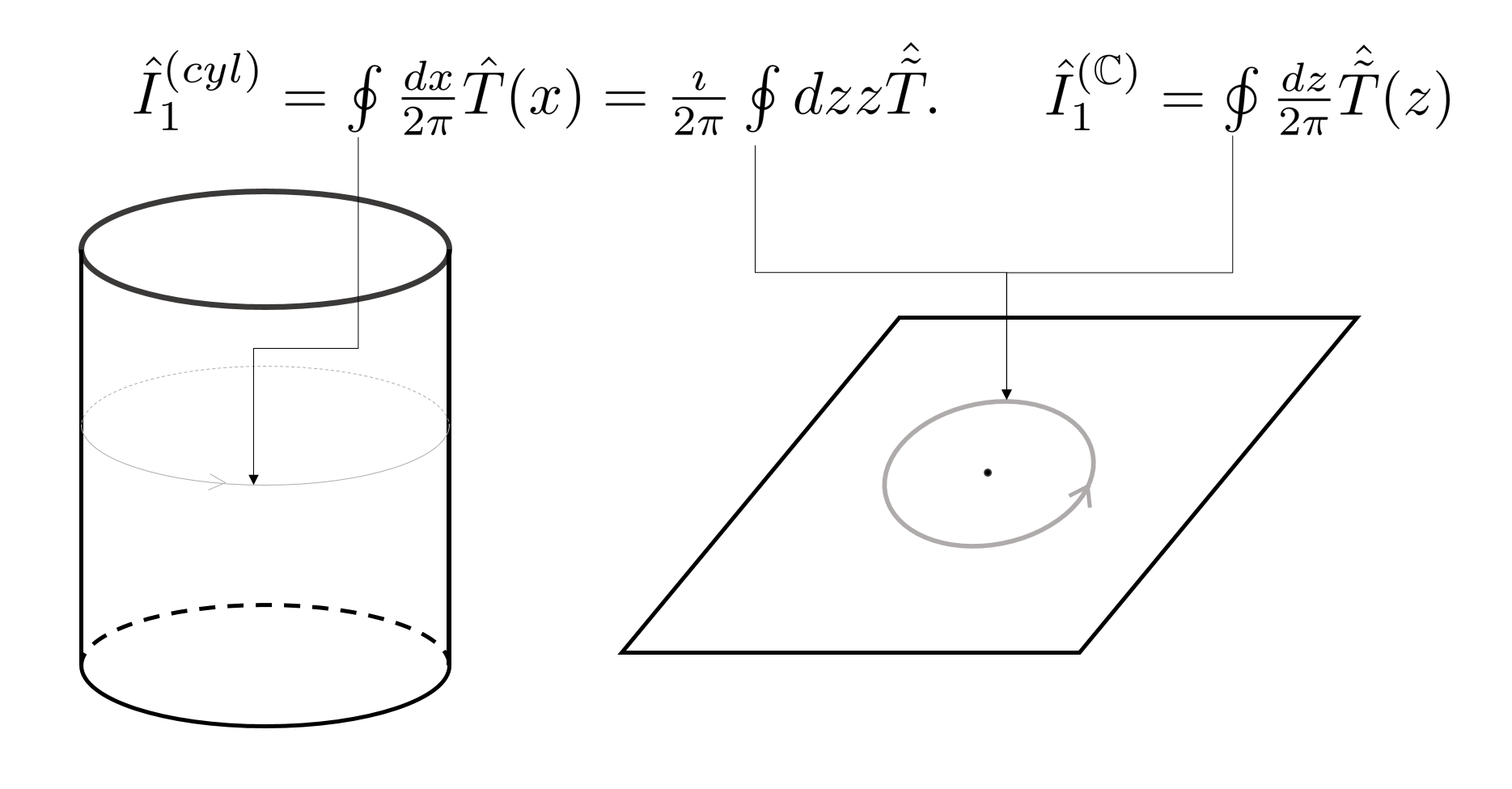}
\caption{The contours of integration of the conserved quantities $\hat I_1^{(cyl)}$ and $\hat I_1^{(\mathds{C})}$. The contours of integration on the cylinder or the complex plane are indicated by arrows.  \label{TContours}}
\end{figure}

There is another way to obtain conserved quantities on the Riemann sphere, which yields different commuting quantities. Note that  the proof that $\hat I^{(cyl)}_{2k+1}$ are commuting quantities relies only on the form of the commutation relations and on the fact that the integration contour, which defines the conserved quantity, is closed. This means that if we take the quantities in $I^{(cyl)}_{2k+1}$ and replace $dx$ by $dz,$  $\hat T$ by $\hat{\tilde{T}}$ and the integration contour around the cylinder by a closed contour in the complex plane surrounding the origin,  we will get a mutually commuting set of quantities.  We denote these conserved quantities by $\hat I_{2k+1}^{(\mathds{C})}$. This procedure is demonstrated in Fig. \ref{TContours}.

As mentioned above, the mutual eigenvalues of the commuting operators $\hat I_{2k+1}^{(cyl)}$ are the primary fields and descendants of conformal field theory. To find possible spectra and the properties of these eigenstates, one uses the representation theory of the Virasoro algebra, Eq. (\ref{Virasoro}). We shall find opportunity in the next section to quote some results from this representation theory, but we will not review it here. 

The diagonalization of $\hat I^{(\mathds{C})}_{2k+1}$ follows a different route. The reason for the different approach in comparison to $\hat I^{(cyl)}_{2k+1}$ is the fact that while $\hat{I}^{(cyl)}_{2k+1}$ are scale invariant, the operator $\hat I^{(\mathds{C})}_{2k+1}$ scale as $L^{-(1+2k)}$, where $L$ is sum large scale cut-off. For example, if we think of $\mathds{C}$ as the cylinder of radius $L$ in the limit $L\to\infty$, we expect that eigenvalue of $\hat I^{(\mathds{C})}_1$ to scale to zero in the limit $L\to\infty$ as $L^{-1}$, while $\hat I_1^{(cyl)}$  becomes a universal number independent of $L$ in the limit. The latter behavior (relating to  $\hat I_1^{(cyl)}$) is more tractable using representation theory than the former behavior  (relating to  $\hat I_1^{(\mathds{C})}$). To diagonalize  $\hat I_1^{(\mathds{C})}$, one rather puts the system on a cylinder of radius $L$ (to provide a cutoff) and solves the problem using the Bethe ansatz\cite{Al:Zamolodchikov:TBA:MAssless,Al:Zamolodchikov:Bethe:Ansatz:CFT,Zamolodchikov:Integrability:CFT:3State}. Indeed, treating the complex plane as a cylinder with the radius sent to infinity, the method of the quantum separation of variables discussed in section \ref{QSOV}, produces the set of states, which simultaneously  diagonalizes the quantum Hamiltonians $\hat I^{(\mathds{C})}_1$.  

Regardless of how the two  bases, which diagonalize  the two commuting sets  $\hat I^{(cyl)}_{2k+1},$   $\hat{I}^{(\mathds{C})}_{2k+1}$, are found, one may consider the expansion of a basis vector belonging to one set, in terms of the basis vectors belonging to the other. These objects, called form factors, are the main object we wish to introduce in this section. They hold special importance for the message this paper wishes to convey, as we shall argue in the next section that the modulus square of such  form factors are an appealing candidate to describe the statistics of the Hele-Shaw interface and in particular may hold the key to analytically obtaining the fractal properties of the interface, including the fractal dimension.  

To be able to more precisely introduce the form factors, let us first note that $\hat I^{(cyl)}_{2k+1}$ are usually diagonalized in terms  operators, rather than directly by states, the two points of view being equivalent due to the operator-state correspondence of conformal field theory. One finds operators, $\hat \Phi(z),$ such that:
\begin{align}
[\hat I^{(cyl)}_{2k+1},\hat \Phi(0)] = \tilde \Delta_{2k+1}\hat\Phi(0).\label{PhiEigenoperator}
\end{align}  
In addition, a vacuum, $|0\>$ is defined as a special highest weight state. Namely, this state  satisfies:
\begin{align}
\hat L_{n}|0\>=0,
\end{align}
 for $n\geq-1$.  An eigenstate of  $I^{(cyl)}_{1}=L_0-\frac{c}{24}$ with eigenvalue $\tilde \Delta_1=\Delta_{1}-\frac{c}{24}$,can thus be written as:
\begin{align}
\hat \Phi(0)|0\>.
\end{align}
The $z$ dependence of $\hat\Phi(z)$ is defined by:
\begin{align}
[\hat I_1^{(\mathds{C})},\hat\Phi(z)]=\partial_z \hat \Phi(z),
\end{align}
which may be more familiar  when one realizes $I_1^{(\mathds{C})}=L_{-1}$.
In Addition, it can easily be show,  based on Eqs. (\ref{Texpansion}), (\ref{PhiEigenoperator}) and the Baker-Campbell-Hausdorff lema, that the following relation holds:
\begin{align}
e^{\imath\theta \oint z\hat {\tilde T}(z)} \hat \Phi(z_0)e^{-\imath\theta \oint z\hat {\tilde T}(z)} = e^{\imath\Delta_1\theta} \hat \Phi(z_0e^{\imath\theta}).\label{ShiftedL0Invariance}
\end{align}

As for  the eigenstates of $\hat I_{2k+1}^{(\mathds{C})}$, these are obtained by quantizing the Korteweg-de Vries problem, as described in section \ref{QSOV}. In that section it was shown that the separated  variables of multi-phase solutions of classical Korteweg-de Vries become separated quantum operators. The eigenstates so obtained are then in one to one correspondence with multi-phase solutions of Korteweg-de Vries, which may be labelled by the hyper-elliptic surface $\mathcal{A}.$ We thus denote such eigenstates by $|\mathcal{A}\>$. In particular:
\begin{align}
\hat I^{(\mathds{C})}_{2k+1}|\mathcal{A}\> = H_{2k+1}(\mathcal{A})|\mathcal{A}\>\label{AlgebraicEigenvalue}
\end{align}

We may now introduce the form factor. This is given by:
\begin{align} 
\<\mathcal{A}|\hat\Phi(z)|0\>.\label{Einfachformfactor} \end{align}  
The form factors have had an important role in the study of integrable field theories, and a full understanding of them, would allow also for a full understanding of the integrable field theory beyond the spectrum. Important advances in this direction have been made by F. Smirnov (see  \cite{Smirnov:Book} and references therein)\ and others (r.g., Refrs.  \cite{Bernard:LecLaid:RSG,Smirnov:RSG}) . How these objects stands in relations to the quantum Korteweg-de Vries and its semiclassical limit are described in Refrs. \cite{Smirnov:Babelon:Bernard:Qauntization:Solitons,Smirnov:Babelon:Bernard:Null:Vectors,Smirnov:Babelon:Defoned:Hyperlliptic,Smirnov:Quasi-Classical:qKdV}.

The importance of the form factors lies in the fact that the set of states, $|\mathcal{A}\>,$ is, by definition, a complete basis and as such one can write a resolution of the identity as:
\begin{align}
\mathds{1}= \oint |\mathcal{A}\>\<\mathcal{A}| d\mathcal{A}.\label{AlgebraicResolutionof1}
\end{align}
The form factors then appear naturally when one inserts the resolution of the identity into correlators of the fields $\hat \Phi$. We must note, however, that in (\ref{AlgebraicResolutionof1}), the measure $d\mathcal{A}$ must be specified. In addition, it is not clear that the set of states described by the quasi periodic solutions, to which $\mathcal{A}$ serves as a label, are complete. Thus, it should be assumed that in (\ref{AlgebraicResolutionof1}), the label $\mathcal{A}$ may have to be generalized to go beyond labelling  hyper-elliptic Riemann surface. We shall, however, in the sequel, be interested mainly in the semi-classical interpretation of Eqs. (\ref{AlgebraicResolutionof1},\ref{Einfachformfactor}) and in a context where such distinctions are less crucial. 

\section{Quantum-Classical Distribution\label{QCDostrbution}}

We shall now discuss some properties of a certain sesquilinear of the form factor in Eq. (\ref{Einfachformfactor}). These properties will eventually clarify the reason for proposing a possible connection with the probability distribution $P(\mathcal{A})$. To do so, we need first to consider a semiclassical version of the form factor. We imagine  taking a wave packet of quantum states such that the packet saturates the Heisenberg uncertainty bound and denote the packet by $  |\mathcal{A}\!\!\succ,$ and define the object:
\begin{align}
p(\mathcal{A}) =  \prec\!\! 0|\hat \Phi(0)|\mathcal{A} \!\!\succ  \prec\!\!\mathcal{A} |\hat \Phi(0)|0\!\!\succ. \label{pADef}
\end{align}
Here we allow to take for the state $ \prec\!\! 0|\hat \Phi(0)$ some semiclassical approximation, and thus denote it as   $ \prec\!\! 0|\hat \Phi(0)$ rather than as $ \<0|\hat \Phi(0)$.

\subsection{Time-Translation and Scale Invariance}
We would like to show that $p(\mathcal{A})$ is time translation invariant, just as $P(\mathcal{A})$, Eq. (\ref{PTimeTranslation}). First, from (\ref{AlgebraicEigenvalue}) we have for any sufficiently well behaved generic (namely, one that cannot be written as a linear combination of $\hat I_{2k+1}^{(\mathds{C})}$) operator, $\hat O$:\begin{align}
\lim_{\delta\to0}e^{-\imath\Delta t\left(\hat I_1^{(\mathds{C})} +\delta \hat  O\right)}|\mathcal{A} \>=e^{-\imath\Delta tH_1(\mathcal{A})}|\mathcal{A} \>.\label{APerturbedEigenvalue}
\end{align}
Now semiclassically, and by using the Whitham method, which states that any perturbation will lead, in zeroth order, to the Whitham evolution, we may write:
\begin{align}
\lim_{\delta\to0}e^{-\imath\Delta t\left(\hat I_1^{(\mathds{C})} +\delta \hat  O\right)}|\mathcal{A} \> \sim|\mathcal{A}^{\Delta t}\!\!\succ\label{PerturbationEvolvesA}
\end{align}
Assuming that the object:
\begin{align}
\<0|\hat\Phi(0)e^{-\imath\Delta t\left(\hat I_1^{(\mathds{C})} +\delta \hat  O\right)}|\mathcal{A} \>
\end{align}
is smooth as $\delta\to 0,$ we obtain, by employing (\ref{APerturbedEigenvalue}):
\begin{align}
\lim_{\delta \to 0} \prec\!\!0|\hat\Phi(0)e^{-\imath\Delta t\left(\hat I_1^{(\mathds{C})} +\delta \hat  O\right)}|\mathcal{A} \!\!\succ \prec\!\!\mathcal{A}|e^{\imath\Delta t\left(\hat I_1^{(\mathds{C})} +\delta \hat  O\right)}\hat\Phi(0)|0 \!\!\succ = p(\mathcal{A}),
\end{align}
which when compared with Eq. (\ref{PerturbationEvolvesA}) gives:
\begin{align}
p(\mathcal{A})= p(\mathcal{A}^{\Delta t}).\label{pTimeTranslation}
\end{align}

 To show that $p$ also has the requisite scale invariance property, we must first consider the correlation function:
\begin{align}
\<0|\hat\Phi( 0)\hat\Phi( z)|0\>=\frac{C}{ z^{2\Delta_1}},\label{CFTcorrelIsPowerLaw}
\end{align}
where
\begin{align}
C=\<0|\hat\Phi(0)\hat\Phi(1)|0\>.
\end{align}
this equality
being easily obtained by making use of (\ref{ShiftedL0Invariance}).  On the other hand by employing the resolution of identity, Eq. (\ref{AlgebraicResolutionof1}), one obtains:
\begin{align}
\<0|\hat\Phi(0)\hat\Phi( z)|0\>=\int\<0|\hat\Phi(0)|\mathcal{A}\>\<\mathcal{A}|\hat\Phi( z)|0\> d\mathcal{A}.
\end{align} 
This equation may be written through $p(\mathcal{A})$:
\begin{align}
\<0|\hat\Phi(0)\hat\Phi( z)|0\>=\int p(\mathcal{A}) e^{-zH_1(\mathcal{A})}d\mathcal{A},
\end{align} 
which when combined with (\ref{CFTcorrelIsPowerLaw}) yields:
\begin{align}
p(\mathcal{A})d\mathcal{A} \sim d \left( H_1(\mathcal{A})\right)^{2\Delta_1},
\end{align}
where $H_1(\mathcal{A})\ $ is the dispersionless Korteweg-de Vries first Hamiltonian, Eq. (\ref{HkDispersionless}), for the Riemann surface $\mathcal{A}$. 
If we interpret the algebraic Riemann by means of the Hele-Shaw interface, we have the relation $H_1(\mathcal{A})\sim\sqrt{R}$ such that we may write:
\begin{align}
p(R)dR \sim d  R^{\Delta_1}.\label{pScaleInvariance}
\end{align}

We  also make a note on the measure, $d\mathcal{A}$. We should demand that this measure is time translation invariant. We believe that this is indeed the case, as the measure of integration for the semiclassical wave-function is often assumed to be the Liouville measure\cite{Smirnov:Babelon:Bernard:Qauntization:Solitons,Smirnov:Babelon:Bernard:Null:Vectors,Smirnov:Quasi-Classical:qKdV}. This measure is time-translation invariant, by a well-known theorem. The fact that it is time translation invariant with respect to the Whitham evolution, may abe surmised by noting that Whitham evolution is the zeroth order limit in any well-behaved perturbation. In the zeroth order the time translation invariance of the metric must be recovered, and so the measure must be invariant with respect to the Whitham evolution as well.

\subsection{Some  Results from Virasoro Representation Theory}
\begin{figure}[b!!!]
\centering
\includegraphics[width=0.8\columnwidth]{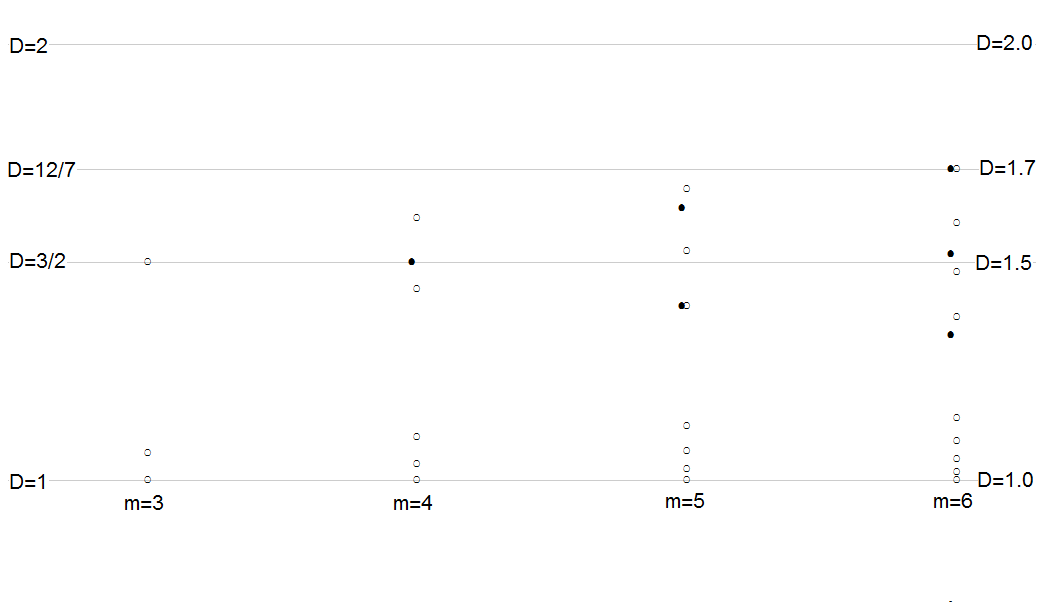}
\caption{The spectrum of dimensions, $D$, for the first non-trivial unitary minimal models. Only dimensions between $1$ and $2$ are shown. Empty circles represent first descendent of primary fields (namely, fields with dimensions $D=\Delta^{(r,s)}+1),$ higher descendants do not have fractal dimensions between $1$ and $2$,  filled circles represent primary  fields (namely, fields with dimensions $D=\Delta^{(r,s)}$). The minimal unitary model with $m=6,$  corresponding to $c=6/7$, contains a field with dimension $12/7.$  This field is the $(5,3)$ field in the Kac classification. The first descendent of the  Kac $(5,4)$ field has the same dimension.  Only a few fields, and fewer primaries, are above the line $D=3/2$ corresponding to Kesten's bound \cite{Kesten:DLA:Bound}, $D\geq3/2. $  \label{Dimensions}}
\end{figure}
Both the  proposed time translation invariance and the scale invariance of $p(\mathcal{A}),$  Eqs. (\ref{pTimeTranslation})  and (\ref{pScaleInvariance}), compare well, with the properties expected from the Hele-Shaw probability function Eqs. (\ref{PTimeTranslation}) and (\ref{PScaleInvariance}) , respectively, if we identify the $\hat I_1^{(cyl)}$ eigenvalue associated with $\hat\Phi$,  namely,  $\Delta_1,$ with the fractal dimension,  $D$. It turns out, that the representation theory for the Virasoro algebra (the algebra given in Eq. (\ref{Virasoro})),  gives a discrete spectrum (again under some assumptions) for the possible values of $\Delta_1$, and hence for the fractal dimension $D$.

At this point, it is perhaps advisable to reiterate that a more satisfactory connection between Laplacian growth and the representation theory of conformal field theory, would start from  the two dimensional Toda lattice hierarchy rather than the Korteweg-de Vries\ hierarchy. As it stands, the relation between the Korteweg-de Vries limit of Laplacian growth and conformal field theory
may not be satisfactory in order to make even  conjectural statements. Thus, the purpose of the current exercise linking the fractal dimension, $D$, to the eigenvalue, $\Delta_1,$ in conformal field theory, is give the gist of how the computation may work within a framework  of more comprehensive theory. 

After these words of caution, let us remind the reader that the representation theory of the Virasoro algebra, Eq. (\ref{Virasoro}), admits representations which depend on $c$, which appears in (\ref{Virasoro}) on the right hand side. Among those representation there are distinguished values of $c$, at which the especially simple representations exist, the field theories at those values of $c$ are called minimal models. These representations are simple in the fact that finite number of fields $\hat \Phi$ satisfying condition (\ref{PhiEigenoperator}) can be chosen that generate the entire representation. These are called primary fields and are labelled by two number $r,s$. We write $\hat\Phi_{r,s}$ for such a field. We concentrate here on such representations which are unitary. For unitary representations,  the possible values of $c$ are given by:
\begin{align}
c=1-\frac{6}{m(m+1)},
\end{align}   
where $m$ is an integer larger or equal to $3$,  $m\geq3$. Within these theories and for   the operators  $\hat\Phi_{r,s},$ the labels $r$ and $s$ can take the values $1\leq r<m$, $1\leq s \leq r$. The value of $\Delta_1$ for  $\hat\Phi_{r,s}$, which we denote by $\Delta_1^{(r,s)}$, is given by:
\begin{align}
\Delta_1^{(r,s)}=\frac{\left[(m+1)r-ms\right]^2-1}{4m(m+1)}.
\end{align}
Other solutions of 
 (\ref{PhiEigenoperator}) have a dimension $\Delta_1$ which may be  larger by an integer, $\Delta_1=\Delta_1^{(r,s)} +n$, for some $n\geq0$. 

We have focused on the unitary models, since they seem to have the kind of properties which are needed to have a consistent description of the modulus square of the form factors as the Laplacian growth probability distribution. But of course,  this statement must be taken with a grain of salt. At any rate, one can focus on these models and see whether the fractal dimension $D$ which is observed in Laplacian growth \cite{Procaccia:Levermann:Correct:HEle-Shaw:D,Swinney:Praud:Hele:Shaw:1.71}, $D=1.71\pm0.3,$  is reproduced in this approach as  $D=\Delta_1^{(r,s)} +n$, for some admissible values  $n$, $r$ and $s$. The observed fractal dimension have long been recognized as being equal to $D=12/7,$ within the accuracy of the observation. As shown in Fig. \ref{Dimensions}, such a fractal dimension appears at $m=6$ for $r=5$, $s=3$ and $n=0$ and for $r=5$, $s=4$ and $n=1$. 

Many other fractal dimension, $D,$ can be found for, say, the first four unitary models, $3\leq m\leq 6.$ However, given that the fractal dimension of cluster embedded in two dimensions must be between $1$ and $2$, and moreover given Kesten's work who showed that $3/2\leq D \leq 2$, only a few possible fractal dimension appear for the first four unitary minimal models, and   $D=12/7$ may be found among their numbers.  We  should note that Kesten's bound was found for the closely related process of diffusion limited aggregation\cite{Witten:Sander}, believed to be in the universality class of  Laplacian growth \cite{Procaccia:Levermann:Correct:HEle-Shaw:D}.
We should also note that the number of fractal dimensions respecting the known bounds, increases rapidly as $m$ increases, such that without further insight into how to determine the correct conformal field theory, labelled by $c$, from which to take the fractal dimension, the approach will not yield unambiguous results, this of course assuming the validity of the approach in the first place.

\section{Conclusion\label{Discussion}}
We have presented a possible connection between Laplacian growth and conformal field theory, which relies on the classical and quantum integrability of the  problems, respectively, after reviewing some necessary material regarding the classical integrability of the Laplacian growth problem and the quantum inverse scattering method applied to conformal field theory and the quantum Korteweg-de Vries equation. It is perhaps advisable to stress again the tentative nature of the proposed connection. At the same time, it may be reasonably said that the proposed connection seems to be offer an interesting paradigm for how non-equilibrium integrable systems may be treated analytically. Namely, that a sort of spectral expansion of field theory operators  may provide the probability density for a non-equilibrium system. The prospect of such an approach is appealing because it allows to deviate from the more common analytic device of proposing Gibbs or Gibbs-like measures for the solution of non-equilibrium problems (for example, in problems of large deviations  in which  probability densities can be understood and computed as the exponent of some non-equilibrium action), while  allowing for progress within an analytic approach.

\appendix

\section{The Baker-Akhiezer Function:\ Existence and Explicit Form\label{BAexplicitSection}}We give here an expression  Baker-Akhiezer functions defined on some Schottky double. To do so it is necessary to describe the Schottky double by the `Jacobi variety'. This description is very familiar in the case of a genus $1$ surface. In that case There are two interfaces and two cycles $a$, $b$.  (see Fig. (\ref{SchottkyCycles})). a genus $1$ surface has the topology of a torus and a torus is well known to be representable in terms of a rectangle with opposite sides identified. This latter is the Jacobi variety for a genus $1$ surface. The extension of this two higher genus surfaces, when there more than two interfaces, is described below. The construction is then used in order to write an explicit formula for the Baker-Akhiezer function on the Schottky double. 

A genus $g$ surface admits $g$ independent holomorphic differentials. A natural basis for such differentials is given by  $\zeta_i$ required to be holomorphic one-forms satisfying:
\begin{align}
\oint_{a_k} \zeta_j = \delta_{jk}. 
\end{align}
Let us consider the Abel map\cite{Farkas:Kra:Riemann:Book,84:Dubrovin:Algebr:Geome,Teschl} $\vec u (\bm z):\mathcal{D} \longrightarrow \mathds{C}^g/(\mathds{Z}^g+\bm B\mathds{Z}^g ),$ where \begin{align}
 \bm B _{kj}= \oint_{b_k} \zeta_j.
 \end{align}
The Abel map is defined by:
\begin{align}
u_j(\bm z) = \int^{\bm z}_{\bm z_0} \zeta_j \mod \vec v,\quad \vec v\in\mathds{Z}^g+\bm B\mathds{Z}^g,
\end{align} 
where $\bm z_0$ is some point on the Schottky double satisfying $\tau(\bm z)=\bm z.$  Namely, $\vec u(\bm z)$ is a vector whose $j$ element is $\int^{\bm z} \zeta_j$, whereby the freedom in drawing the  contour of integration to surround $m$ times around cycle $a_k$ and $n$ times around cycle  $b_l$ leads to an additive  ambiguity in the value of $\vec u$  of the form  $ m  \hat e_k +n\bm B\hat e_l .$ To remove this ambiguity $\vec u(\bm z)$ is considered   to take values in the Jacobian $\mathds{C}^g/(\mathds{Z}^g+\bm B\mathds{Z}^g )$.  For $g\geq1$ the image of the Abel map is a two dimensional surface within the Jacobian.

The matrix $\bm B$ turns out to be symmetric and, because of the existence of an anti-holomorphic involution, $\bm B$ is purely imaginary.
Furthermore, the matrix $\bm B$ is positive definite. This allows to define the Riemann theta as follows:
\begin{align}
\theta(\vec u)  = \sum_{\vec n \in \mathds{Z}^g} \exp2\pi\imath \left[\vec n \cdot \vec u+\frac{\vec n^t\bm B\vec n}{2} \right] .\label{ThetaFunctionDef}
\end{align}
The Riemann Theta function is quasi-periodic on the Jacobian:
\begin{align}
\theta(\vec u +\hat e_k) = \theta (\vec u), \quad \theta (\vec u + \bm B \hat e_j)=e^{-\pi\imath B_{jj}-2\pi\imath u_j} \theta(\vec u). 
\end{align} 
The function itself is, thus, {\it not} single valued on the Jacobian. Nevertheless, it may be used to construct all single valued holomorphic functions on the Riemann surface by taking proper combinations of it. For example, the logarithmic derivative of $\theta$ is already single valued. 

The Baker-Akhiezer function can be written in terms of the Riemann theta function, as well. However, to write the explicit expression, we shall first need, for each $k$ to define  two $g$-dimensional vectors $\vec \omega^a_k ,\vec \omega^b_k $, as follows :

\begin{align}
\left( \vec{\omega}_k^a  \right)_j=\imath \oint_{a_j} dH_k , \quad \left( \vec{\omega}_k^b  \right)_j=-\imath \oint_{b_j} dH_k. 
\end{align}
In addition, let us define a set of  points $\bm \gamma^\pm_j$ as a set of $g$ points on the Schottky double, such that $\bm \gamma^+_j =\tau(\bm \gamma_i^-),$ and such that there exist a differential $d\Omega$  with first order poles at $\infty_\pm$ with residues $\pm 1$, respectively, and with zeros at the points $\bm \gamma_j^+$ and $\bm \gamma_j^-$.
We let $\vec u_{(i)}^\pm =\vec u(\bm \gamma^\pm_i)$, respectively.

We denote from here on 
\begin{align}
H_0=H/2, \quad t_0=t=n.
\end{align}

The explicit  expression for the Baker Akhiezer function, $\psi^+$, and a function $\psi^-$, which may be treated as  its adjoint, reads as follows:
\begin{align} 
\psi_n^\pm (\bm z)=e^{ \pm \sum _{k\geq0} t_k  {H}_k+\bar t_k\bar{  {H}}_k} \Phi^\pm\left(\left.\sum_k  \Re\left[t_k(\bm B \vec{\omega}_k^a +\vec{\omega}_k^b) \right]\right|\vec u(\bm z)\right) .\label{BAExplicit} \end{align}
where
$\Phi^\pm(\vec w|\vec u)$ is a function of two complex $g$-dimensional vectors, $\vec w $ and  $\vec u$. The function $\Phi^\pm$ takes the following explicit form:\begin{align}
\Phi^\pm(\vec{w}|\vec u)=A^\pm(\vec w) e^{\pm2\pi\imath \vec{u} \cdot\bm B^{-1}\Im(\vec w)}\frac{\theta\left(\vec u\pm\vec w-\sum_i\vec u_{(i)}^\pm+ \vec K\right)}{\theta(\vec u-\sum_i\vec u_{(i)}^\pm+ \vec K)},
\end{align}
where $\vec K$ is the vector of Riemann constants, which  is given by:
\begin{align}
K_i = \frac{1+B_{ii}}{2} -\sum_{j\neq i}\oint_{a_j} u_i \zeta_j. 
\end{align}

Expression (\ref{BAExplicit}) is single valued on the Riemann surface even though, as noted before, the Hamiltonians are multi-valued function. In fact, the explicit form for the Baker-Akhiezer function is constructed such as the quasi-periodicity of the Riemann theta function exactly cancels the quasi periodicity of the exponent of the Hamiltonians as they appear in (\ref{BAExplicit}).

The amplitudes, $A^\pm$, are, for now, arbitrary.  The fact that the Baker-Akhiezer function defined in Eq. (\ref{BAExplicit}) indeed solves the spectral problem, Eq. (\ref{BAsimplistic}), will be shown in  appendix \ref{BADefPropAppen}
based on  the following analytic properties that this functions obeys:

\begin{itemize}
\item{As $z\to\infty_{\uparrow}$, the Baker-Akhiezer functions have following expansion: 
\begin{align}
\psi^\pm_n= z^{\pm n} e^{\pm \sum_{k>0} t_kz^k} \sum_{j\geq0} \frac{\xi^{\pm,\uparrow}_{jn}}{z^k} ,\label{BAasymptitesUp} 
\end{align}

while as $z\to\infty_\downarrow$, the Baker Akhiezer functions have the following expansion
\begin{align}
\psi^\pm_n= \bar z^{\mp n} e^{\mp \sum_{k>0} \bar t_k {\bar z^k}} \sum_{j\geq0} \frac{\xi^{\pm,\downarrow}_{jn}}{\bar z^j} .\label{BAasymptitesDown} 
\end{align}
The parameters $\xi^{\pm,\uparrow/\downarrow}_{jn}$ are functions of the times, $t_k$, $\bar t_k$.  }
\item{The functions $\psi_n^\pm(z) $ have poles at the points $\bm \gamma_j^\pm$, respectively. This fact is based the Riemann theta function participating in the definition of $\Phi^\pm$ has zeros at those points\cite{Farkas:Kra:Riemann:Book,Teschl,84:Dubrovin:Algebr:Geome}. }
\item{The amplitudes $A^\pm$ may be chosen such that the  Baker-Akhiezer functions are  normalized by requiring $\xi^{+,\uparrow}_{0n}$ to be real and positive along with  the following  condition  
\begin{align}
\label{normalizationIntermediate}\xi_{0n}^{+,\uparrow}= \frac{1}{\xi_{0n}^{-,\uparrow}}=\frac{1}{\xi_{0n}^{+,\downarrow}}. \end{align}
 } 
\end{itemize}

These analytic properties can be ascertained from the explicit form of the Baker-Akhiezer function, Eq. (\ref{BAExplicit}).

It is important to note that the Riemann theta function has $g$ zeros, and thus the Baker-Akhiezer function vanishes at $g$ points on the Riemann surface. As discussed in Ref. \cite{Flaschka:McLaughlin:Canonical:Conjugate:Variables} and reviewed in section \ref{KdVInveseScatteringSection}, the $z$ coordinate of the points on the Riemann at which the Baker-Akhiezer function vanishes, can be considered as set of $g$ dynamical variables. One may take  the Bloch multiplier associated with the Baker-Akhiezer function on the  opposite sheet of the  Schottky double  as another set of $g$ dynamical variables. Together we have $2g$ dynamical variables, which, as discussed in section \ref{KdVInveseScatteringSection} in the context of the  Korteweg-de Vries equation, are separated variables .

\section{The Baker-Akhiezer Function: Definition and Properties\label{BADefPropAppen}}

Assume a Schottky double, with anti-holomorphic involution, $\tau$. We assume that as $z\to\infty$, the function $\tau(z)$\ may be expanded as $\tau(z) = z^J + O(z^{J-1}),$ for some positive integer, $J$, which one should assume to be large.   

We have defined the Baker-Akhiezer function as a function satisfying (\ref{BAsimplistic}). We need a somewhat more explicit and detailed definition of the spectral problem. We write this definition here as follows:
\begin{align}
&\sum_m L_{nm} \psi^+_m = z \psi^+_n,\quad \sum_m  \psi^-_mL_{mn} = z \psi^-_n\label{SpectrakBA} \\
&\sum_m L^\dagger_{nm} \psi^+_m = \overline{\tau (z)} \psi^+_n,\quad \sum_m   \overline{L_{nm}}\psi^-_m(z)=  \overline{\tau (z)}  \psi^-_n\label{SpectrakBAbar}\\
&\partial_{l} \psi^+_n =\sum_m L_{nm}^{(l)} \psi^+_m,\quad \partial_{l} \psi^-_n =-\sum_m ^{(l)} \psi^-_mL^{(l)}_{mn},\label{LaxBA} \\ 
&\bar\partial_{l} \psi^+_n =\sum_m L^{(l)\dagger}_{nm} \psi^+_m,\quad \bar \partial_{l} \psi^-_n =-\sum_m \overline{L^{(l)}_{nm}} \psi^-_m, \label{LaxBAbar}
\end{align}
 Equations (\ref{LaxBA}), (\ref{LaxBAbar}) should be understood as written on the front side, and may be analytically continued to the back side. Thus for example $z$ in (\ref{LaxBA}) is to be replaced by $\tau(z)$ on the back side.

Given that the analytic conditions the  Baker-Akhiezer functions obeys, detailed in appendix \ref{BAexplicitSection}, one may prove that these functions obey a list of properties enumerated below. The central item in the list is the fact that the Baker-Akhiezer function solves the Lax spectral problem. The properties enumerated as \ref{BAuniquenessProperty}- \ref{BAreflectionProperty} below may be treated as lemmas helping to prove this central fact, which itself is enumerated as property \ref{BALaxProprty} below.   
\begin{enumerate}
\item{{\bf Uniqueness}:\label{BAuniquenessProperty}\\A\ function $f^\pm_{n}$ satisfying the same conditions as  $\psi^\pm_n$, Eqs.  (\ref{BAasymptitesUp}-\ref{normalizationIntermediate}), is equal to $\psi^\pm_n$. A function $f^\pm_n$ satisfying  the same conditions as  $\psi^\pm_n$, except the normalization condition, Eq. (\ref{normalizationIntermediate}), is proportional to $\psi^\pm_n.  $   
}
\item{\bf\ Orthonormality}:{\label{Orhotgonality}
\begin{align}
\frac{1}{2\pi\imath}\oint_\mathcal{C} \psi^-_n(z) \psi^+_m(z)d \Omega = \delta_{n,m},\label{OrthoEq}
\end{align}
where  $d\Omega$  is a differential with first order poles at $\infty_\pm$ with residues $\pm 1$, respectively, and with zeros at the points $\bm \gamma_j^+$ and $\bm \gamma_j^-$.
The differential is normalized such that its integral over any cycle is purely imaginary. 
}
\item{{\bf\ Completeness}\label{BACompletenessProperty}: \\Consider functions, $f^\pm$, on the Schottky double meromorphic everywhere except at $\infty_{\uparrow/\downarrow}$, where they have the expansions:
\begin{align}
f^\pm=\left\{\begin{array}{lr} 
z^{m^\pm_\uparrow}  e^{ \sum_{k>0} t_kz^k} \sum_{j\geq0} C^{\pm,\uparrow}_{j}z^{-j}, & z\to\infty_\uparrow \\
\bar z^{-m^\pm_\downarrow}  e^{ -\sum_{k>0} \bar t_k\bar z ^k} \sum_{j\geq0} C^{\pm,\downarrow}_{j}\bar z^{-j}, & z\to\infty_\downarrow
\end{array} \right.,\label{fExpansion} 
\end{align}   
with poles at $\bm \gamma_i^\pm$, respectively. The completeness property states that such functions are expandable in terms of Baker-Akhiezer functions. Namely, there exist coefficients $\alpha^\pm_j$ such that:
\begin{align}
f^\pm= \sum_{j=n_1^\pm}^{n_2^\pm} \alpha^\pm_j\psi^\pm_j,\label{CompletenessEq}
\end{align}
where $n^+_1= m_\downarrow^+, n^-_1=m^-_\uparrow, n^+_2=m^+_\uparrow, n_2^-=m_\downarrow^-,  $ respectively. The expansion (\ref{CompletenessEq}) holds if $n_2^+\geq n_1^+$, otherwise $f^+=0$. Similarly, $f^-=0$ unless $n_2^-\geq n_1^-$.  
}
 \item\ {\bf\ Reflection}:{\label{BAreflectionProperty}
\begin{align}
\overline{\psi_n^+(\tau(z))}=\psi_n^-&(z),\quad  \overline {\psi_n^-(\tau(z))}=\psi_n^+(z)\label{ReflectionEq},\\& -\overline{d\Omega(\tau( z))} = {d\Omega}(z)\label{OmegaReflection} 
\end{align}
}   
\item{{\bf Spectral problem}:\label{BALaxProprty} \\The Baker-Akhiezer functions, $\psi_n^+,$ are  wave-functions associated with the Lax spectral problem. Namely, Eqs. (\ref{SpectrakBA}-\ref{LaxBAbar}) hold, with (\ref{Ltothepowerl+}).} 
\end{enumerate}

We now give brief, semi-rigorous `proofs' of these properties\\
{\bf\ Proof of \ref{BAuniquenessProperty}.}

If two such functions $\psi_n$ and $\tilde\psi_n$ exist, then their ratio $\frac{\psi_n}{\tilde \psi_n}$ is a meromorphic function which has zeros  at the poles of $\tilde \psi_n$ and the zeros of $\psi_n$. The Riemann-Roch theorem\cite{Farkas:Kra:Riemann:Book,Teschl,84:Dubrovin:Algebr:Geome} states that, generically, a function cannot be found with arbitrarily chosen positions for the zeros. Thus, the zeros and poles of $\tilde \psi$ and $\psi$ coincide, namely   $\frac{\psi_n}{\tilde \psi_n}$ is entire on the compact Riemann surface,  and thus $\frac{\psi_n}{\tilde \psi_n}=c, $ where $c$ is some constant independent on $z$. The normalization condition (\ref{normalizationIntermediate}) fixes this constant to be $ 1$. 
\\
{\bf Proof of  \ref{Orhotgonality}}: 

Consider the integrand $\psi^+_n(\{t'_k\},z) \psi^-_m(\{t_k\},z)d \Omega$ has no singularities except at $\infty_{\uparrow/\downarrow}$ where it has the form $z^{(n-m)-1}\xi_0^{-,\uparrow} dz $ and  $-z^{(m-n)-1}\xi_0^{+,\downarrow} dz$, respectively. If $n>m$ we may deform the contour of integration to surround $\infty_\downarrow$  and obtain the the integral is zero and if $n<m$ we may deform the contour of integration to surround $\infty_-$ with the same conclusion. If $n=m,$ we may deform the contour of integration to surround $\infty_+$ to obtain that the integral is given by $1$, by using Eq. (\ref{normalizationIntermediate}). If for $n=m$ we deform the contour of integration around $\infty_-$  instead of around $\infty_+$ and apply the condition that the result must not depend on the choice of how to deform he contour, one obtains $\xi_{0n}^{-,\downarrow}=\frac{1}{\xi_{0n}^{+,\downarrow}}. $  The latter relation naturally joins the normalization condition Eq. (\ref{normalizationIntermediate}), and so we record  here the   full normalization condition, which includes this relation, for future reference:
\begin{align}
\label{normalization} \xi_{0n}^{+,\uparrow}= \frac{1}{\xi_{0n}^{-,\uparrow}}=\frac{1}{\xi_{0n}^{+,\downarrow}}=\xi^{-,\downarrow}_{0,n}. \end{align}
{\bf Proof of \ref{BACompletenessProperty}}:

Let us focus on $f^+$,  only. The proof with regards to $f^-$  goes along the same lines, so we shall omit it. 

Let us assume first that  $m^+_\uparrow\geq  m_\downarrow^+$. Given the expansion of $\psi_j^+$ around $\infty_\uparrow$, Eq. (\ref{BAasymptitesUp}), one  sees that one may cancel the coefficients $C^{+,\uparrow}_j$ in (\ref{fExpansion}), one by one for $j=m^+_\uparrow,\dots, m_\downarrow^+ +1$ by adding to $f^+$ a  linear combination of $\psi^+_j,$ and this without affecting the leading order behavior around $\infty_\downarrow$.  More formally, there exists a set of numbers $\{\alpha_j^+\}_{j=m_\downarrow^++1}^{m_\uparrow^+}$ such that the following expansion for the function $ $ $\tilde\psi\equiv f^+ - \sum^{m^+_\uparrow}_{j=m^+_\downarrow+1} \alpha_j^+\psi^+_j$ holds around $\infty_{\uparrow/\downarrow}$:
\begin{align}
\tilde \psi=\left\{\begin{array}{lr}
z^{m_{\downarrow}^+} e^{\sum t_k z^k}(c_\uparrow+O(1/z)) & z\to\infty_\uparrow\\
\bar z^{-m_{\downarrow}^+}e^{-\sum \bar t_k \bar z^k}(c_\downarrow+O(1/\bar z)) & z\to\infty_\downarrow
\end{array} \right.,
\end{align}
for some $c_\uparrow$ and $c_\downarrow$. Since $\tilde\psi$ satisfies all the analytic requirements defining the Baker-Akhiezer function $\psi^+_{m_\downarrow^+},$ except perhaps the normalization condition, it follows from  uniqueness (property \ref{BAuniquenessProperty} above), that $\tilde \psi$ must be proportional to  $\psi^+_{m_\downarrow^+}.$ Let us denote the constant of proportionality by $\alpha^+_{m_\downarrow^+}$ and write $$\tilde \psi =f^+ - \sum^{m^+_\uparrow}_{j=m^+_\downarrow+1} \alpha_j^+\psi^+_j=\alpha_{m^+_\downarrow}^+  \psi^+_{m_\downarrow^+},$$ from which (\ref{CompletenessEq}) follows immediately for $f^+$ and with $n_1^+=m^+_\downarrow$ and $n_2^+=m^+_\uparrow$.  

If $m^+_\uparrow<m_\downarrow^+$, then it follows that $f^+$ obeys all the analytic conditions required of $\psi^+_{m_\downarrow^+}$, except the normalization condition. Indeed, the expansion of   $f^+$ around $\infty_\downarrow $ is  given by the form on the right hand side of Eq. (\ref{BAasymptitesDown}), while the expansion around $\infty_\uparrow$, is given by the right hand side of Eq. (\ref{BAasymptitesUp}) with $\xi^{+,\uparrow}_{0m_\downarrow}=0 $, and the existence of a normalized Baker-Akhiezer function. Thus, due to uniqueness, $f^+=\alpha \psi^+_{m_\downarrow^+}$, for some $\alpha$. The fact that $\alpha=0$ follows from   $\xi^{+,\uparrow}_{0m_\downarrow}=0 $, showing that if $n_2^+<n_1^+,$ then $f^+=0$.\\
{\bf Proof of \ref{BAreflectionProperty}}: 

If one compares the asymptotic expansion of $\psi^-,$ given in (\ref{BAasymptitesUp}),(\ref{BAasymptitesDown}) with the expansion of $\overline{\psi^+}(\overline{\tau(z)})$ one sees that they match, based on (\ref{normalization}), and the fact that $p(z)$ maps $\infty_{\uparrow/\downarrow}$ to $\infty_{\downarrow/\uparrow}$, respectively. The location of the poles match as well. The uniqueness of the Baker-Akhiezer function, then dictates that the two functions must be equal, proving the first equality in (\ref{ReflectionEq}). The proof of the second equality follows much the same lines.

 The equality, (\ref{OmegaReflection}),  may be proved by comparing the analytic properties of the left and right hand sides. The poles and residues are the same. The $a$ and $b$ cycle integrals are purely imaginary on both right and left hand side. Two differentials having the same singularities normalized to have imaginary periods  over any cycle are equal, since their difference is a holomorphic differential with imaginary periods. The latter is necessarily zero, since the requirement of zero imaginary period represents $2g$ real linear constraints on the $2g$ dimensional space of holomorphic differentials (viewed as a vector space over $\mathds{R}$).\\
{\bf Proof of \ref{BALaxProprty}}:

The analytic function on the Schottky double,  which on the front side is given by $z$ is given on the back side by $\tau(z)$, which as $z\to\infty_\downarrow$ has the expansion $\tau(z)=\bar z^J +O(\bar z^{J-1})$, thus the front-side function $z \psi_n^+$, when analytically continued  to the back side, has the expansion:
\begin{align}
z \psi_n^+ = \left\{\begin{array}{lr}
z^{n+1} e^{\sum t_k z^k }(C_\uparrow + O(1/z)) & z\to\infty_\uparrow \\
\bar{z}^{-(n-J)}e^{-\sum \bar t_k \bar z^k }(C_\downarrow + O(1/\bar z)) & z\to\infty_\downarrow
 \end{array} \right. .
\end{align}  
By the completeness property (\ref{BACompletenessProperty} above), we obtain that there exist coefficients $L_{nj}$, which are non-zero for $j=n-J,\dots,n+1$ and vanish otherwise, such that:
\begin{align}
z\psi_n^+ = \sum_{j=n-J}^{n+1} L_{nj} \psi^+_j.\label{zasL}
\end{align}
This defines the matrix $L$ with the property (\ref{semiinfiniteness}) and proves the first equality in (\ref{SpectrakBA}). By orthonormality (Eq. (\ref{OrthoEq})), we have:
\begin{align}
L_{nj} =\frac{1}{2\pi\imath} \oint_{\mathcal{C}} \psi^-_j z \psi^+_n d\Omega.\label{LasOverkap}
\end{align}
Similarly, considering $z\psi^-$, the completeness property shows:
\begin{align}
z\psi^-_n = \sum^{n+J}_{j=n-1} \tilde L_{nj} \psi^-_j,
\end{align}
and orthonormality gives: 
\begin{align}
\tilde L_{nj} = \frac{1}{2\pi \imath} \oint_{\mathcal{C}} \psi^-_n z \psi^+_j d\Omega = L_{jn},
\end{align}
such that $\tilde L = L^t$, and the second equality in (\ref{SpectrakBA}) is obtained. 

The same considerations allow one to write:
\begin{align}
\overline{\tau(z)} \psi^+_n = \sum_{j=n-1}^{n+J} \tilde{\tilde{L}}_{nj} \psi^+_j ,
\end{align}
with
\begin{align}
\tilde{\tilde{L}}_{nj} =\frac{1}{2\pi\imath} \oint\psi^-_j \overline{\tau(z)}\psi^+_n d\Omega 
\end{align}
We note that the analytic continuation of the front-side function $\overline{\tau(z)}$  is equal to $\bar z$ on $\mathcal{C}$. Thus taking the complex conjugate of this equation and, in addition, using the reflection property (\ref{BAreflectionProperty}), we obtain:
\begin{align}
\overline{\tilde{\tilde{L}}_{nj}} =\frac{1}{2\pi\imath} \oint\psi^+_j z\psi^-_n d\Omega = L_{jn},
\end{align}  
such that $\tilde{\tilde{L}}=L^\dagger$. Proving the first equation in (\ref{SpectrakBAbar}). The proof of the second equation follows the same lines and we omit it. 

To prove the first equation in (\ref{LaxBA}) we examine the expansion of $\partial_l \psi^+_n$ around $\infty_{\uparrow/\downarrow}$:
\begin{align}
\partial_l \psi^+_n = \left\{\begin{array}{lr}
z^{ n+l} e^{ \sum t_kz^k} \left(\sum_{j=0}^{l-1} \frac{\xi_{jn}^{+,\uparrow}}{z^j}+\frac{\xi_{jn}^{+,\uparrow}+\partial_l \xi^{+,\uparrow}_{0n}}{z^l}+\dots\right) & z\to \infty_\uparrow \\
\bar z^{ -n} e^{- \sum \bar t_k\bar z^k} (\partial_l \xi_{0n}^{+,\uparrow}+\dots) & z\to\infty_\downarrow 
 \end{array}\right. .\label{paritialLparitlaPsiExpand}
\end{align}
Thus we have from completeness:
\begin{align}
\partial_l \psi^+_n - \psi^+_n \frac{\partial_l {\xi^{+,\downarrow}_{0n}}}{{\xi^{+,\downarrow}_{0n}}}= \sum_{j=n+1}^{n+l} L^{(l)}_{nj}\psi^+_j,
\end{align}
which entails:
\begin{align}
\partial_l \psi^+_n = \sum_{j=n}^{n+l} L^{(l)}_{nj}\psi^+_j,
\end{align}
with 
\begin{align}
L^{(l)}_{nj}= \left\{ \begin{array}{lr}
\frac{1}{2\pi\imath} \oint_{\mathcal{C}} \psi^-_j \partial_l \psi^+_n & n+l \geq j\geq n \\
\frac{\partial_l {\xi^{+,\downarrow}_{0n}}}{{\xi^{+,\downarrow}_{0n}}}& j=n \\
 0& {\rm otherwise}
 \end{array} \right.
.\label{LlIntermediate} \end{align} 
To obtain the expression (\ref{Ltothepowerl+}), for $L^{(l)},$
compare (\ref{paritialLparitlaPsiExpand}) with the expansion of $z^l \psi^+_n$ to obtain:
\begin{align}
\partial_l \psi^+_n - z^l \psi^+_n -\psi^+_n\frac{\partial_l {\xi^{+,\uparrow}_{0n}}}{{\xi^{+,\uparrow}_{0n}}}= \sum_{j=n-J}^{n-1} \alpha_j \psi^+_j,
\end{align}
for appropriately chosen $\alpha_j$. Applying the orthonormality property, Eq. (\ref{OrthoEq}),  to this equation, while taking into account Eq. (\ref{LlIntermediate}), one obtains:
\begin{align}
L^{(l)}_{nj}= \left\{ \begin{array}{lr}
\frac{1}{2\pi\imath} \oint_{\mathcal{C}} \psi^-_j z^l\psi^+_n & n+l \geq j\geq n \\
\frac{\partial_l {\xi^{+,\uparrow}_{0n}}}{{\xi^{+,\uparrow}_{0n}}}+\frac{1}{2\pi\imath} \oint_{\mathcal{C}} \psi^-_n z^l\psi^+_n &j=n\\
 0& {\rm otherwise}
 \end{array} \right.
.\label{LlIntermediate2}
\end{align}
Comparing the element $L^{(l)}_{nn}$ in (\ref{LlIntermediate}) and (\ref{LlIntermediate2}) and taking into account  $\frac{\partial_l {\xi^{+,\uparrow}_{0n}}}{{\xi^{+,\uparrow}_{0n}}} = -\frac{\partial_l {\xi^{+,\downarrow}_{0n}}}{{\xi^{+,\downarrow}_{0n}}},$ which follows from Eq.  (\ref{normalization}), one obtains  \begin{align} L_{nn}^{(l)}=\frac{1}{2}\times\frac{1}{2\pi\imath} \oint_{\mathcal{C}} \psi^-_n z^l\psi^+_n.\label{LlIntermediate3}\end{align} But (\ref{SpectrakBA}) one has $ \frac{1}{2\pi\imath} \oint_{\mathcal{C}} \psi^-_j z^l\psi^+_n =(L^l)_{nj}$, which together with  (\ref{LlIntermediate2})  and (\ref{LlIntermediate3}) proves (\ref{Ltothepowerl+}) .

The remaining equalities in (\ref{LaxBA}) and (\ref{LaxBAbar}) are proved by repeating the same methods used above. The only new ingredient is that one differentiates the orthonormality condition (\ref{OrthoEq}) to obtain:
\begin{align}
\frac{1}{2\pi\imath} \oint _{\mathcal{C}}\psi_m^- \partial_l \psi^+_n d\Omega =-\frac{1}{2\pi\imath} \oint _{\mathcal{C}}\psi^+_n\partial_l  \psi_m^-  d\Omega,
\end{align}
which turns out to be useful in deriving the second equalities in (\ref{LaxBA}) and (\ref{LaxBAbar}) from the first equalities, respectively.
We thus omit these further proofs.

\section{Algebro-Geometrical Whitham Averaging Method\label{WhithamAppendix}}

We now describe the algebro-geometrical version\cite{83:Krichever:Averaging} of the Whitham averaging method\cite{Whitham:1966:NDW,Whitham:Book}. The method requires thinking of the variation of time as coming in two sorts, long and short. This can be formalized by the two-timing approach, which we overview in the next subsection,  this followed first by a treatment of the averaging procedure and finally by a derivation of the Whitham equations.  

\subsection{A two-timing approach}

Due to (\ref{LasOverkap}) we have the following expression for $a^{(1)}_k$ in Eq. (\ref{LasShifts}):
\begin{align}
a^{(1)}_k(n) = \frac{1}{2\pi\imath}\oint_\mathcal{C}\psi^-_{n} \psi^+_{n-k}d\Omega =\frac{1}{2\pi\imath}\oint_\mathcal{C}e^{-2k  {H}_0}\Phi^-_{n} \Phi^+_{n-k}d\Omega,
\end{align}
where in the last equality we have made use of (\ref{BAExplicit}) and we have included explicitly the $t_0=n$ dependence of $\Phi^\pm$ on the index, writing $\Phi^\pm_n$. The expression on the right hand side of this equation depends only on $\Phi^\pm$, which themselves depend on the times $\{t_k\}$ only through the variable, $\vec w(\{t_k\}) ,$ given by 
\begin{align}
\vec w(\{t_k\}) = \sum_k  t_k\vec \omega_k +  \bar t_k\vec{\bar\omega}_k,
\end{align}
where
\begin{align}
\vec\omega_k \equiv\bm B \vec{\omega}_k^a +\vec{\omega}_k^b.
\end{align}
It may also be ascertained that $a^{(1)}_k(n)$\ is bounded and that  the functions $\Phi^\pm,$ and thus $a^{(1)}_k,$ respect the periodicity of the Jacobian, $\vec w \to \vec w+\hat e_i$, $\vec w \to \vec w +\bm B \hat e_i$, as noted in Appendix \ref{BAexplicitSection}. We shall call  bounded functions which depends only on time through $\vec w(\{t_k\})$ and which respect the periodicity of the Jacobian, `multi-periodic'. 

Due to multi-periodicity it is convenient to define $[\vec w]$ as being equal to $\vec w$ modulo the periods of the Jacobian, such that $[\vec w]$ lies in the primitive cell of the Jacobian lattice, $\sum_j\mathds{Z}\hat e_j + \mathds{Z} \bm B \hat e_j.$ A multi-periodic function is then, in fact, a function of $[\vec w]$. Now consider some initial times $\{t_j^0\}_{j=0}^\infty$ and then consider what happens to $ [\vec w]$ as $t_k,$   for given $k$, continuously increases, by an amount which we shall denote as  $\Delta t_k$. Namely, we consider  $[\vec w(\{t_j\})]$ for $t_j=t_j^0 + \delta_{jk} \Delta t_k$.  It is easy to see that,$[\vec w] $ will cover ergodically the primitive cell of the Jacobian lattice, unless $\vec{\omega}_k$ happens to be commensurate with one of the periods of the that lattice (which is not to be expected generically). Thus a multi-periodic function is a function from the Jacobian to $\mathds{C}$ (or $\mathds{R}$), the argument of which generically covering ergodically the Jacobian with the increase of any of the times. 

The ergodic property will be important in what follows, as it will allow to replace time averages by averages over the Jacobian. Multi-periodicity and thus ergodicity  hold for the matrix elements of  $L$ (since they are expressible through the multi-periodic  $a_k(n)$) and for any of the operator $L^{(l)}$(due to the way  the matrix elements of $L^{(l)}$ are expressible through the matrix elements of $L$, Eq. (\ref{Ltothepowerl+}). We can thus write:
\begin{align}
L = L\left(\vec w\left(\{t_k\}\right),\hat p,\bm I\right),\label{LAsFunctionReadyForWhitham}
\end{align}
where $\bm I$ denotes the Riemann surface on which the function the Baker-Akhiezer function is defined, from which $L$\ is derived. The inclusion of the dependence of $L$\ on  $\hat{p}$ is made here for future convenience. It connotes the form (\ref{LasShifts}), which includes this derivative operator. The convenience  for the inclusion of $\bm I$ in (\ref{LAsFunctionReadyForWhitham}) will also only be revealed below, where we shall want to consider different multi-periodic solutions. Indeed, since for every Riemann surface of the form considered above we have constructed a multi-periodic solution, we may consider different solutions by considering different  Riemann surfaces. The matrix $L$ thus depends on the Riemann surface chosen. This is encoded in the dependence on the somewhat abstract parameters $\bm I$.  To make $\bm I$ less abstract, one may use the coefficient of the polynomial in two variables  $Q(z,S(z))=0$ as the data contained within $\bm I$, as this data uniquely determines the Riemann surface, by specifying the complex curve underlying it. We shall not, however, need such an explicit representation of $\bm I$. 

The Whitham theory considers finding approximate solutions to integrable nonlinear equations by taking slowly modulated multi-periodic solutions. To define what constitutes as `slow' as opposed to `fast',   one must first determine a scale for $\vec\omega_k.$ Let fix $\tau_k$ is the typical time associated with $\vec \omega_k$: $|\vec \omega_k| \sim \frac{1}{\tau_k}.$  Choose a small number $\varepsilon$. A long  time scale, $\tau,$ is one such that it is of the order or greater than the scale $\varepsilon^{-1}\tau_k$. Namely,  $\tau \gtrsim \varepsilon^{-1}\tau_k.$  A short time scale is a scale $\tau$ such that $\varepsilon^{-1}\tau_k\gg\tau$. Often one considers short time scales such that $\varepsilon^{-1}\tau_k\gg\tau\gg\tau_k.$  On such time scale ergodicity takes hold, but time is still considered short. 

To make use of the separation of time scales formally, it is useful to utilize the two-timing method. In this method one  replaces the time variables  as $t_k $ by $ t_k+\varepsilon^{-1} T_k$, and considers only situations where both $t_k$ and $T_k$ vary on the scale $\tau_k$ or much larger than $\tau_k$ but definitely smaller than $\varepsilon^{-1} \tau_k$. The variation with respect to $T_k$ are then considered slow, and thus $T_k$ is called the `slow time', while $t_k$  is the fast time, corresponding to the fast variations with respect to it.    A modulated solution is one in which the Riemann surface is allowed to vary with the slow time, $T_k,$ while the phase $\vec w$ depends on the fast time, $t_k$ : 
\begin{align}
L(t_k+\varepsilon^{-1} T_k )= L\left(\vec w\left(\{t_k\}\right),\hat P,\bm I(T_k)\right),\label{NaiveWhitham}
\end{align}
where 
\begin{align}
\hat P \equiv \frac{\partial}{\partial t_0} + \varepsilon \frac{\partial}{\partial T_0}.\label{Plong}
\end{align}
The matrix elements of such an $L$ are slowly modulated solutions of two dimensional Toda lattice.

The form (\ref{NaiveWhitham}) conveys the fact that the multi-periodic $L$ varies only on the long time scale, $T_k$. This form is an approximate solution to the two dimensional Toda lattice\ equations, which becomes exact in the limit $\varepsilon\to0$. Indeed, in this limit a variation of the argument of $L$ by $d\tau$  is accompanied by variation of the argument of $\bm I$ by $\varepsilon d\tau \to 0,$  and the multi-periodic wave becomes unmodulated. The form (\ref{NaiveWhitham}) is, however, rather naive, in that it is not a good basis as a zeroth order term in an expansion in the small parameter $\varepsilon$.  A better ansatz is given by:

\begin{align}
L = L(\varepsilon^{-1}\vec{ \mathcal{W}}\left(\{T_k\}\right),\hat P,\bm I(T_k)),\label{lessNaive}
\end{align}
with
\begin{align}
\frac{\partial \vec{ \mathcal{W}}}{\partial T_k} =\vec \omega_k, \quad \frac{\partial \vec{\mathcal W}}{\partial \bar T_k} = \overline{\vec \omega_k}.\label{VectorActionDerivative}
\end{align}
This ansatz also coincides with an unmodulated solution in the limit $\varepsilon\to0$, but takes better account of the change of the phase of the multi-periodic wave as the wave is modulated. 

Note that the integrability of  equations (\ref{VectorActionDerivative})  imply:
\begin{align}
\frac{\partial \vec \omega_k}{\partial T_l} = \frac{\partial \vec\omega_l}{\partial T_k}, \quad\frac{\partial \vec \omega_k}{\partial \bar T_l} = \frac{\partial\overline{ \vec\omega_l}}{\partial  T_k} ,\label{ConservationOfWaves}
\end{align}
and the complex conjugates of these equations. These equations called `conservation of waves' constitute constraints on the way the Riemann surface may change. These constraints are not enough to determine the dynamics, but shall be useful below to obtain the modulation equations in full. 

We shall need two more ingredients in order to proceed. First, we shall want to include the first order correction term to (\ref{lessNaive}), which we shall denote by $\varepsilon L_1$. Second, we shall need in the sequel derivative of $L_0$ with respect to the slow times which appears in $\bm I$ only. Namely, neglecting the derivative of the slow times in  $\vec{\mathcal{W}}$. It is thus useful to introduce the times $\tilde T_k$ as follows
\begin{align}
L&=L_0\left(\varepsilon^{-1}\vec{ \mathcal{W}}\left(\{T_k\}\right),\hat P,\bm I\left(\{\tilde T_k\}\right)\right)+\varepsilon L_1 +\dots, \label{FinalWhithamExpansionL}
\end{align}
and take $T_k=\tilde{T}_k$ at the end of the calculation. The operator $\hat P$\ is still given by (\ref{Plong}), as the shift operator in (\ref{LasShifts}) does not depend on $\{\tilde T_k\}$, since it does not hit $\bm I$.  

Note that, as a result of the three-timing method, a derivative with respect to time, $\partial_k$, which was taken before the slow times were introduced and before these slow times were split into $T_k$ and $\tilde T_k$, becomes the expression $\left.\partial_k +\varepsilon \frac{\partial}{\partial T_k} +\varepsilon \frac{\partial}{\partial \tilde T_k}\right|_{\tilde T_k=T_k},$ after the introduction of the new times. 

Finally, note also that a similar expansion holds for $L^{(l)}$. We write this expression for future reference:
\begin{align}
L^{(l)}&=L_0^{(l)}\left(\varepsilon^{-1}\vec{ \mathcal{W}}\left(\{T_k\}\right),\hat P,\bm I\left(\{\tilde T_k\}\right)\right)+\varepsilon L_1^{(l)} .\label{FinalWhithamExpansionLl} \end{align}
We shall use the boldface $\bm T_l$ to denote both times $T_l$ and $\tilde T_l$. Furthermore we shall write:
\begin{align}
\frac{\partial}{\partial \bm T_l}= \frac{\partial}{\partial T_l} + \frac{\partial}{\partial \tilde T_l}.
\end{align}

To obtain (\ref{WhithamEqsforHs}) as the semiclassical limit of the zero curvature conditions, Eq. (\ref{ZeroCurvatureDispersionfull}), we start from the latter equations and expand them in $\varepsilon$. We have:
\begin{align}
\left[\frac{\partial}{\partial t_l} +\varepsilon\frac{\partial}{\partial\bm  T_l} - L^{(l)},\frac{\partial}{\partial t_k} +\varepsilon\frac{\partial}{\partial \bm T_k} - L^{(k)}  \right]=0.
\end{align} 
Substituting form $L^{(l)}$ and $L$ the expansions (\ref{FinalWhithamExpansionL},\ref{FinalWhithamExpansionLl}), and $\hat P=\hat p+\varepsilon \partial_0,$ and then expanding in $\varepsilon$, one obtains that the zero order terms in $\varepsilon$ vanish due to (\ref{ZeroCurvatureDispersionfull}), while the first order term in $\varepsilon$\ yields the equation:
\begin{align}
F=G,\label{WhithamF=Garbage}
\end{align}
with 
\begin{align}
F&=\frac{\partial L^{(k)}_0}{\partial \bm T_l} -\frac{\partial L^{(l)}_0}{\partial\bm  T_k} - \{L_0^{(l)},L^{(k)}_0\}\label{Fdef}\\
G&=\frac{\partial L^{(k)}_1}{\partial t_l}   -\frac{\partial L^{(l)}_1}{\partial t_k}-[L^{(l)}_0,L^{(k)}_1]-[L^{(l)}_1,L^{(k)}_0].\label{Gdef}
\end{align}
The Poisson bracket on the left hand side of (\ref{WhithamF=Garbage}) is defined for any two function of $\hat P$ and $T_0$ as follows:
\begin{align}
\{A(\hat p,\bm T_0),B(\hat p,\bm T_0)\} = \frac{\partial A}{\partial\hat p}\frac{\partial B}{\partial \bm T_0} -\frac{\partial B}{\partial \hat p}\frac{\partial A}{\partial\bm  T_0} ,
\end{align}
and  it assumed that the operators
act  on  functions of the fast variables only. Note also that $\hat P \equiv \frac{\partial}{\partial t_0} + \varepsilon \frac{\partial}{\partial T_0}$ is equal to leading order to  $\hat p \equiv \frac{\partial }{\partial t_0}$one may replace $\hat P$ by $\hat p$ in (\ref{WhithamF=Garbage}).

The appearance of the Poisson brackets in (\ref{WhithamF=Garbage}) is related to the expansion with respect to $\varepsilon$ of functions of  $\hat P\equiv \hat p + \varepsilon \frac{\partial}{\partial \bm T_0}$ . Indeed, given $A(\hat P,\bm T_0)$ and $B(\hat P,\bm T_0)$, one may expand $A(\hat P, \bm T_0) = A(\hat p,\bm T_0)+\varepsilon\frac{\partial A(\hat P,\bm T_0)}{\partial \hat P}  \frac{\partial}{\partial \bm T_0} +O(\varepsilon^2),$ with a similar expansion for $B$. One then obtains:
\begin{align} 
&[A(\hat P,\bm T_0),B(\hat P,\bm T_0)] =[A(\hat p,\bm T_0),B(\hat p,\bm T_0)]+\varepsilon\{A(\hat p,\bm T_0),B(\hat p,\bm T_0)\}+\\
&+\varepsilon\left(\left[A,\frac{\partial B}{\partial\hat P}\right]- \left[B,\frac{\partial A}{\partial\hat P}\right]\right)\frac{\partial}{\partial \bm T_0}+O(\varepsilon^2).\nonumber
\end{align}
Since we assume the operators act on a function of fast times only, we can drop the third term on the right hand side. 

 \subsection{Averaging}
 We define the averaging of any matrix, $A$, as:
\begin{align}
&\<A\> =\lim_{L\to\infty}\frac{1}{(2L)^J} \sum_{n=n_0-L}^{n_0+L}\sum_{i,j \in\mathds{Z}}\int_{[-L,L]^J} A_{nn}\prod_{k=1}^J dt_k.
\end{align}
The scale $L$ is to be intermediate between the fast and slow scale. Namely, $L\varepsilon$, must be thought as a small number. Since this small number does not, however, scale with $\varepsilon$, for the purpose of the expansion in $\varepsilon$, one may assume  $L^{-1}\sim\varepsilon.$  Note that in the limit $L\to \infty$, the sum over $n$ becomes a trace, and thus one expects:
\begin{align}
\<[A,B]\> \sim \frac{1}{L} \sim \varepsilon. 
\end{align}
For this to be true, it is necessary that the object being varied over, $[A,B],$ is bounded within the domain being averaged over. 

An important average that we shall encounter will be of the form: $\<[\hat t_0 A,B]\>,$ with $\hat t_0$ being a matrix with elements given by: 
\begin{align}
(\hat t_0)_{nm} = n \delta_{nm} .\label{hatt0Def} 
\end{align}
We shall need an alternative representation of such an average. For this purpose note that the average can explicitly written as:
\begin{align}
\<[\hat t_0 A,B]\> =\sum_j \sum_{n=n_i}^{n_f} \left\< nA_{n,n+j}B_{n+j,n} -(n-j)A_{n-j,n}B_{n,n-j}\right\>_{\{t_k\}},
\end{align}
where $n_i=n_0-L$ and $n_f=n_0+L$. Here the average on the right hand side, denoted by $\<...\>_{\{t_k\}},$ is an average over the times $t_k,$  with $k>0$, only. This expression may be further simplified by writing:
\begin{align}
\<[\hat t_0 A,B]\> =\sum_j \left(\sum_{n=n_f-j}^{n_f} -\sum_{n=n_i}^{n_i+j} \right)\left\< nA_{n,n+j}B_{n+j,n} \right\>_{\{t_k\}}.
\end{align}
Since $j\ll L$, one may replace $n$ by $n_f$ in the first sum and $n$ by $n_i$ in the second sum. In addition, if $A$ and $B$\ are multi-periodic, all the terms in the first sum are equal to each other in leading order, due to ergodicity, and the same is true for the second sum. Taking all these considerations into account, we may write:
\begin{align}
&\<[\hat t_0 A,B]\> = \sum_jj\left\< n_fA_{n_f,n_f+j}B_{n_f+j,n_f} -n_iA_{n_i,n_i+j}B_{n_i+j,n_i}   \right\>_{\{t_k\}} +O(\varepsilon), 
\end{align}
which in turn can be written as:
\begin{align}
\<[\hat t_0 A,B]\> = \sum_{n=n_i}^{n_f-1}\left\<  (n+1)(A[\hat t_0,B])_{n+1,n+1}-  n(A[\hat t_0,B])_{n,n}\right\>_{\{t_k\}}  +O(\varepsilon). 
\end{align} 
The fact that this equation is equal to the one preceding it easily follows by returning to a representation of $(A[\hat t_0,B])_{n,n}$ as $\sum_j jA_{n,n+j}B_{n+j,n},$ and similarly for  $(A[\hat t_0,B])_{n+1,n+1}$.  Ergodicity implies that, to a high precision, $\left\<  (A[\hat t_0,B])_{n+1,n+1}\right\>_{\{t_k\}}=\left\< (A[\hat t_0,B])_{n,n}\right\>_{\{t_k\}}$, which leads to:
\begin{align}
\<[\hat t_0 A,B]\>=\<A[\hat t_0,B]\> + O(\varepsilon).
\end{align} 
Finally, we may use the equation $[\hat t_0 ,B]=-\frac{\partial B}{\partial \hat p }$, which holds generally to write:
\begin{align}
\<[\hat t_0 A,B]\>=-\left\<A \frac{\partial B}{\partial\hat p}\right\> + O(\varepsilon)=-\left\< \frac{\partial B}{\partial\hat p}A\right\> + O(\varepsilon)\label{t0averageing},
\end{align} 
where in the last equation we used the fact that the average over the commutator of two mulit-periodic functions is small. This equation, which holds for quasi-periodic, $A$\ and $B$, will be useful in the following. 

\subsection{Krichever's derivation of Whitham}
  The algebro-geometric version, due to Krichever \cite{83:Krichever:Averaging} (see also \cite{Fucito:Gamba:NonlinearWKB}) of the Whitham averaging method, consists of averaging the nonlinear equations with the Baker-Akhiezer function. Thus, we shall consider objects such as $\<A\vec\psi^+ \vec\psi^{-t}\>$. Here $t$ denotes the transpose, such that $A\vec\psi^+ \vec\psi^{-t} $ is a matrix whose $mn$  element is $\sum_i A_{mi} \psi^+_{i} \psi^-_n.$ Furthermore, the Baker-Akhiezer function is assumed to depend on the fast times through the variables $\{t_k\}$ that appear explicitly in  (\ref{BAExplicit}), the dependence on the slow times is implicit in this explicit formula, through the meromorphic differentials, $H_k$, the frequencies $\vec \omega_k$, and indeed the definition of the Riemann theta function, Eq. (\ref{ThetaFunctionDef}), which depends on $\bm B$, which is itself a function of the slow times. Note that 
\begin{align}
(\vec \psi^+ \vec \psi^{-t})_{n,n+j} = \Phi^+_n \Phi^-_{n+j} e^{j  {H}_0},
\end{align}
and following the multi-periodicity of $\Phi^\pm$,  the matrix elements of $\psi^+ \vec \psi^{-t}$ are also multi-periodic.

Applying this averaging procedure to  (\ref{WhithamF=Garbage}), we shall show that  one obtains:
\begin{align}
0=\<G\vec\psi^+\vec\psi^{-t}\>=\<F\vec\psi^+\vec\psi^{-t}\>,\label{AvarageFis0}
\end{align}
with $F$\ and $G$\ defined in (\ref{Fdef}) and (\ref{Gdef}), respectively. We shall show this by showing that $\<G\vec\psi^+\vec\psi^{-t}\>=0 $, and then $F=G$ gives the second equality. Indeed, consider  averaging  applied to $G$. One easily obtains from (\ref{LaxBA}):
\begin{align}
\frac{\partial}{\partial t_l}  \left( {L_1^{(k)}} \vec\psi^+  \vec\psi^{-t}\right) =\frac{\partial{L_1^{(k)}} }{\partial t_l} \vec\psi^+ \vec \psi^{-t}+\vec  L_1^{(k)}L_0^{(l)} \vec\psi^+ \vec \psi^{-t}-L_1^{(k)} \vec\psi^+ \vec \psi^{-t}  L_0^{(l)}   .
\end{align}
Averaging this equation, the left hand side vanishes to leading order as an average over a full derivative, and one obtains:
\begin{align}
0=\left\< \frac{\partial{L_1^{(k)}} }{\partial t_l} \vec\psi^+ \vec \psi^{-t}+L_1^{(k)}L_0^{(l)} \vec\psi^+ \psi^{-t}-L_1^{(k)} \vec\psi^+ \vec \psi^{-t}  L^{(l)}_0\right\> .
\end{align}
This can be written as:

\begin{align} 
0=\left\< \frac{\partial{L_1^{(k)}} }{\partial t_l} \vec\psi^+ \vec \psi^{-t}+\left[L_1^{(k)},L_0^{(l)}\right] \vec\psi^+ \psi^{-t}+\left[L_0^{(l)},L_1^{(k)} \vec\psi^+ \vec \psi^{-t}  \right]\right\> .\label{Firstordering}
\end{align}
As an average over a  commutator of multi-periodic functions, the last term on the right hand side vanishes in the limit $L\to\infty  $ (or rather is of higher order in $\varepsilon$).  If we follow the same procedure with $k$ and $l$ interchanged, one obtains $\<G\vec\psi^+\vec\psi^{-t}\> =0$ and thus $\<F\vec\psi^+\vec\psi^{-t}\> =0$

The specific terms in the equation $\<F\vec\psi^+\vec\psi^{-t}\> =0,$  when one inserts the definition of $F$, Eq. (\ref{Fdef}), can be further developed to obtain the modulation equations we seek, Eq. (\ref{WhithamEqsforHs}). We start with the first term in the definition of $F$, that is the first term on the right hand side of  Eq. (\ref{Fdef}). To be able to re-write the average over it with the Baker-Akhiezer functions, one first considers the following objects:
\begin{align} 
&\frac{\partial}{\partial\bm T_l}\frac{\partial}{\partial t_k}\left( \vec\psi^+ \vec\psi'^{-t}\right)
=\\
&=\frac{\partial}{\partial\bm T_l}\left(L^{(k)}\vec\psi^+ \vec\psi'^{-t}-\vec\psi^+ \vec\psi'^{-t} L'^{(k)}\right)\nonumber=\\
&=\nonumber\frac{\partial}{\partial\bm T_l}\left(\left(L^{(k)}-L'^{(k)}\right)\vec\psi^+ \vec\psi'^{-t}-\left[ \vec\psi^+ \vec\psi'^{-t} ,L'^{(k)}\right]\right),
\end{align} 
where the prime denotes taking the function at slow times $\{\bm T'_j\}$ rather than at $\{\bm T_j\}$.  
We may now take $\{\bm T_j\}=\{\bm T_j'\}$ in this equation to obtain:
\begin{align}
\frac{\partial}{\partial\bm T_l}\frac{\partial}{\partial t_k}\left(\vec\psi^+ \vec\psi^{-t}\right)=\frac{\partial L^{(k)}}{\partial\bm T_l}\vec\psi^+ \vec\psi^{-t}+\left[\frac{\partial \vec\psi^+ }{\partial \bm T_l} \vec\psi^{-t} ,L^{(k)}\right] .\label{}
\end{align}
The derivative $\frac{\partial \vec\psi^+}{\partial\bm T_l}$ reads explicitly:
\begin{align}
\frac{\partial \vec\psi^+}{\partial\bm T_l} = \sum_{m>0} t_m\left( \frac{\partial {H}_m}{\partial T_l}  + \frac{\partial \vec \omega_m}{\partial T_l} \cdot \vec \nabla^{\vec w} \right)\vec\psi^+ +\hat t_0\left( \frac{\partial {H}_0}{\partial T_l}  + \frac{\partial \vec \omega_0}{\partial T_l} \cdot \vec \nabla^{\vec w} \right)\vec\psi^++ \frac{\partial\vec\psi^+}{\partial \tilde T_l},\label{SlowTimeBADervi}
\end{align}
where $\hat t_0$ is defined in (\ref{hatt0Def}) and the gradient $\vec \nabla^{\vec w}$ denotes taking the gradient derivative of $\Phi^\pm$ with respect to its first, vector, argument. Namely, 
\begin{align}
\vec\nabla^{\vec w}\psi_n^+ = e^{2\Re\left(\sum t_k  {H}_k\right)} \left.\vec\nabla^{\vec w}\Phi^+(\vec w)\right|_{\vec w =\vec w(\{t_k\})} .
\end{align}

Inserting (\ref{SlowTimeBADervi}) into (\ref{Firstordering}) and making use of  (\ref{t0averageing}) gives:
\begin{align}
&\frac{\partial}{\partial\bm T_l}\frac{\partial}{\partial t_k}\left\<\vec\psi^+ \vec\psi^{-t}\right\>=\left\< \frac{\partial L^{(k)}}{\partial\bm T_l}\vec\psi^+ \vec\psi^{-t}-\frac{\partial L^{(k)}}{\partial \hat p} \left( \frac{\partial {H}_0}{\partial T_l}  + \frac{\partial \vec \omega_0}{\partial T_l} \cdot \vec \nabla^{\vec w} \right)\vec\psi^+ \vec\psi^{-t}\right\>.\label{firstorderingaveraged}
\end{align}

On the other hand taking the opposite order of the derivatives in (\ref{Firstordering}) one obtains:
\begin{align}
&\frac{\partial}{\partial t_k}\frac{\partial}{\partial\bm T_l}\vec\psi ^+ \vec{\psi}^{-t}=\frac{\partial}{\partial t_k}\left(  \left(  \sum_mt_m\left( \frac{\partial {H}_m}{\partial T_l}  + \frac{\partial \vec \omega_m}{\partial T_l} \cdot \vec \nabla^{\vec w} \right)+\frac{\partial}{\partial \tilde T_l}\right)\vec\psi ^+ \vec{\psi}^{-t} \right).
\end{align}
Averaging over this equation, and discarding averages over full derivatives (note that one averages over all times and thus, e.g. $\<t_m\partial_k\dots\>=0,$ by ergodicity, even for $k=m$.), one arrives at:
\begin{align}
\frac{\partial}{\partial t_k}\frac{\partial}{\partial\bm T_l}\left<\vec\psi^\dagger \vec\psi\right\> =\left\<  \left( \frac{\partial {H}_k}{\partial T_l}  + \frac{\partial \vec \omega_k}{\partial T_l} \cdot \vec \nabla^{\vec w} \right)\vec\psi ^+ \vec{\psi}^{-t}\right\>\label{secondorderingaveraged}
\end{align} 
Equating (\ref{firstorderingaveraged}) and (\ref{secondorderingaveraged}) one obtains:
\begin{align}
&\left\<  \frac{\partial L^{(k)}}{\partial\bm T_l}\vec\psi ^+ \vec{\psi}^{-t}\right\>=\\&=\left\<   \left( \frac{\partial {H}_k}{\partial T_l}  + \frac{\partial \vec \omega_k}{\partial T_l} \cdot \vec \nabla^{\vec w} \right)\vec\psi ^+ \vec{\psi}^{-t}\right\>-\left\< \frac{\partial L^{(k)}}{\partial \hat p}\left( \frac{\partial {H}_0}{\partial T_l}  + \frac{\partial \vec \omega_0}{\partial T_l} \cdot \vec \nabla^{\vec w} \right)\vec\psi ^+ \vec{\psi}^{-t}\right\>, \nonumber
\end{align}
giving a representation of the average over the first term in the definition of $F$, Eq. (\ref{Fdef}). In that respect note that exchanging $L_0^{(k)}$ by $L^{(k)}$ has no consequence to leading order.  

We may subtract this equation the same equation where the roles of $k$ and $l$ are reversed. Making use of the conservation of waves, Eq. (\ref{ConservationOfWaves}), one obtains:
\begin{align}\label{AveragedMixedDerivatives}
&\left\<  \left(\frac{\partial L^{(k)}}{\partial\bm T_l}-\frac{\partial L^{(l)}}{\partial\bm T_k}\right)\vec\psi ^+ \vec{\psi}^{-t}\right\>=\\
&=\left(\frac{\partial {H}_k}{\partial T_l}-\frac{\partial {H}_l}{\partial T_k}\right)\left\< \vec\psi ^+ \vec{\psi}^{-t}\right\>-\frac{\partial {H}_0}{\partial T_l}  \left\<\frac{\partial L^{(k)}}{\partial \hat p} \vec\psi ^+ \vec{\psi}^{-t}\right\>+\frac{\partial {H}_0}{\partial T_k}  \left\<\frac{\partial L^{(l)}}{\partial \hat p}\vec\psi ^+ \vec{\psi}^{-t}\right\>\nonumber+\\
&+\frac{\partial \vec \omega_0}{\partial T_k}\left\<   \frac{\partial L^{(l)}}{\partial \hat p} \vec \nabla^{\vec w}\vec\psi ^+ \vec{\psi}^{-t}\right\> -\frac{\partial \vec \omega_0}{\partial T_l}\left\<   \frac{\partial L^{(k)}}{\partial \hat p} \vec \nabla^{\vec w}\vec\psi ^+ \vec{\psi}^{-t}\right\> \nonumber 
\end{align}

To deal with the average of the third term in the definition of $F$, Eq. (\ref{Fdef}), one considers:
\begin{align}
&\frac{\partial}{\partial T'_0}\left\<\frac{\partial}{\partial t_l}\left( \frac{\partial L^{(k)}}{\partial \hat p}\vec{\psi}'^+ \vec{\psi}^{-t}\right) -\frac{\partial}{\partial t_k}\left( \frac{\partial L^{(l)}}{\partial \hat p}\vec{\psi}'^+ \vec{\psi}^{-t}\right) -G\vec{\psi}'^+ \vec{\psi}^{-t}\right\>=\\&=
\frac{\partial}{\partial T'_0}\left\< \frac{\partial L^{(k)}}{\partial \hat p} \left(L'^{(l)}-L^{(l)} \right) \vec{\psi}'^+ \vec{\psi}^{-t} -\frac{\partial L^{(l)}}{\partial \hat p} \left(L'^{(k)}-L^{(k)} \right)  \vec{\psi}'^+ \vec{\psi}^{-t} \right\> ,\nonumber
\end{align}
where $G$ is defined in (\ref{Gdef}). Taking $\{\bm T\}=\{\bm T'\},$ one obtains:
\begin{align}
 \left\< \frac{\partial}{\partial t_l}\left(\frac{\partial L^{(k)}}{\partial \hat p}\frac{\partial  \vec\psi^+ }{\partial T_0}\vec{\psi}^{-t}\right)\right\> - \left\<\frac{\partial}{\partial t_k}\left( \frac{\partial L^{(l)}}{\partial \hat p}\frac{\partial  \vec\psi^+ }{\partial T_0} \vec{\psi}^{-t}\right)\right\> =\<\{L^{(k)},L^{(l)}\} \vec \psi ^+ \vec{\psi}^{-t} \>,\label{AveragePoissonFirstway}
\end{align}
where  that fact that $\<G\vec\psi^+\vec\psi^{-t}\>=0$  for all $T_0$ has been used. One may compute the left hand side of (\ref{AveragePoissonFirstway})  by first applying the derivatives with respect to $T_0$ and only then taking the derivative with respect to $t_l$ or $t_k$. Much the same methods as were applied to obtain (\ref{AveragedMixedDerivatives}) allow one to write:
\begin{align} 
&\label{AveragedPoisson}\<\{L^{(k)},L^{(l)}\}\vec \psi^+\vec \psi^{-t} \> =\frac{\partial {H}_l}{\partial T_0}  \left\<\frac{\partial L^{(k)}}{\partial \hat p} \vec\psi ^+ \vec{\psi}^{-t}\right\>-\frac{\partial {H}_k}{\partial T_0}  \left\<\frac{\partial L^{(l)}}{\partial \hat p}\vec\psi ^+ \vec{\psi}^{-t}\right\>+\\
&+\frac{\partial \vec \omega_l}{\partial T_0}\left\<   \frac{\partial L^{(k)}}{\partial \hat p} \vec \nabla^{\vec w}\vec\psi ^+ \vec{\psi}^{-t}\right\> -\frac{\partial \vec{\omega}_k}{\partial T_0}\left\<   \frac{\partial L^{(l)}}{\partial \hat p} \vec \nabla^{\vec w}\vec\psi ^+ \vec{\psi}^{-t}\right\> \nonumber 
\end{align} 

We are now in a position  to write down $\<F\vec\psi^+\vec\psi^{t-}\>,$by combining  (\ref{AveragedMixedDerivatives}) and (\ref{AveragedPoisson}) with the definition of $F$, Eq. (\ref{Fdef}). Taking into account the conservation of waves, Eq. (\ref{ConservationOfWaves}), yields:
\begin{align}
&0\label{almostKricheverWhitham}=\left(\frac{\partial {H}_k}{\partial T_l}-\frac{\partial {H}_l}{\partial T_k}\right)\left\< \vec\psi ^+ \vec{\psi}^{-t}\right\>+\left(\frac{\partial {H}_l}{\partial T_0}-\frac{\partial {H}_0}{\partial T_l}   \right)\left\<\frac{\partial L^{(k)}}{\partial \hat p} \vec\psi ^+ \vec{\psi}^{-t}\right\>+\\&\nonumber+ \left(\frac{\partial {H}_0}{\partial T_k}-\frac{\partial {H}_k}{\partial T_0}\right)  \left\<\frac{\partial L^{(l)}}{\partial \hat p}\vec\psi ^+ \vec{\psi}^{-t}\right\> 
\end{align}
To further simplify this equation consider:
\begin{align}
&0=\left.\left\<\frac{\partial}{\partial z'} \frac{\partial}{\partial t_k } \left(\vec \psi'^+ \vec \psi^{-t}\right)\right\>\right|_{\tiny\begin{array}{c}z=z'\\ \{\bm T\}=\{\bm T'\} \end{array}}-\left\< \frac{\partial}{\partial t_k}\left(\frac{\partial \vec \psi^+}{\partial z} \vec \psi^{-t}\right)\right\>=\\&=\left\< \left[\frac{\partial \vec \psi^+}{\partial z} \vec \psi^{-t},L^{(k)}\right]\right\>-{H}_k \left\< \vec\psi^+\vec\psi^{-t}\right\> =\nonumber \\&=\nonumber -H_0 \left\<\frac{\partial L^{(k)}}{\partial \hat p} \vec\psi ^+ \vec{\psi}^{-t}\right\>-{H}_k \left\< \vec\psi^+\vec\psi^{-t}\right\>,
\end{align}
which combined with (\ref{almostKricheverWhitham}) gives:
\begin{align}
0=\left(\frac{\partial {H}_k}{\partial T_l}-\frac{\partial {H}_l}{\partial T_k}\right)H_0-\left(\frac{\partial {H}_0}{\partial T_k}  - \frac{\partial {H}_k}{\partial T_0}\right)H_l-\left(\frac{\partial {H}_l}{\partial T_0}-\frac{\partial {H}_0}{\partial T_l}\right) H_k. \label{KricheverWhithamasSum}
\end{align}
These equation must be true for all $z$ on the Riemann surface. Examining the analytic properties of each term of this equation, one obtains that an over-determined set of conditions on the functions  multiplying $H_0$, $H_l$ and $H_k$ must be satisfied if the right hand side is to be equal to zero, implying that each summand is zero by itself. Leading to:
\begin{align}
\frac{\partial {H}_i}{\partial T_j}-\frac{\partial {H}_j}{\partial T_i}=0.
\end{align}
for any $i$ and $j$ (including any two  chosen from the set $\{k,l,0\}),$ which is equivalent to the first equation in  (\ref{WhithamEqsforHs}). The same methods may be applied to obtain the second equation, namely:
\begin{align}
\frac{\partial {\bar H}_i}{\partial T_j}-\frac{\partial {H}_j}{\partial \bar T_i}=0.
\end{align}
We shall give details of how (\ref{StringEq}) is obtained from (\ref{QStringEq}) by averaging since much the same methods are employed here as well.

\bibliographystyle{unsrt}

\bibliography{mybib}

\end{document}